\documentclass{aa}  

\usepackage{graphicx}
\usepackage{natbib}
\bibpunct{(}{)}{;}{a}{}{,} 

\usepackage{txfonts}

\usepackage{xcolor,colortbl}
\definecolor{Gray}{gray}{0.85}

\begin{document}

   \title{Stellar chemo-kinematics of the Cetus dwarf spheroidal galaxy \thanks{Based on observations made with ESO telescopes at the La Silla Paranal Observatory as part of the program 090.B-0284(B).}
        }
   
   \subtitle{}

   \author{S. Taibi \inst{1}\fnmsep \inst{2} \fnmsep \thanks{\email{staibi@iac.es}}, 
        G. Battaglia\inst{1}\fnmsep \inst{2}, 
        N. Kacharov\inst{3}, 
        M. Rejkuba\inst{4,5}, 
        M. Irwin\inst{6},
        R. Leaman\inst{3},
        M. Zoccali\inst{7,8}, 
        E. Tolstoy\inst{9},
        P. Jablonka\inst{10,11}
   }
   
   \institute{Instituto de Astrofisica de Canarias, C/ Vía Láctea s/n, E-38205, La Laguna, Tenerife, Spain
        \and
        Departamento de Astrofísica, Universidad de La Laguna, E-38205, La Laguna, Tenerife, Spain
        \and
        Max Planck Institute for Astronomy, K\"{o}nigstuhl 17, D-69117 Heidelberg, Germany
        \and
        European Southern Observatory, Karl-Schwarzschild Strasse 2, D-85748 Garching, Germany
        \and
        Excellence Cluster Origin and Structure of the Universe, Boltzmannstr. 2, D-85748 Garching bei München, Germany       
        \and
        Institute of Astronomy, University of Cambridge, Madingley Road, CB3 0HA, Cambridge, UK
        \and
        Instituto de Astrofísica, Pontificia Universidad Católica de Chile, Av. Vicuña Mackenna 4860, 782-0436 Macul, Santiago, Chile
        \and
        Millennium Institute of Astrophysics, Av. Vicuña Mackenna 4860, 782-0436 Macul, Santiago, Chile
        \and
        Kapteyn Astronomical Institute, University of Groningen, 9700AV Groningen, The Netherlands
        \and
        Institute of Physics, Laboratory of Astrophysics, École Polytechnique Fédérale de Lausanne (EPFL), 1290 Sauverny, Switzerland
        \and
        GEPI, CNRS UMR 8111, Observatoire de Paris, PSL Research University, F-92125, Meudon, Cedex, France
   }

   \date{Received M DD, YYYY; accepted M DD, YYYY}

 
  \abstract
         {The great majority of early-type dwarf galaxies, in the Local Group as well as in other galaxy groups, are found in the vicinity of much larger galaxies, making it hard to disentangle the role of internal versus external effects in driving their evolution.}
        {In order to minimize environmental effects and gain an insight into the internal mechanisms that shape the properties of these systems, we study one of the few dwarf spheroidal galaxies of the Local Group found in isolation: Cetus. This system is of particular interest since it does not follow the Local Group morphology-density relation.}
        {We obtained Very Large Telescope (VLT) FORS2 spectra ($R\sim2600$) in the region of the nIR CaII triplet lines for 80 candidate red giant branch (RGB) stars. The analysis yielded line-of-sight velocities and metallicities ([Fe/H]) for 54 bona fide member stars.}
        {The kinematic analysis shows that Cetus is a mainly pressure-supported ($\sigma_v = 11.0_{-1.3}^{+1.6}$ km/s), dark-matter-dominated system ($M_{1/2}/L_V = 23.9_{-8.9}^{+9.7} M_\odot/L_\odot$) with no significant signs of internal rotation. 
         We find Cetus to be a metal-poor system with a significant metallicity spread (median [Fe/H] = -1.71 dex, median-absolute-deviation = 0.49 dex), as expected for its stellar mass. We report the presence of a mild metallicity gradient compatible with those found in other dwarf spheroidals of the same luminosity; we trace the presence of a stellar population gradient also in the spatial distribution of stars in different evolutionary phases in ancillary SuprimeCam photometric data. There are  tentative indications of two chemo-kinematically distinct sub-populations, with the more metal-poor stars showing a hotter kinematics than the metal-richer ones. Furthermore, the photometric dataset reveals the presence of a foreground population that most likely belongs to the Sagittarius stream.}
        {This study represents an important step forward in assessing the internal kinematics of the Cetus dwarf spheroidal galaxy as well as the first wide-area spectroscopic determination of its metallicity properties. With our analysis, Cetus adds to the growing scatter in stellar-dark matter halo properties in low-mass galactic systems. The presence of a metallicity gradient akin to those found in similarly luminous and pressure-supported systems inhabiting very different environments may hint at metallicity gradients in Local Group early-type dwarfs being driven by internal mechanisms.}

   
   \titlerunning{Chemo-dynamics of Cetus dSph}
   \authorrunning{S. Taibi et al.}
   
   \maketitle
%
   
\section{Introduction} \label{sec:intro}
The study of dwarf galaxies is of great importance to understand the evolution at the low-mass end of the galaxy mass function. In the Local Group (LG), the great majority of dwarf galaxies are gas-poor spheroidals found to be satellites of the Milky Way (MW) or M31, while the rest are gas-rich systems, with an irregular optical morphology, typically found in isolation. This morphology-density relation (e.g., \citealp{VanDenBergh1994}) has raised the interesting question of whether dwarf irregulars (dIrr) and dwarf spheroidals (dSph) could share similar progenitors, which evolved differently  as a consequence of environmental effects. Numerical simulations (see \citealp{Mayer2010} review) have shown that strong environmental effects such as close and repeated encounters with a large host galaxy may induce a transformation both of the system morphology and internal kinematics, driven by tidal and ram-pressure stripping and aided by ultraviolet (UV) heating due to re-ionization ("tidal stirring", see e.g., \citealp{Mayer2001,Mayer2006}); removal of only the gaseous component, however, does not require very strong interactions.  In general, the inclusion of internal mechanisms in the simulations, such as stellar feedback, can enhance the impact of environmental effects (e.g., \citealp{Sawala2010, Zolotov2012, Arraki2014, Kazantzidis2017, Revaz&Jablonka2018}).
There have been many observational studies of the star formation history (SFH) of dSph systems (e.g., \citealp{Hidalgo2009,Monelli2010,deBoer2012,Cole2014,Weisz2014}). The  comprehensive analysis by \citet{Gallart2015}, focusing on the systems with the most accurate SFHs back to the earliest times, suggested that the different morphological types may be imprinted by early conditions at the time of formation rather than being exclusively the result of a more recent transformation driven by the environment. The recent determination of orbital parameters for almost all known MW satellites from Gaia DR2 data (e.g., \citealp{Helmi2018, Simon2018, Fritz2018}) suggests that the low orbital eccentricities and relatively large pericenter distances of dSphs of comparable luminosity to dIrrs disfavor a tidal-stirring origin of dSphs \citep{Helmi2018}.

Detailed studies of the internal kinematics and chemical properties of the various morphological types of dwarf galaxies placed in different environments are crucial for identifying similarities and differences between these classes of systems, which can then be used to understand their evolutionary paths. Thanks to their vicinity, MW's dSphs have been studied in great detail (see \citealp*{Tolstoy2009} and references therein, but also \citealp{Walker2009, Battaglia2011, Kirby2011, Lemasle2012, Lemasle2014, Hendricks2014}). However, it remains difficult  to constrain to what extent their present-day observed properties are mainly caused by internal or environmental mechanisms. 
The study of isolated systems allows us to minimize the impact of external effects and gain insight into the internal mechanisms that shape the properties of dwarf galaxies (see e.g., \citealp{Leaman2013, Kirby2014, Kacharov2017}).

In the LG, the great majority of isolated dwarf galaxies are dIrrs and transition-types. Just a handful of dSphs are found in isolation, breaking the general LG morphology-density relation: Cetus - which is the subject of this study, Tucana and And~XVIII. 
It is possible that during their evolution, these dSphs, which appear isolated at present, may have passed once near a large host, but without becoming bound (so-called "backsplash galaxies", e.g., \citealp{Sales2007, Teyssier2012}).  
Nevertheless, they offer a unique opportunity to contrast systems which have spent the great majority of their life in isolation against those that are likely to have experienced repeated environmental interactions, such as the MW and M31 satellites.

The Cetus dSph was discovered by \citet*{Whiting1999} from visual inspection of ESO/SRC survey photographic plates. Follow-up observations disclosed an early-type system, placed in isolation ($D_\odot > 600$ kpc) from both the MW and M31 galaxies. \cite{McConnachie2005} calculated a heliocentric distance of $755\pm23$ kpc using the tip of the red giant branch (TRGB) method, a result subsequently confirmed by \cite{Bernard2009} using RR-Lyrae stars ($D_\odot = 780\pm40$ kpc).
Structural parameters estimated by \cite{McConnachie+Irwin2006}, revealed an extended galaxy with a half-light radius almost double that of MW dSphs of similar luminosity, and one of the largest nominal tidal radii of the LG\footnote{We note that here ``nominal tidal radius'' is to be understood as simply the relevant parameter resulting from a King profile fit to Cetus surface density profile; Cetus in fact shows no evidence of tidal truncation.}.

From deep HST/ACS observations reaching below the oldest main sequence turnoff (oMSTO), \cite{Monelli2010} were able to derive the SFH of the Cetus dSph, finding that it is an old and metal-poor system ($\left \langle \textrm{[Fe/H]} \right \rangle \sim -1.7$ dex), with the SFH peaking around 12 Gyr ago and lasting approximately 2 Gyr, with no stars being formed in the last 8 Gyr; this makes Cetus comparable in age to some of the oldest dSph satellites of the MW, such as, for example, Sculptor. On the spatial scale of the HST/ACS data ($\lesssim$1.5 half-light radii $R_e$, see Table~\ref{table:1}), no age gradient is detected \citep{Hidalgo2013}. 
The analysis of a more spatially extended photometric dataset (\citealp{Monelli2012}, based on VLT/VIMOS observations) revealed the presence of a radial gradient in the RGB and horizontal branch (HB) morphologies, both of which become bluer when moving away from Cetus' center (approximately beyond the half-light radius). On the other hand, from the same VLT/VIMOS dataset, the RR-Lyrae stars did not show any spatial variation of their mean period properties. It appears then that the age and metallicity properties of Cetus are fairly homogeneous in the central regions, but might change in the outer parts. 

The first spectroscopic study of individual stars in Cetus was conducted by \cite{Lewis2007}, using Keck/DEIMOS data of $\sim70$ stars selected from the red giant branch (RGB) of the galaxy and mainly distributed along its optical projected major axis.
The spectroscopic analysis led to the first determination of the systemic velocity ($-87\pm2$ km/s) and velocity dispersion ($17\pm2$ km/s) of the galaxy, revealing also a hint of rotation ($<$ 10 km/s). Furthermore the kinematic analysis excluded any association with nearby HI clouds, thus establishing that Cetus is devoid of gas given the actual observational limits. Metallicity estimates from Ca~II triplet (CaT) lines resulted in agreement with the expectations based on photometry of finding a metal-poor stellar population ([Fe/H] $\sim -1.9$ dex). 

A subsequent spectroscopic study conducted by \cite{Kirby2014} on a larger sample ($\sim 120$ RGB targets), also obtained with the Keck/DEIMOS spectrograph, led to lower values of the systemic velocity ($-83.9\pm1.2$ km/s) and velocity dispersion ($8.3\pm1$ km/s), excluding at the same time any presence of internal rotation. The discrepant results with the previous work of \cite{Lewis2007} were attributed to the different membership selection method implemented and the higher signal-to-noise ratio (S/N) of their sample.

In this work, we present results of a new chemo-kinematic study of the stellar component of the Cetus dSph, based on wide-area VLT/FORS2 MXU spectroscopic observations in the region of the nIR CaT for a sizable sample (80) of individual RGB stars. This is the first study that makes [Fe/H] estimates of stars in the Cetus dSph publicly available. The article is structured as follows. 
In Sect. \ref{sec:acqred} we present the data acquisition and reduction process. In Sect.~\ref{sec:velocities} we describe the determination of line-of-sight velocities for the whole sample. Section~\ref{sec:membership} is dedicated to the criteria applied to select likely member stars to the Cetus dSph. Section~\ref{sec:kinematics} presents the kinematic analysis, where we determine the galaxy systemic velocity and the internal dispersion and search for the possible presence of rotation. In Sect.~\ref{sec:metallicity} we describe the determination of metallicities ([Fe/H]) and the subsequent chemical analysis. In Sect.\ref{sec:struct} we determine the structural properties of the Cetus dSph using ancillary Subaru/SuprimeCam photometric data, while in Sect.~\ref{sec:stream}  we analyze a portion of the Sagittarius stream found in the foreground to the Cetus dSph in the same photometric dataset. Finally, Sect.~\ref{sec:summary} is dedicated to the summary and conclusions.
The parameters adopted for the Cetus dSph are summarized in Table~\ref{table:1}.

\begin{table}
        \caption{Parameters adopted for the Cetus dSph.}             
        \label{table:1}      
        \centering          
        \begin{tabular}{c c c}    
                \hline\hline
                Parameter & Value  & Reference \tablefootmark{$\star$}\\ 
                \hline           
                $\alpha_{J2000}$ & $00^h26^m10.5^s$ & (1) \\
                $\delta_{J2000}$  & $-11^{\circ}02'32''$ & (1) \\
                ellipticity\tablefootmark{a} & 0.33$\pm$0.06 & (1) \\
                P.A. ($^{\circ}$)& $63\pm3$ & (1) \\
                $R_{core} \ (')$ & $1.3 \pm 0.1$ & (1) \\
                $R_{tidal} \ (')$ & $32.0\pm6.5$ & (1) \\
                $R_{e} \ (')$ & $2.7\pm0.1$ & (1) \\
                $M_V$ & $-11.3\pm0.3$ &(1) \\
                $I_{TRGB}$ & $20.39\pm0.03$ & (2) \\
                E(B-V) & 0.029 & (2) \\
                $(m-M)_0$ & $24.39\pm0.07$ & (2) \\
                $D_\odot$ (kpc) & $755\pm23$ & (2) \\
                \hline        
        \end{tabular}
        \tablefoot{\tablefoottext{a}{$\epsilon=1-b/a$} 
                \tablefoottext{$\star$}{\textbf{References: }(1) \cite{McConnachie+Irwin2006}; (2) \cite{McConnachie2005}} 
}
\end{table}

\section{Data acquisition and reduction process} \label{sec:acqred}

 \begin{figure*}
        \centering
        \includegraphics[width=\hsize]{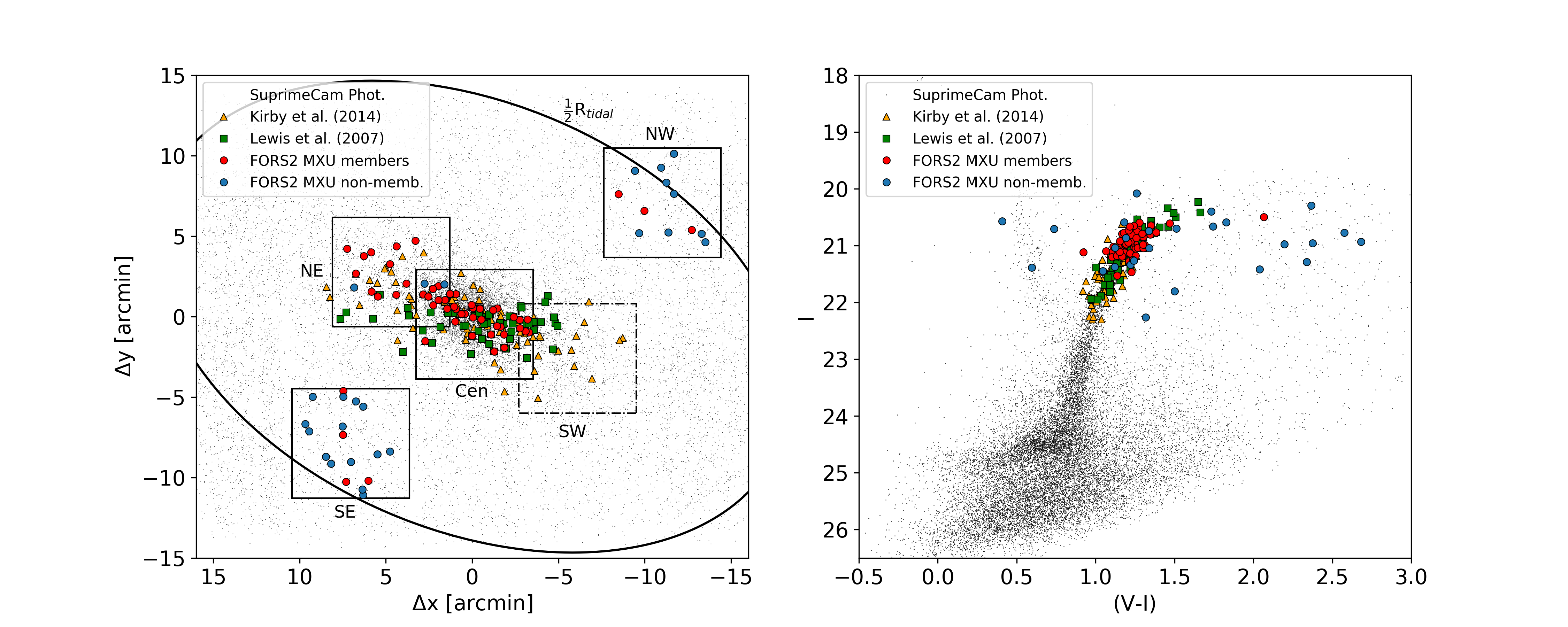}
        \caption{Spatial distribution (left) and color-magnitude diagram (right) of stars along the line-of-sight to the Cetus dSph. Black points represent the objects classified as stars in the Subaru/SuprimeCam photometric data (see main text); FORS2 MXU targets classified as members are marked with red dots, while the non-members are marked with blue dots. The large squares indicate the 4 observed FORS2 pointings, together with the not-observed SW pointing marked as a dot-dashed box. The ellipse indicates the galaxy half tidal radius. Overplotted are also spectroscopic targets classified as probable members in \cite{Kirby2014} (orange triangles) and \cite{Lewis2007} (green squares).}
        \label{FigTarg}
\end{figure*}

 \subsection{Target selection and observations} \label{subsec:targsel}
 The data were obtained using the FORS2 instrument \citep{Appenzeller1998} mounted at the Cassegrain focus of the Very Large Telescope's (VLT) UT1 (Antu) at ESO Paranal Observatory in service mode over several nights of observations between December 2012 and November 2014, as part of the ESO Program 090.B-0284(B), PI: M. Zoccali. 
 
 The FORS2 instrument was set up with the Mask eXchange Unit (MXU), a solution that allows multi-object spectroscopy employing selectable masks with custom cut slits. The targets were selected from Subaru/SuprimeCam imaging data in the Johnson V- and I- band (Subaru Program S05A-015, PI: N. Arimoto), covering Cetus out to more than half its tidal radius ($0.5 \times R_{tidal}= 16'$).
 We selected all the sources flagged as stellar or probably stellar, and with magnitudes and colors compatible with RGB stars at the distance of the Cetus dSph (taken to be $D_\odot = 755\pm23$ kpc, \citealp{McConnachie2005}). Slits to which we could not assign likely RGB stars belonging to Cetus were allocated to random stars within the same magnitude range ($20.5\lesssim I \lesssim 21.5$). To ensure precise slit allocations to our targets, we used short pre-imaging exposures obtained with FORS2 within the same program.
 
Figure~\ref{FigTarg} shows the targets spatial distribution and location on the color-magnitude diagram (CMD)\footnote{Magnitudes and colors reported in this article were not corrected for extinction and reddening, since the photometric information for the RGB stars is only needed in the estimation of [Fe/H], that already takes these effects into account (see Sect.~\ref{sec:metallicity}). However we did consider extinction and reddening when looking at which isochrones were compatible with the Sagittarius stream feature spotted in the Subaru/SuprimeCam photometry (see Sect.~\ref{sec:stream}).}, respectively. 
We have observed 83 objects distributed over four FORS2 pointings (see Fig.~\ref{FigTarg} left), of which the northeast (NE) is aligned along the projected major axis of the galaxy and partially overlaps with the Central field, while the southeast and northwest (SE and NW) are aligned with the projected minor axis. 
Among these objects, three were repeated on purpose on the overlapping Central-NE fields; therefore 80 different stars were observed. We had planned a further pointing along the major axis (namely the southwest one), which, however, was not observed.  
We refer the reader to Table~\ref{table:2} for the observing log: each reported science exposure corresponds to an individual observation block (OB). Several identical OBs were defined for each pointing in order to accumulate S/N necessary for velocity and metallicity measurements.

The instrumental setup and observing strategy we adopted is the same as in \citet{Kacharov2017} (hereafter K17), where the chemo-dynamical properties of the stellar component of the Phoenix transition type galaxy have been studied. Mask slits were designed to be $1''$ wide per $8''$ long ($7''$ in rare cases, to avoid overlap between two adjacent slits) for the Cen and NE fields, while $10''$ long for both the SE and NW. The instrumental setup included a mosaic of two red-sensitive 2k$\times$4k MIT CCDs (pixel size of $15\times15 \mu m$) that together with the Standard Resolution Collimator and a $2\times2$ binning granted a pixel-scale of $\sim 0.25 ''/$pxl and a field of view of $6.8'' \times 6.8''$. We then used the 1028z+29 holographic grism in conjunction with the OG590+32 order separation filter to cover a wavelength range of $7700-9500 \AA$ with a binned spectral dispersion of 0.84 \AA/pxl and a resolving power of $R=\lambda_{cen}/\Delta \lambda =2560$ at $\lambda_{cen}=8600$\AA. The two component chips worked in standard operation mode (high gain with 100kHz readout) having a gain of 0.7 ADUs/$e^-$ and readout noises of 2.9$e^-$ and 3.15$e^-$   for chips 1 and 2, respectively. 

Calibration data (biases, arc lamp, dome flat-field frames) and slit acquisition images were acquired as part of the FORS2 standard calibration plan.
 
\begin{table*}
  \caption{Observing log of VLT/FORS2 MXU observations of RGB targets along the line-of-sight to the Cetus dSph. From left to right, column names indicate: the pointing field name; the field center coordinates; observing date and starting time of the scientific exposure; the exposure time in seconds; the starting airmass; the average DIMM seeing during the exposure in arcsec; the ESO OB fulfillment grades (a full description is reported in the notes below); the number of slits/observed objects per mask. For each field, the mask design remained identical in each OB.
  The total number of slits (83) is reported in the last row of the table. }             
\label{table:2}      
\centering          
\begin{tabular}{c c c c c c c c}    
\hline\hline
Field & Field center (RA, Dec)  & Date / Hour & Exp. time & Airmass & DIMM Seeing & Grade\tablefootmark{$\star$} & Slits\\ 
 & (J2000) & (UT) & (sec) & & (arcsec) & & \\
\hline           
   Central & 00:26:10.43, -11:03:09.0 & 2012-12-11 / 01:11 & 2614 & 1.09 & 0.84 & A & 35\\
    &  & 2012-12-11 / 01:56 & 2614 & 1.20 & 0.84 & A & \\ 
    &  & 2012-12-12 / 00:55 & 2614 & 1.08 & 0.88 & A & \\
    &  & 2012-12-12 / 01:40 & 2614 & 1.16 & 0.94 & A & \\  
    &  & 2013-08-11 / 06:30 & 2614 & 1.09 & 0.71 & A & \\  
    &  & 2013-08-30 / 04:59 & 2614 & 1.11 & 0.94 & A & \\   
   NE & 00:26:30.16, -10:59:53.9 & 2013-09-06 / 05:45 & 2614 & 1.03 & 0.80 & A & 18\\
    & & 2013-09-06 / 06:33 & 2614 & 1.04 & 0.83 & A & \\ 
    & & 2013-09-14 / 02:03 & 2614 & 1.63 & 0.87 & C\tablefootmark{a} & \\  
    & & 2013-09-14 / 02:48 & 2614 & 1.34 & 0.82 & B & \\ 
    & & 2013-09-14 / 03:33 & 2614 & 1.18 & 0.71 & B & \\ 
    & & 2013-09-30 / 02:14 & 2614 & 1.23 & 0.64 & A & \\
    & & 2013-09-30 / 06:56 & 2614 & 1.25 & 0.58 & A & \\  
   SE & 00:26:39.71, -11:10:16.4 & 2014-09-30 / 02:58 & 2614 & 1.12 & 0.62 & A & 18\\
    & & 2014-09-30 / 03:42 & 2614 & 1.06 & 0.61 & A & \\
    & & 2014-10-02 / 01:18 & 2614 & 1.45 & 0.83 & B & \\
    & & 2014-10-02 / 02:06 & 2614 & 1.23 & 0.85 & B & \\
    & & 2014-10-02 / 02:52 & 2614 & 1.11 & 0.98 & B & \\   
   NW & 00:25:26.11, -10:56:12.1 & 2014-10-02 / 03:53 & 2614 & 1.04 & 1.08 & A & 12\\
    & & 2014-10-02 / 04:38 & 2614 & 1.03 & 1.55 & A & \\
    & & 2014-10-29 / 01:48 & 2614 & 1.05 & 1.02 & B & \\ 
    & & 2014-10-29 / 02:34 & 1277 & 1.03 & 1.08 & C\tablefootmark{b} & \\
    & & 2014-11-23 / 02:02 & 2900 & 1.07 & 1.19 & B & \\ 
    & & 2014-11-23 / 02:52 & 2900 & 1.15 & 0.90 & B\tablefootmark{$\dagger$} & \\  
\hline    
   Total & &  &  &  &  & & 83 \\  
    \hline    
\end{tabular}
\tablefoot{\tablefoottext{$\star$}{ESO OB fulfillment Grades:
                A) Fully within constraints -- OB completed;
                B) Mostly within constraints, some constraint is ~10\% violated -- OB completed;
                C) Out of constraints -- OB must be repeated:}
        \tablefoottext{a}{airmass out of constraints -- OB repeated;}
        \tablefoottext{b}{at 02:57 seeing increased to $>1.0''$ during execution -- OB aborted.}
        \tablefoottext{$\dagger$}{Although classified as completed, the OB did not have visible stellar continua, and was therefore discarded and not reduced.}
}
\end{table*}
 
\subsection{Data reduction} \label{subsec:datared}
We adopt the same data-reduction process as in K17, based on IRAF\footnote{IRAF is the Image Reduction and Analysis Facility distributed by the National Optical Astronomy Observatories (NOAO) for the reduction and analysis of astronomical data. \url{http://iraf.noao.edu/}} routines and custom-made \textit{python} scripts. 
We have also explored the dedicated FORS2 pipeline available for download from ESO\footnote{http://www.eso.org/sci/software/pipelines/}. Although the ESO pipeline allows for a faster and more automatic data reduction leading to satisfactory results, we put it aside in favor of the custom-made pipeline, because the latter allowed us more flexibility over the intermediate steps that are part of the reduction process, such as cosmic-ray removal and sky-subtraction methodology. 

For our custom-made pipeline we have managed standard IRAF tasks in a \textit{python} environment in order to organize and reduce each OB dataset independently. 
Bias and flat-field corrections were performed on each of the two-dimensional (2D) scientific and arc-lamp calibration images; master bias and normalized master flat-field were created by combining five individual bias and five screen flat-field frames, respectively, which were typically taken during the morning after the night observing run. Science frames also needed to be corrected for bad-rows and cleaned from cosmic rays. 
The former step was necessary since several 2D spectra, especially in chip-2 frames, were affected by bad rows. We therefore used the IRAF \textit{fixpix} task to replace bad regions linearly interpolating with nearby rows, a solution that improved the subsequent sky subtraction and spectral extraction.
To deal with cosmic rays instead we used several iteration of the \textit{L.A. Cosmic} algorithm \citep{VanDokkum2001} adapted to the spectroscopic case.

It is known that images of 2D multi-object slit spectra can show significant distortions both in the spatial direction, where slit traces appear curved (the so-called \textit{S-distortion}), and in the dispersion one, along which the instrument disperser tends to impose a wavelength-dependent curvature of the spectral lines (\textit{C-distortion}). In our case, this is particularly evident in the red spectral range ($\lambda > 7000$\AA), which presents numerous OH telluric emission lines. These distortions needed to be taken into account for our dataset in order to perform a correct wavelength calibration and obtain well-extracted 1D spectra with minimal sky residuals. 
We used a custom made IRAF script (a combination of IRAF \textit{identify}, \textit{reidentify} and \textit{fitcoords} tasks acting along the spatial direction) to trace slit apertures in the science and arc-lamp images in order to correct for the S-distortion. We then cut the individual rectified 2D spectra and perform the wavelength calibration on each of them separately. We used in sequence IRAF \textit{identify}, \textit{reidentify} and \textit{fitcoords} tasks to identify the Hg-Cd-Ar-He-Ne emission lines in the arc-lamp spectra and find the wavelength calibration function to be applied on the science spectra using the task \textit{transform}. We used a spline3 function of order 2 in \textit{identify / reidentify} tasks, while in \textit{fitcoords} we made use of a chebyshev function of orders 4 and 2 along the x- and y- axis respectively. The typical RMS accuracy of the wavelength solution was of the order of 0.03\AA. Performing the wavelength calibration in the 2D spectra leads correction of the C-distortion and thus to straightened sky lines orthogonal to the stellar continuum, important for limiting sky-subtraction residuals in the extracted 1D spectra.

\begin{figure}
        \centering
        \includegraphics[width=\hsize]{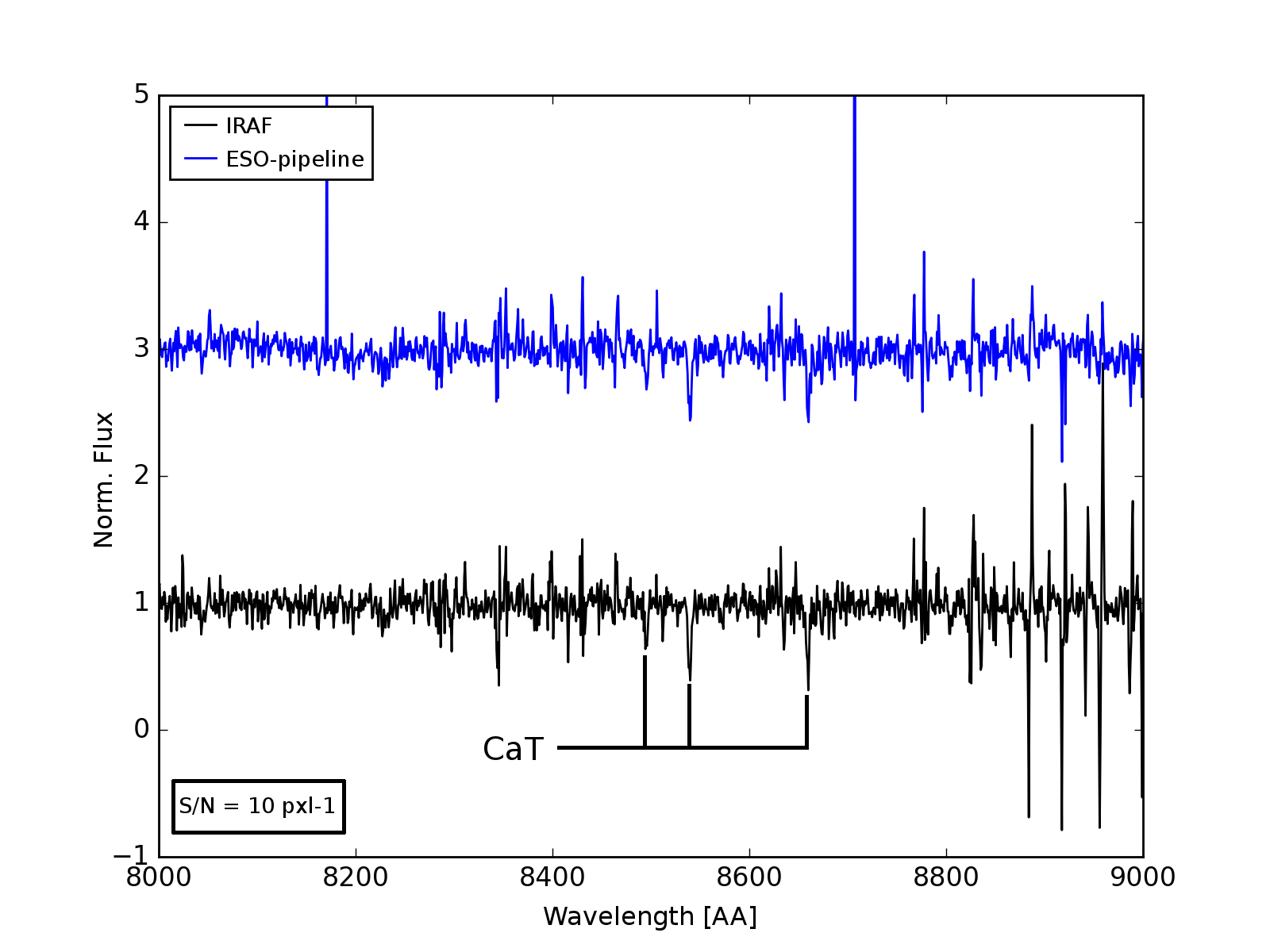}
        \caption{Example of output single-exposure normalized spectra obtained using the ESO-pipeline (upper blue spectrum) vs. IRAF tasks (lower black spectrum), for the same target star. The two spectra were offset on purpose for direct comparison. Although the IRAF reduced spectrum presents higher residual at redder wavelengths (where we note the presence of a telluric absorption band), around the CaT it turned out to be cleaner and less noisy than the ESO-pipeline reduced one.}
        \label{FigZ}
\end{figure}

For the last part of the reduction process we made use of the IRAF \textit{apall} task to obtain background-subtracted and optimally extracted 1D spectra. Outputs included also the extracted sky-background and the error spectrum (i.e. the flux uncertainty at each pixel). Finally, we applied the IRAF \textit{continuum} task to normalize the flux distribution of the extracted 1D spectra fitting a high-order polynomial to the stellar continuum. The median S/N calculated around the Ca~II triplet for the individual exposures resulted in $\sim$9 pxl$^{-1}$ for the Cen and NE fields and $\sim$7 pxl$^{-1}$ for the SE and NW ones. An example of a single exposure extracted spectrum (with a S/N = 10 pxl$^{-1}$), obtained using both IRAF tasks and the ESO-pipeline, can be seen in Fig.~\ref{FigZ}.

Below are some notes on the reduced data products.
\begin{itemize}
        \item The NE field had an aborted OB due to airmass out of constraints. Since the exposure time was complete, we have decided to reduce it anyway. The extracted spectra were suitable for the subsequent analysis.
        \item Each OB of aperture 2 in chip-2 of the NE field suffered from bad rows on the stellar continuum that we could not fix. The extracted spectra were therefore compromised and were excluded from the analysis. 
        \item The last OB of the NW field (date/time: 2014-11-23/02:52h), although classified by the observer as complete, did not have visible stellar continua and was therefore discarded and not reduced.
\end{itemize}

\section{Determination of line-of-sight velocities} \label{sec:velocities}
We determined the line-of-sight (l.o.s.) velocities of the target stars on the combined spectra from the multiple science exposures. Prior to that, the spectra from the individual exposures had to be placed on a common zero-point by correcting them for offsets due to slight differences of the wavelength calibration, slit-centering shifts, and the different observing date.

We took advantage of having numerous OH emission lines in the extracted sky background to refine the wavelength calibration of the individual spectra. Scientific exposures, in fact, can suffer from instrument flexure that may introduce an offset with respect to calibration arc-lamp spectra, usually taken in daytime with the telescope pointing at the zenith. We used the IRAF \textit{fxcor} task to evaluate the offset between a reference sky spectrum and the object ones performing a Fourier cross-correlation over 8250-9000\AA. The associated errors were calculated based on the fitted correlation peak height and the antisymmetric noise \citep{Tonry+Davis1979}. The calculated offsets, $v_{\lambda}$, varied between 5 and 22 km/s with a mean error of 2 km/s (equivalent to 0.2-0.8 pxl; 1pxl = 28 km/s).

\begin{figure}
        \centering
        \includegraphics[width=\hsize]{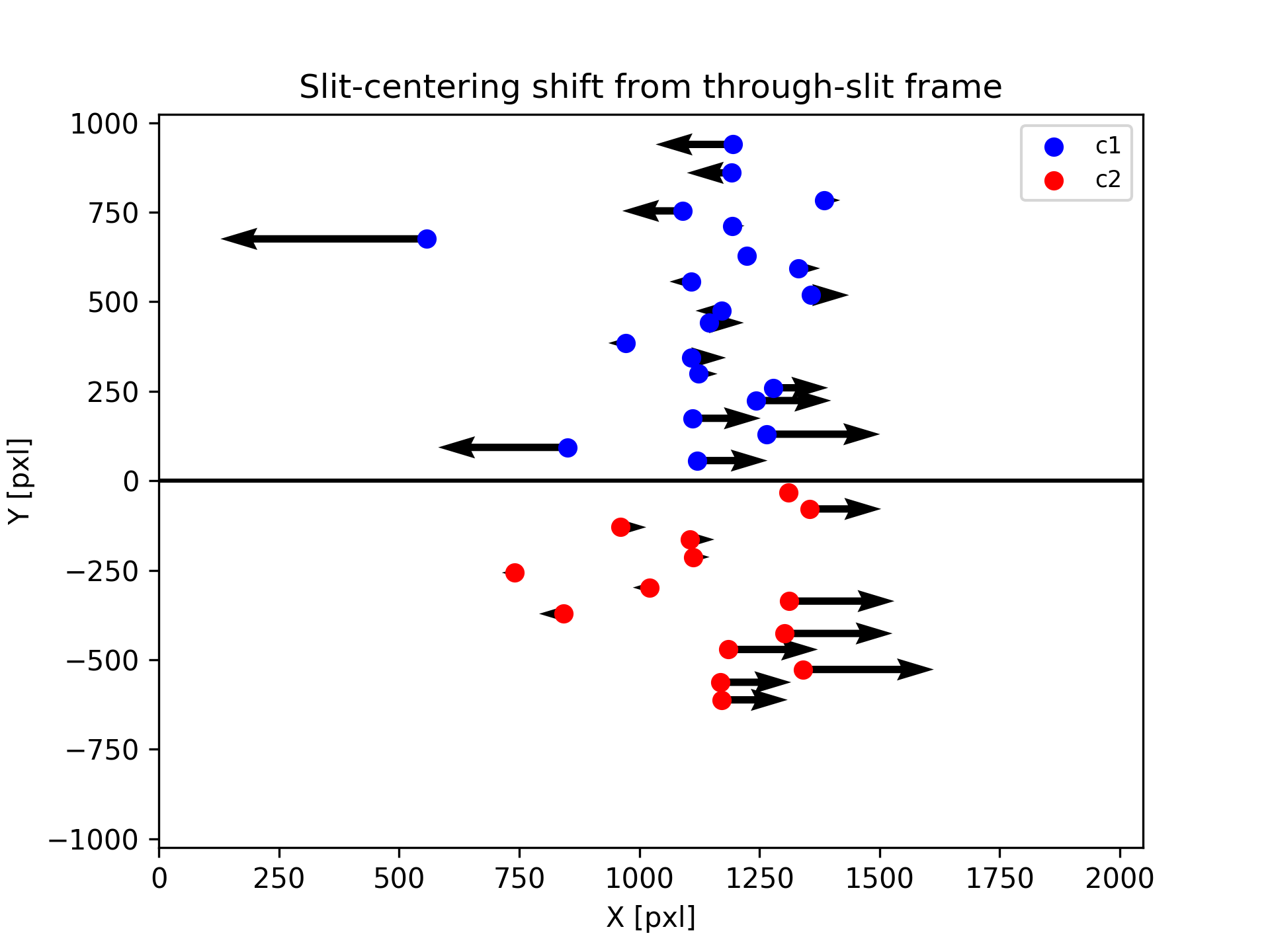}
        \caption{Arrow diagram showing slit-centering shifts from the through-slit frame associated with the first exposure of the central field: arrow lengths are the slit-centering shifts multiplied by a factor of $1000$.}
        \label{Fig:arrow.mask}
\end{figure}

Most observations were taken under very good seeing conditions, with seeing smaller than the slit-width. If targets are not perfectly centered on their slits, there will be a systematic offset in wavelength calibration, and therefore on the velocity measurement.
In order to calculate the slit-centering shift of the target stars, we made use of the through-slit images typically taken before each science exposure. The slit-offset was calculated as the difference in pixels between the center of the slit and the star centroid for every target in each mask. 
As shown in Fig.~\ref{Fig:arrow.mask}, for example, significant deviations were found at the borders of the frames, which were systematically present also on the other exposures, a fact that would make the mean shift per through-slit image an incomplete description of the situation. 
Therefore, we calculate an offset for each target as the median value of all the slit-shifts obtained for that target. We associate to it as error the scaled median absolute deviation (MAD) \footnote{$MAD(X)=1.48\ median(\left |  X-median(X) \right |)$} of those values.
Median shifts, $v_{slit}$, were found to be in the range $\pm 0.1$-$9$ km/s with errors of $\pm$2-5 km/s.

Finally, we used the IRAF \textit{rvcorrect} task to calculate the heliocentric correction, $v_{\phi}$, to apply to the individual spectra.

The spectra from the individual science exposures of each star were corrected for $\Delta v= v_{\phi}-v_{\lambda}-v_{slit}$ with the IRAF \textit{dopcor} task and finally averaged together weighting them by their associated $\sigma$-spectra. In order to associate error spectra to the stacked ones we have divided the individual $\sigma$-spectra by the polynomial used for the continuum normalization of the science spectra and combined them according to the formula for the error of the weighted mean \footnote{$\sigma^2(\lambda)= 1/\Sigma_i[1/\sigma^2_i(\lambda)]$}. We did this for all the targets presented in our dataset.

The heliocentric velocity $v_{hel}$ of each stacked spectra was obtained using the \textit{fxcor} task by cross-correlating with an interpolated Kurucz stellar atmospheric model resembling a low-metallicity RGB star, similar to what we expect for our Cetus targets -- log(g) = 1.0, T$_{eff}=$ 4000 K, [Fe/H] = -1.5 dex, convolved to have the same dispersion of our spectra and wavelength range between 8400 and 8700 \AA\ (as in K17). The resulting final velocity errors have a mean value of $\pm6$ km/s, with shift-related errors added in quadrature. Typical S/N values resulted $\sim20$ pxl$^{-1}$.

A complete table with the velocity determinations for each target in our sample is reported at the end of this article (see Table~\ref{table:X}), along with the corresponding field and slit information, RA-Dec coordinates, V- and I-band magnitudes, metallicity values obtained from the CaT lines as explained in Sect.~\ref{sec:metallicity}, the S/N per pixel, and the membership status according to the criteria provided in Sect.~\ref{sec:membership}.

\subsection{Sanity checks} \label{subsec:vel-sanitycheck}
Our data reduction process has already been adopted and well tested in K17 using the same instrumental setup. 
However, our methodology differs from K17 in that the target velocities we obtained are from stacking the spectra of the individual exposures, rather than from the weighted mean of the velocities calculated on the individual science exposures. This choice was due to the fainter magnitudes of our targets with respect to those in the Phoenix dwarf studied in K17, which yielded a lower S/N on the individual exposure spectra. We then performed a series of tests in order to assess the efficacy of our methodology.

We found that in the lower S/N regime, as for our individual exposure spectra, the velocity errors from \textit{fxcor} appear underestimated. This was tested by injecting Poisson noise on a synthetic spectrum obtained from the \cite{Munari2005} library of spectra based on the Kurucz’s code -- log(g) = 2.5, T$_{eff}=$ 4000 K, [M/H] = -2.5 dex to obtain a S/N $\sim8$ pxl$^{-1}$, then shifting the spectrum of a known quantity (2\AA $\equiv 2.5$ pxl $\equiv 70$ km/s) to simulate a l.o.s. velocity and finally cross-correlating with the CaT template. The process was repeated for 500 random realizations of the noise: while the median value of the recovered velocity was in excellent agreement with the input velocity (70.1 km/s), the median of associated errors calculated by \textit{fxcor} was 8.2 km/s, much smaller than the MAD of the distribution of velocities (12.3 km/s). Therefore the errors calculated by \textit{fxcor} were underestimated compared to the scatter in the velocity distribution from the simulated spectra.

We then repeated the previous test by stacking six synthetic spectra with the same shift and S/N as before, in order to reach a S/N$_{sum} = \sqrt{6}$ S/N $\sim$ 20 pxl$^{-1}$,  i.e., similar to that of our combined spectra. The shift was also very well recovered (71 km/s) and the median velocity error was now in much better agreement with the scatter of the velocity distribution (3.5 km/s and 4 km/s, respectively). 

Furthermore, in the great majority of cases, each science exposure had a through-slit image associated to it, and taken immediately before. There were however a couple of exceptions among the consecutive exposures. In order to assess the reliability of velocity measurements from combined spectra with different (unknown) individual shifts, as for the exposures missing their own through-slit image, we repeated the previous test by stacking individual spectra, but this time assigning a slightly different velocity shift to each of them: the velocities are correctly recovered as long as the difference between individual shifts is less than 1\AA, which appears to be the case for our observations, as estimated from those consecutive exposures that {\it did} have their own through-slit frame associated. 
The above results then motivated our choice to derive the velocities, as well as the metallicities, directly from the stacked spectra.

We  also verified the internal accuracy of our velocity measurements between the different pointings. For the Cen and NE fields, we used the three stars that they have in common: targets 17, 20, and 21 from Cen field corresponding to targets 10, 8, and 6 from the NE field, respectively (see Table \ref{table:X} for further details). The calculated heliocentric velocities for two of these stars resulted in very good agreement within 1-$\sigma$. However the heliocentric velocities of the third star (target Cen-17/NE-8) were found to to agree only within 3.5-$\sigma$. We cannot exclude that the source of this discrepancy is due to the fact that this star is part of a binary system. Taking a look at the velocities obtained from single-exposure spectra as a function of the observing date, we obtain two blocks of measurements: those taken on December 2012 (with exposure from the Cen field only) show in general lower velocities than those taken between August and September, 2013 (comprising exposures from both the Cen and NE fields). Furthermore, this star is one of those in common with the \citet{Kirby2014} dataset (see Sect.~\ref{subsec:vel-comparison}): their observations were taken on the first two days of September 2013 and their reported heliocentric velocity was found to be in agreement with those we took in the same period. Nevertheless we have decided to not reject this star and average together the velocities from the two fields. The inclusion or exclusion of this target in the following kinematic analysis (see Sect.~\ref{sec:kinematics}) did not have any significant impact on the final results.
We did the same for the other two stars, since the velocity measurements between the two pointings were compatible with each other.

On the other hand, the SE and NW pointings do not overlap with the other ones.
As another check of our internal zero points between different pointings, we derived the velocities of our stars also on the spectra reduced with the ESO pipeline, which is an entirely independent data reduction approach. We find that the velocities compare very well, except for a systematic offset of $\sim 2-3$ km/s  ($\sim 0.07-0.1$ pxl) for all the fields. Since our reduction method has been consistently the same for all the exposures and fields, from this comparison we conclude that, at most, systematics between different pointings are negligible.

\subsection{Comparison with other works} \label{subsec:vel-comparison}
We have compared our l.o.s. velocities with those  obtained in other works (\citealp{Lewis2007}, and \citealp{Kirby2014}). Primarily we compare with the work of \cite{Kirby2014} (hereafter K14), since the analysis made by \cite{Lewis2007}, although performed using data of lower S/N, resulted in agreement with the former author. 

\begin{figure}
        \centering
        \includegraphics[width=\hsize]{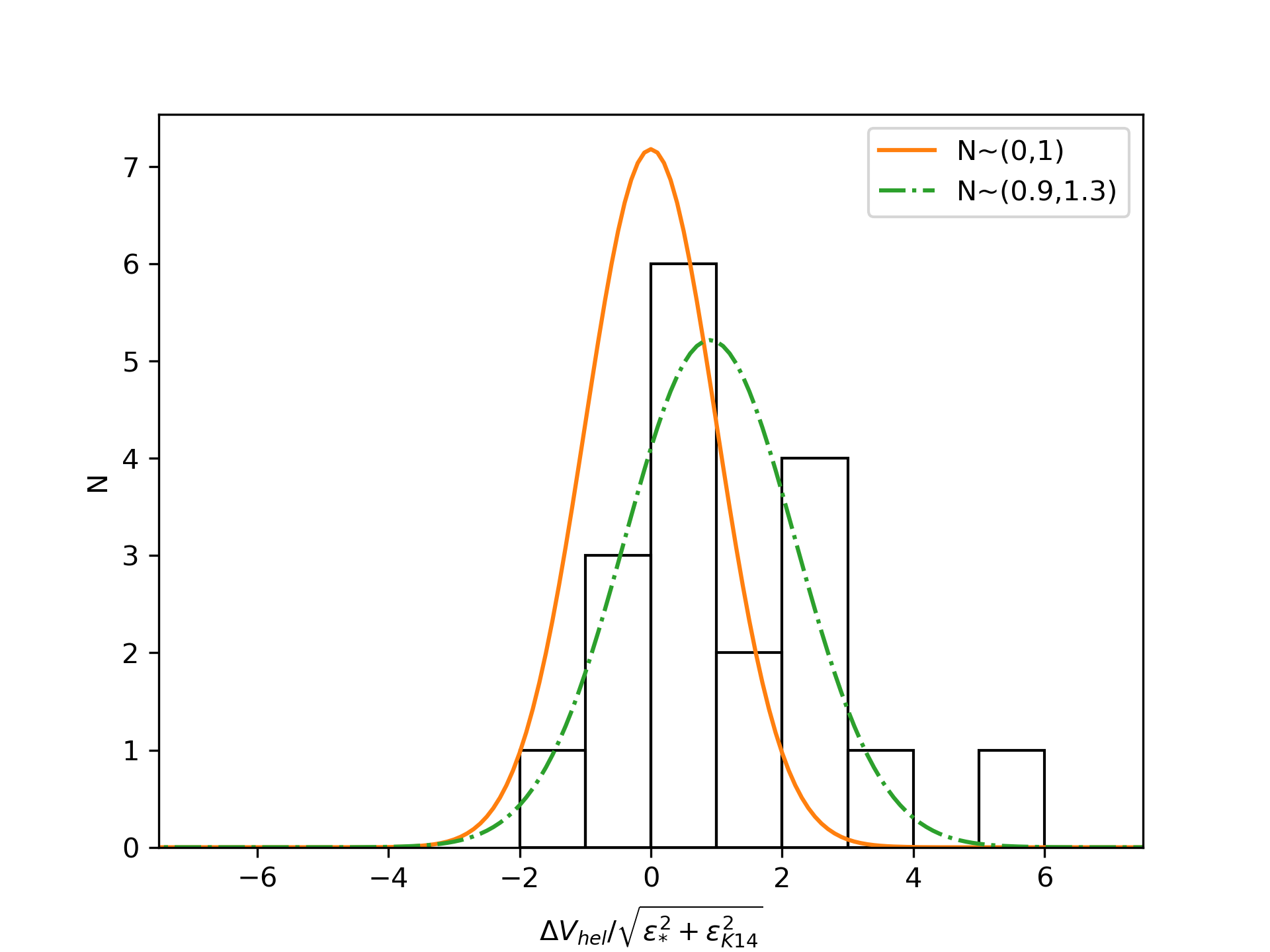}
        \caption{Distribution of velocity differences for stars in common between our dataset and that of \cite{Kirby2014}. A normal distribution with mean and standard deviation $N(\mu,\sigma)\sim(0,1)$ is overplotted for direct comparison (continuous line), together with a normal distribution $N(\mu,\sigma)\sim(0.9,1.3)$ (dot-dashed line) fitted to the velocity differences after discarding the most deviant measurement.}
        \label{Fig:Kirby3}
\end{figure}

We find 18 stars in common with K14, all of them classified as Cetus members. The l.o.s. velocity measurements overlap for 9 targets within the $1\sigma$ level of the quadratic sum of the errors, i.e. only 50\% of the expected 68\% if errors were perfectly estimated. Specifically, we examined the quantity $v/\varepsilon = (v_{hel,*} - v_{hel,K14})/\sqrt{\varepsilon^2_{*} +\varepsilon^2_{K14}}$, which for random errors should resemble a normal distribution with mean and standard deviation $N(\mu,\sigma)\sim(0,1)$. In our case, the mean and standard deviation were found to be 1.2 and 1.6, respectively (see Fig.~\ref{Fig:Kirby3}). If we exclude the highest deviant point with $v/\varepsilon > 5.0$, the $v/\varepsilon$ mean and dispersion reduce to 0.9 and 1.3, respectively. This was a justified choice since this measure in the K14 dataset has a low S/N ($\sim5$\AA$^{-1}$) compared to the average S/N of the other stars ($\sim20$\AA$^{-1}$), despite having a velocity error comparable to the dataset average value. We conclude that we mainly observe a systematic shift in the velocity measurements between the two datasets ($\sim 5$ km/s), but estimate the velocity errors fairly well.

Moreover, we detected a slope when examining the velocity difference $\Delta V_{hel}= v_{hel,*} - v_{hel,K14}$ versus our measured values $v_{hel,*}$. In order to understand if this slope is due to a statistical fluctuation, we simulated a Gaussian velocity distribution with an intrinsic dispersion of 10 km/s from which we randomly selected 18 values. We then created two sets of values by further reshuffling the selected velocities according to the velocity error distributions measured in our sample and K14, respectively. We examined the two sets like we did for the common targets and calculated the slope by means of a least-square (LSQ) linear fit. We repeated this process 500 times, finding a median slope of $0.30\pm0.15$, where the error indicates the MAD scatter of the 500 measurements. The LSQ linear fit to the original data yields a slope of $0.67\pm0.12$, that is, $\sim$2-$\sigma$ away from the simulated one. Also in this case, excluding the highest deviant point from the observed sample would bring the data and the mock sets into much better agreement ($\sim$1.5-$\sigma$). This appears to confirm that the main source of discrepancy between the two datasets is a systematic shift in the velocity measurements. 

In the following kinematic analysis (see Sect. 6.3) we therefore expect to find at most a systematic displacement in the determination of the systemic velocity parameter between our sample and that of K14.

\section{Membership selection} \label{sec:membership}
In order to perform an analysis of the properties of the Cetus stellar component, we need to identify the probable member stars and weed out possible contaminants from the sample. Our membership selection is based on the following criteria, applied step by step:
\begin{itemize}
\item In order to select only stars with magnitude and colors compatible with being RGB stars at the distance of the Cetus dSph, we compare the location of our targets on the color-magnitude diagram to expectations from theoretical isochrones shifted at Cetus' distance and broadly bracketing the range of stellar ages and metallicities expected from Cetus' stellar population from SFH determinations \citep{Monelli2010}. This step is necessary because not all targets fall on the Cetus RGB (see Fig.~\ref{FigTarg} right) due to the allocation to random objects for some of the slits remaining otherwise empty. We selected all targets located between Padova isochrones \citep{Girardi2000} with age $t_{age}=10$ Gyr and [Fe/H] $\sim -2.3$ dex (which sets the "blue" limit at (V-I)$\sim$0.9), and age $t_{age}=8$ Gyr and [Fe/H] $\sim -0.4$ dex (which sets the "red" limit at (V-I)$\sim$2.0).
Our sample was reduced from 80 to 69 targets.
        
\item We performed an initial kinematic selection on the sample of 69 targets, excluding all those with evidently outlying l.o.s. velocities, imposing the following velocity cut: $\left | v_{hel,i} - median(v_{hel}) \right | \leq 5\ MAD(v_{hel})$. 
We also excluded one target to which we could not associate a metallicity value (see Sect.~\ref{sec:metallicity}):  this choice was motivated by the fact that this target also has one of the highest velocity errors ($\sim20$ km/s) and the lowest S/N ($\sim 5$ pxl$^{-1}$) in our sample. Our dataset was therefore reduced from 69 to 58 targets.

  We further performed a more strict kinematic selection, iteratively retaining those objects $\left | v_{hel,i} - \bar{v}_{hel}\right | \leq 3\sigma_v + \varepsilon_i$, where the heliocentric systemic velocity $\bar{v}_{hel}$ and intrinsic velocity dispersion $\sigma_v$, are derived as explained in Sect.~\ref{sec:kinematics}. This step finally reduced the sample to 54 most probable members. Since the spread in the observed distribution of l.o.s. velocities is the result of both the intrinsic l.o.s. velocity dispersion of the system and the uncertainties in the velocity measurements, it could be argued that this last step is too strict and excludes genuine members. We have double-checked that making an iterative selection on $\left | v_{hel,i} - \bar{v}_{hel}\right | \leq 3\ MAD(v_{hel})$ would lead to the same result.
\end{itemize}

According to the Besan\c{c}on model \citep{Robin2003}, simulated in the direction of Cetus on a solid angle of $0.05 \ deg^2$ (equivalent to the summed area of the four FORS2 pointings) and over a distance range up to 100 kpc, five MW contaminants could still have passed our photometric and kinematic selection criteria. This number can be considered as an upper limit since on the area surveyed the surface density of Cetus is higher than that of the Galaxy, therefore when allocating slits onto objects we are more likely to have sampled Cetus' population than the MW one.

\section{Kinematic analysis} \label{sec:kinematics}
\subsection{Systemic velocity and velocity dispersion} \label{subsec:kin-disp}
We have performed a Bayesian analysis to measure kinematic parameters for our dataset (such as Cetus' heliocentric systemic velocity and velocity dispersion) and investigate the possible presence of rotation.
Due to the small angular scales we are exploring, no significant velocity gradient is expected due to the projection of the Sun and Local Standard of Rest (LSR) motion onto the l.o.s. of the individual stars (nor of the 3D motion of the galaxy), therefore in the following we continue working with velocities in the heliocentric reference frame, even if not explicitly mentioned.

First, we have carried out an initial analysis considering our system as being supported by  dispersion only. Following \cite{Walker2006}, we have assumed that the likelihood  function for a distribution of N member stars with l.o.s. velocities $v_{hel,i}$ and associated errors $\varepsilon_i$ has the following form:
\begin{equation}
L\left ( \left \{ v_{hel,1},\dots ,v_{hel,N} \right \} \right ) =\prod_{i=1}^{N} \frac{1}{\sqrt{2\pi(\varepsilon_{i}^{2}+\sigma_{v}^{2})}} exp\left [ - \frac{1}{2} \frac{(v_{hel,i}-\bar{v}_{hel})^2}{(\varepsilon_{i}^{2}+\sigma_{v}^{2})} \right ]
\label{Eq.likeli}
,\end{equation}
where $\sigma_v$ is the intrinsic l.o.s. velocity dispersion of the system and $\bar{v}_{hel}$ the l.o.s. systemic velocity.

These last two are the parameters of interest that we have numerically estimated using the \textit{emcee} code \citep{Foreman-Mackey2013}, a \textit{python} implementation of the \cite{Goodman-Weare2010} affine-invariant Monte-Carlo Markov chain (MCMC) ensemble sampler. The code allowed us to calculate the posterior distributions associated to the parameters. As priors, we demanded positive and negative values, respectively, for the velocity dispersion and the systemic velocity (the latter is justified because we know from previous works \citep{Lewis2007, Kirby2014} that Cetus is approaching the Sun). We initialized the sampler setting applying the median and the MAD of a Gaussian fit to the l.o.s. velocity distribution of probable member stars as starting guesses for our free parameters.

\begin{figure}
        \centering
        \includegraphics[width=\hsize]{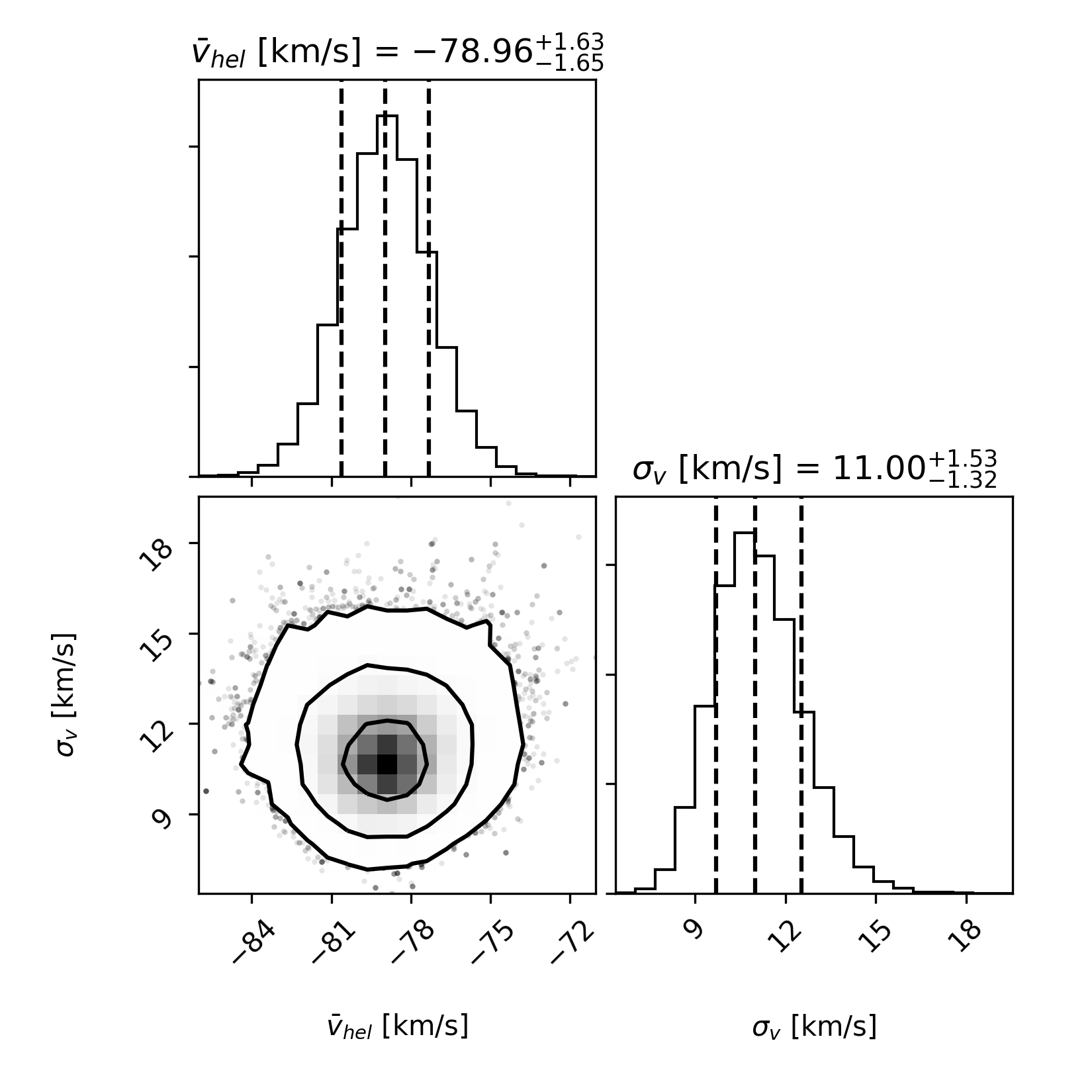}
        \caption{MCMC 2D and marginalized posterior probability distributions for the systemic velocity and velocity dispersion parameters. Dashed lines in the histograms indicate the 16th, 50th, and 84th percentiles. Contours are shown at 1, 2, and 3-$\sigma$ level.}
        \label{Fig:mcmc2}
\end{figure}

As explained in the previous section, the estimation of the parameters was an iterative process, repeated until convergence. We obtain a systemic velocity and velocity dispersion of $\bar{v}_{hel}= -79.0_{-1.7}^{+1.6}$ and $\sigma_v=11.0_{-1.3}^{+1.5}$, as shown in Fig.~\ref{Fig:mcmc2}, where  $\bar{v}_{hel}$ and $\sigma_v$ are the median of the corresponding posterior distributions, while the limits enclosing $68 \%$ of each distribution were set as the asymmetric $1\sigma$ confidence intervals.

\subsection{Rotation} \label{subsec:kin-rot}
The next step in the analysis was to search for evidence of rotation in our dataset, investigating different kinematic models. In order to do so, we modified the likelihood function introduced in Eq.~\ref{Eq.likeli}, substituting the systemic velocity parameter for a relative velocity one: 
\begin{equation}
\bar{v}_{hel} \rightarrow v_{rel,i} = \bar{v}_{hel}+v_{rot}(R_i) cos(\theta - \theta_i)
,\end{equation}
 where $(R_i, \theta_i)$ are the angular distance from the galaxy center and position angle (measured from north to east) of the i-target star, $\theta$ is the position angle of the kinematic major axis\footnote{The kinematic major axis indicates the gradient axis, i.e., the axis along which the l.o.s. velocities deviate furthest from $\bar{v}_{hel}$; it is perpendicular to the axis of rotation.} and $v_{rot}(R_i)$ is the observed rotational velocity along this axis. We considered three models for $v_{rot}$: 
\begin{itemize}
        \item linear or solid-body rotation, $v_{rot}(R_i) = \frac{dV}{dR} R_i = k R_i$ with \textit{k} a constant velocity gradient;
        \item flat or constant rotation, $v_{rot}(R_i) = v_c = constant$; 
        \item no rotation, $v_{rot}(R_i) = 0$, with $v_{rel,i}$ reducing to $\bar{v}_{hel}$ i.e. to the dispersion-only case described above.
\end{itemize}

Subsequently we wanted to explore which model was to be preferred over the others.
We recall the Bayes theorem, rewritten here to condition explicitly on the model under consideration:
\begin{equation}
P(\Theta| D,M)=\frac{P(D|\Theta,M) P(\Theta|M)}{P(D|M)}
\label{Eq.bayes} 
,\end{equation}
where $P(\Theta| D,M)$ is the posterior distribution of parameters $\Theta$ for model $M$ given the observed data $D$, $P(D|\Theta,M)$ is the likelihood function accounting for model parameters, $P(\Theta|M)$ is the prior distribution representing our \textit{a priori} knowledge of the considered model, and $P(D|M) = Z$ is the so-called \textit{Bayesian evidence}, a normalization factor usually ignored for parameter estimation but of central importance for model selection. The Bayesian evidence represents the average of the likelihood over the prior for a specific model choice. If we have two models, $M_1$ and $M_2$, we can compare them through the \textit{Bayes factor}, i.e., the ratio between their evidences: $Z_1/Z_2=B_{1,2}$. A positive value tends to favor $M_1$ over $M_2$. The significance of one model with respect to another can be based on the Jeffrey's scale, computing the natural logarithm of $B_{1,2}$: values of (0-1), (1-2.5), (2.5-5), (5+) corresponds to inconclusive, weak, moderate and strong evidence favoring the first model over the second one (see also \citealp{Wheeler2017}).

The evaluation of \textit{Z} is a nontrivial computational task. To do this, we used the \textit{MultiNest} code \citep{Feroz2009}, a fast and efficient multi-modal nested sampling algorithm. The \textit{nested sampling} \citep{Skilling2006} is a Monte Carlo technique that allows to evaluate Bayesian evidence and at the same time provide posterior parameter estimation as a by-product. The MultiNest code is an implementation of this algorithm that produces posterior samples from distributions that could be multi-modal or degenerate at high dimensions.
In our case, we have calculated the evidences for all three models, together with their parameter estimation. We want to stress that the resulting values relative to the $v_{rot}$ parameter represent a lower limit on the intrinsic value of the rotational velocity: this is due to $v_{rot} = v_{rot}^{intrinsic} \textrm{sin} i$, where \textit{i} is the angle between the angular momentum vector and the line of sight direction. Since $\textrm{sin} i$ is unconstrained in dwarf spheroidals, we refer simply to $v_{rot}$.
We specified the following prior ranges: $\bar{v}_{hel}<0$ km/s, $\sigma_{v}>0$ km/s, $-10<k<10$ km/s/arcmin, $-20<v_c<20$ km/s. For the prior over $\theta$ we performed an iterative choice: we initially set the prior range $0<\theta<\pi$, run the MultiNest code once for the rotational models performing a parameter estimation, took the maximum value from $\theta$ posterior distribution and used this value, $\theta_m$, to update the prior range to $-\frac{\pi}{2}<\theta - \theta_m<+\frac{\pi}{2}$, and run again the MultiNest code.

\begin{figure*}
        \centering
        \includegraphics[width=\hsize]{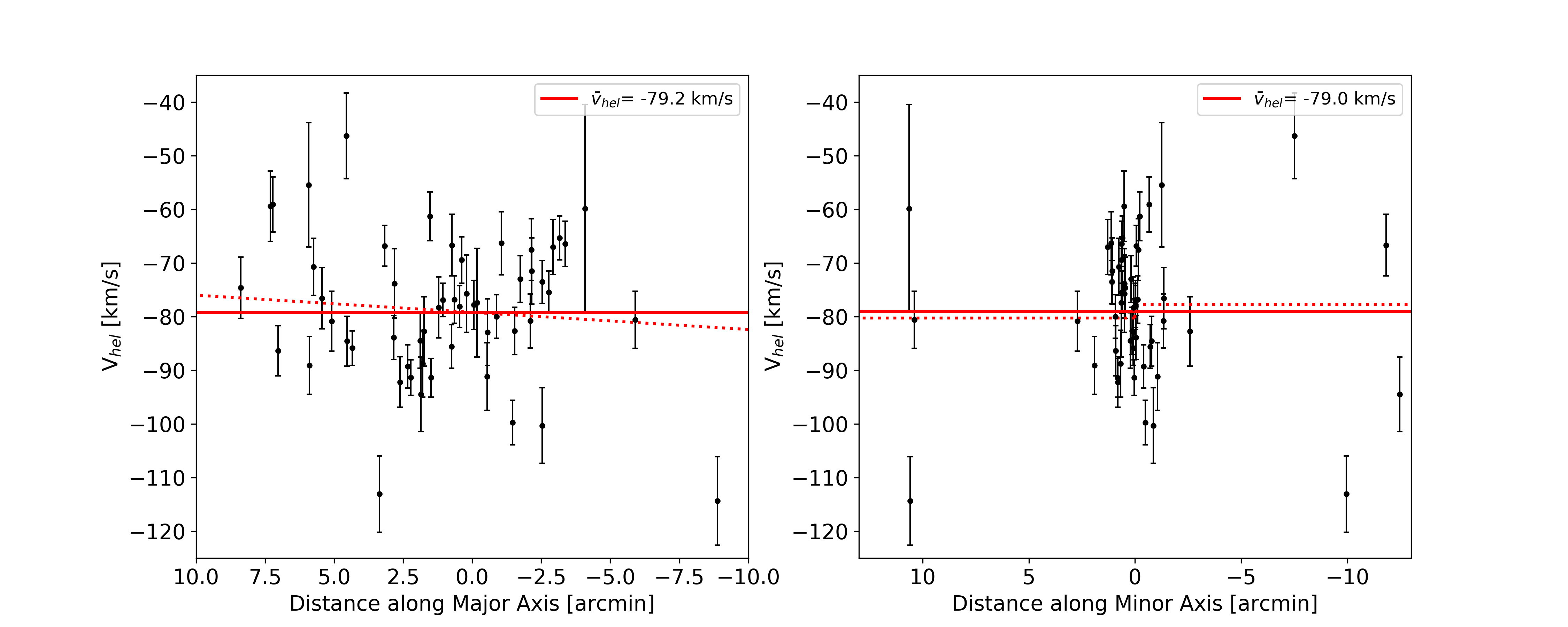}
        \caption{Line-of-sight velocity distributions of the probable members -- \textit{left panel}: along the optical major axis, with the systemic velocity (solid line) and the rotational component (dotted line) resulting from MultiNest run using the linear rotation model overplotted; \textit{right panel}: along  the minor axis, with the systemic velocity (solid line) and the rotational component (dotted line) resulting from MultiNest run using the flat rotation model overplotted. As can be seen from both panels the rotational component is negligible.}
        \label{Fig:Rot}
\end{figure*}

The resulting model evidences and relative estimated parameters are reported in Table~\ref{table:3}. We found that the recovered velocity gradient \textit{k} for the linear rotation model aligns roughly along the major axis, while for the flat rotation model the constant rotational velocity component $v_c$ is recovered instead along the minor axis. We note however that both rotational signals are very weak and compatible with zero. Figure~\ref{Fig:Rot} displays the velocity distribution along the major and minor axis with the velocity estimated parameters from the two models overplotted accordingly.

\begin{figure}
        \centering
        \includegraphics[width=\hsize]{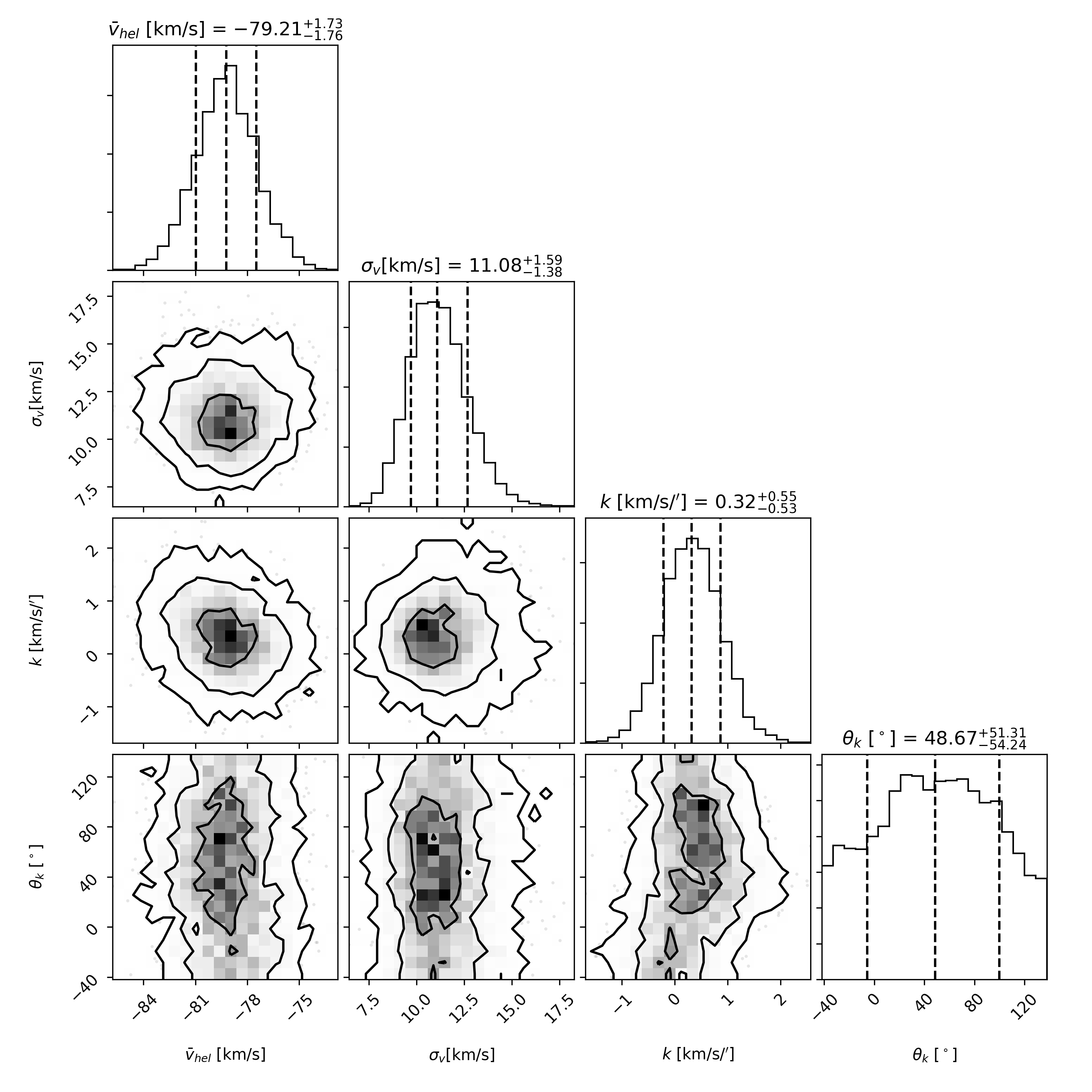}
        \caption{MultiNest 2D and marginalized posterior probability distributions for the solid-body rotational model parameters. Dashed lines in the histograms indicate the 16th, 50th and 84th percentiles. Contours are shown at 1, 2, and 3-$\sigma$ level.}
        \label{Fig:multinest1}
\end{figure}

Comparing the evidences of the linear rotation model against the flat rotation one we have $\textrm{ln}B_{lin,flat} = \textrm{ln}Z_{lin} - \textrm{ln}Z_{flat}= -0.8$, that is, the solid-body model is not favored over the other. If we compare now the evidence of the most favored rotational model (the constant rotation one) with the dispersion-only model, we have $\textrm{ln}B_{rot,disp} = \textrm{ln}Z_{rot} - \textrm{ln}Z_{disp}= -2.0$, that is, the model with rotation is not favored and the simplest dispersion-only model is to be preferred. Estimated parameters for the solid-body rotational model are shown in Fig.~\ref{Fig:multinest1}. We note that the systemic velocity and velocity dispersion are in excellent agreement in all of the three cases analyzed.

Since we have found no evidence of rotation in the Cetus dSph, we can calculate its dynamical mass within the half-light radius using the \cite{Wolf2010} mass-estimator for pressure-supported spherical systems: $M_{1/2}=3G^{-1} \sigma_{v}^2 r_{1/2}$, where $r_{1/2}$ is the 3D de-projected half-light radius that can be well approximated by $\frac{4}{3}R_e$. Substituting the values\footnote{For the velocity dispersion we have used the  \textit{MultiNest} output value from the dispersion-only case.} we obtained $M_{1/2}=67_{-16}^{+19}\times10^6 M_\odot$, that corresponds to a mass-to-light ratio within the half-light radius of $M_{1/2}/L_V = 23.9_{-8.9}^{+9.7} M_\odot/L_\odot$, assuming $L_V=2.8\pm0.8 \times10^6  L_\odot$ (obtained transforming the absolute magnitude value reported by \citealp{McConnachie+Irwin2006}). Our values of the dynamical mass $M_{1/2}$ and velocity dispersion $\sigma_{v}$ were found to be compatible to those found for other galaxies of similar luminosity in the LG (see e.g., \citealp{Kirby2017}).

Using the K14 velocity dispersion value, \cite{Brook+DiCintio2015} found Cetus to be an outlier in its stellar-dark matter halo mass properties, where the latter was calculated taking into account modifications in the dark-matter halo density profile due to stellar feedback. This appears to be due to Cetus being more extended with respect to systems of similar luminosity and internal kinematics. Our new determination of Cetus' velocity dispersion would not imply a significant change in its dark-matter halo mass properties. Therefore Cetus would continue to be an outlier.

\begin{table*}
        \caption{MultiNest output evidences and best-fitting parameters of the kinematic models applied to our dataset.}             
        \label{table:3}      
        \centering          
        \begin{tabular}{l|c|c|c|c|c|c}
                \hline
                Models  & log($Z$) & $\bar{v}_{hel}$ & $\sigma_{v}$  & $k$ & $v_c$ & $\theta$ \\ 
                        &  & [km/s]& [km/s] & [km/s/$'$]& [km/s] & [$^\circ$]\\ \hline
                Linear rotation & -220.0 & $-79.2_{-1.7}^{+1.8}$ & $11.1_{-1.4}^{+1.6}$ & $0.32_{-0.53}^{+0.55}$ & & $48.7_{-54.2}^{+51.3}$ \\
                Flat rotation           & -219.2 & $-79.0_{-1.7}^{+1.7}$ & $11.1_{-1.3}^{+1.6}$ & & $-1.28_{-3.1}^{+2.7}$ & $154.2_{-52.9}^{+46.6}$ \\
                Dispersion-only & -217.2 & $-78.9_{-1.6}^{+1.7}$ & $11.0_{-1.3}^{+1.6}$  & &  & \\ \hline
        \end{tabular}
\end{table*}

\subsection{Comparison with other works} \label{subsec:kin-comparison}
We have also run the MultiNest code on the \cite{Lewis2007} (hereafter L07) and \cite{Kirby2014} datasets. The values we recover for the systemic velocity and velocity dispersion for the dispersion-only case are in agreement (within the 1-$\sigma$ errors) with the values reported by these authors. We therefore refer to their values in the following discussion.

For the systemic velocity, we find a shift of $\sim5$ km/s ($\sim8$ km/s) between the results of K14 (L07) and ours ( $\bar{v}_{hel,K14}= -83.9 \pm1.2$ km/s; $\bar{v}_{hel,L07}= -87 \pm2$ km/s). As reported in Sect.~\ref{subsec:vel-sanitycheck}, we are aware that our velocities might suffer from a 2-3 km/s systematic shift.

The velocity dispersion measured by K14 is $\sigma_{v,K14}=8.3\pm1.0$ km/s, differing from our measured value by approximatively 1.5-$\sigma$. This last discrepancy could be explained by the different spatial distribution of our targets with respect to those observed by K14, and the existence of a mild metallicity gradient in Cetus, with the metal-rich stars displaying a colder kinematics than the metal-poor ones (see Sect.~\ref{subsec:met-pop}).
Indeed, if we select from our catalog only those targets in the central fields (i.e., Cen and NE) and perform again the parameter estimation, we recover $\sigma_{v, inner}=8.9_{-1.2}^{+1.3}$ km/s, which is in perfect agreement with the K14 value. 
However it is harder to reconcile L07 findings with ours and those of K14. Their reported velocity dispersion $\sigma_{v,L07}=17\pm2.0$ km/s deviates significantly from our value and that of K14 in particular, although the two datasets have a similar spatial distribution. We would therefore have expected a lower dispersion value. In their analysis, K14 reported that applying their membership selection criteria they were able to lower the $\sigma_{v,L07}$ value to  $12.0_{-1.9}^{+2.0}$ km/s. They explained the further discrepancy from their reported value with the fact that L07 data had in general a lower S/N. This value instead is in very good agreement with our velocity dispersion result.

When searching for the presence of velocity gradients, we did not find any evidence in support of rotation in the K14 dataset, as already reported by \cite{Wheeler2017} who performed a similar analysis to ours. 
In the case of \cite{Lewis2007} data instead, the flat rotational model is weakly favored over the others, confirming the authors impressions about a hint of rotation in Cetus. 

\subsection{MultiNest mock tests} \label{subsec:kin-multitest}
We performed a series of mock tests to understand what classes of rotational properties we could have expected to detect, given the characteristics of the observational datasets, in terms of number statistics, velocity errors, and spatial coverage.

To this aim we created separate mock catalogs of objects with the spatial positions and velocity error distributions similar to the spectroscopic catalogs of L07, K14, and ours; the velocities were randomly extracted from Gaussian velocity distributions centered around zero and with a fixed $\sigma_v=10$ km/s, to which we add a projected rotational component $v_{rot}$, following the different kinematic models described in Sect.~\ref{subsec:kin-rot}. For our simulations, we tested $v_{rot}/\sigma_v = n$, where $n=\left \{ 0, 0.25, 0.5, 1, 2 \right \}$ at the half-light radius $R_e$ (2.7 arcmin). We note that for $n=0$, we reduce to the dispersion-only case.
These test rotational velocity values were chosen so as to explore different amounts of rotation versus dispersion support of the stellar component, i.e., $v_{rot}/\sigma_v$. For a galaxy of the ellipticity of Cetus ($e=0.3$), the expectation for an oblate isotropic self-gravitating system flattened by rotation corresponds to a $v_{rot}/\sigma_v$ value of 0.5 \citep{Binney1978}; lower values would indicate that the system is flattened by anisotropy, while larger values indicate that the rotational support is dominant over the pressure support. 

Specifically, for the linear rotation model, $v_{rot}= k R_i$, and the \textit{n}-values of the velocity gradient \textit{k} corresponding to the above $v_{rot}/\sigma_v = n$ at the half-light radius, would be $k={n\sigma_v}/{R_e} = \left \{ 0, 0.9, 1.85, 3.7, 7.4 \right \}$ [km/s/arcmin]. 
For the constant rotation model, $v_{rot}= v_c = cte$, and we had $v_c = n\sigma_v =  \left \{ 0, 2.5, 5, 10, 20 \right \}$ [km/s].

We also explored different values for the kinematic axis position angle $\theta =  \theta_C + \left \{ 0^\circ, 45^\circ, 90^\circ \right \}$, where $\theta_C=63^{\circ}$ is the Cetus position angle. We subsequently ran the MultiNest code in order to calculate the evidences of each model (the two rotational and the dispersion-only ones) as done in Sect.~\ref{subsec:kin-rot}, finding $\textrm{ln}B_{lin,flat}$ and $\textrm{ln}B_{rot,disp}$ and estimating the related parameters.
Each case process was repeated N=100 times. All the results are reported in tabulated form in Appendix A.

The analysis showed that the three catalogs have a different sensitivity in the ability to detect rotation with high significance, according to the direction of the input kinematic major axis: for example, the K14 dataset distinguishes gradients along the projected major axis more easily, while our FORS2 dataset is better for detecting gradients along the projected minor axis; this is most likely due to the different spatial coverage.

For all the three spectroscopic catalogs and both rotational models, the analysis of the velocities yielded strong evidence in favor of there being rotation for the $n=2$ case and recovered the correct rotational input model (with moderate evidence for the L07 catalog with constant input rotation).

For the $n=1$ case, evidence in favor of rotation is still strong (or in a minority of cases, moderate or weak) with the three catalogs for both input models. However, in most cases, the correct input rotational model cannot be recovered with conclusive evidence if rotation is constant with radius. Nonetheless, the systemic velocity, internal dispersion, and position angle of the gradient appear reliably determined.

For the $n=0.5$ case and linear model, there is still strong or moderate evidence for rotation in ours and K14's catalog, depending on the axis of the input kinematic gradient.  This indicates that the complementary information provided by our FORS2 and K14's catalogs would have made it possible to establish if Cetus had a rotational velocity at the half-light radius compatible with an isotropic rotator (and have its elliptical shape due to rotational flattening).

In general, rotation constant with radius was more difficult to detect with respect to solid-body rotation. This can be explained by the fact that mock velocities for the flat rotation model were calculated at the half-light radius $R_e$, while the spatial coverage of all the considered catalogs reaches far beyond this value ($r_{max} \sim 15'$ ). Therefore at larger radii than $R_e$, the constant rotational component is more easily hidden by the velocity dispersion with respect to the linear model.

Finally, for case $n=0$, i.e. dispersion-only, all catalogs indeed showed no evidence of rotation. This indicates that the Bayesian analysis using the \textit{MultiNest} code is not biased in favor of rotational models. 

These simulations have shown that if rotation in Cetus were significant, with the current sets of observations we would have already detected it. Instead, our analysis on the observed datasets has shown no evidence in favor of rotation, indicating that Cetus is mainly a dispersion-supported system where an eventual rotational component would be weak, and therefore not likely to cause the observed ellipticity. 
Furthermore, these results should be considered as conservative, taking into account the higher ellipticity value we have found when analyzing the structural properties of Cetus, as explained in Sect.~\ref{sec:struct}:  a higher ellipticity would require the presence of a stronger rotation signal to flatten the system, which is not the case as already described above.

\section{Metallicity properties} \label{sec:metallicity}
We have estimated metallicity ([Fe/H]) values for the probable Cetus members in our FORS2 sample measuring the strength of the CaT lines ($\lambda\lambda =$ 8498.02, 8542.09  and 8662.14 \AA). Our measurements relied on the empirical calibration that connects a linear combination of the CaT lines equivalent widths (EW) and the star magnitude to the corresponding [Fe/H] values. The method has been widely applied in the literature for a variety of stellar systems, from MW globular and open clusters (e.g., \citealp{Rutledge1997, Cole2004, Carrera2012}) to LG dwarf galaxies (e.g., \citealp{Tolstoy2001, Battaglia2008, Ho2015}), and tested and calibrated over a broad range of metallicities and stellar ages (e.g., \citealp{Battaglia2008, Starkenburg2010, Carrera2013, Vasquez2015}).

Given that we share the same stellar target types and instrumental setup with K17, we adopted their approach. The authors examined various calibration methods, alongside testing the most suitable way for measuring the EW of CaT lines in the individual spectra, testing the results on calibrating globular clusters.
Following their conclusions, we have used the \cite{Starkenburg2010} relation, using a Voigt fit for the estimation of the EW of the individual CaT lines (with the flux being integrated over a region of 15\AA\ around the CaT lines of interest) and linearly combining the EW of the two strongest CaT lines; for the $(V-V_{HB})$ term, which allows for comparison of stars of different luminosities, making the calibration also reddening and distance independent, we adopted $V_{HB}=25.03$ (the mean visual magnitude of the RR-Lyrae stars from \citealp{Bernard2009}).

The EWs are determined from the continuum normalized stacked spectra, integrating the flux from the Voigt profile best-fit over a window of 15\AA\ around the CaT lines of interest and adopting the corresponding error-spectra as the flux uncertainty at each pixel in the fitting process.
The errors on the EW measurements were calculated directly from the covariance matrix of the fitting parameters. As a test, we also calculated EW uncertainties using an analytical formula adapted from \cite{Cayrel1988} and \cite{Battaglia2008} based solely on the resolution and S/N of the spectra:
\begin{equation}
\Delta_{EW} = 2.45 \sqrt{\sigma_{Gauss}}\  \textrm{S/N}^{-1}
\label{Eq.analyticErrs} 
,\end{equation}
where S/N is the signal-to-noise ratio per \AA\ calculated in the continuum regions around the CaT and $\sigma_{Gauss}$ is the 1-$\sigma$ width obtained performing a Gaussian fit to the CaT lines, which we can assume as representative of the line broadening of the Voigt profiles. 
Uncertainties on [Fe/H] were calculated propagating the EW errors for both approaches.
The two kind of errors were found to be correlated and have good agreement between them: the median value of those obtained from the analytic estimates resulted in 0.09 dex, while for the uncertainties derived from the line-fitting covariance matrix resulted in a median value of 0.13 dex. We have decided therefore to use the latter as final [Fe/H] errors.

As done for the radial velocity measurements, the relative accuracy of the metallicity values has been assessed using the three stars in common between the Cen and NE pointings: all the calculated values resulted in excellent agreement and were then averaged together (see Table \ref{table:X}).

From the [Fe/H] values we derived, we find that Cetus is a metal-poor system with a significant metallicity spread -- median [Fe/H] $ = -1.71$ dex, standard deviation $=0.45$ dex, MAD $=0.49$ dex \footnote{Median [Fe/H] value obtained transforming the metallicities to their corresponding Z values and then transforming back to [Fe/H] the calculated median Z value.}. This is the first time that metallicities derived from individual RGB stars in Cetus are being made publicly available (see Table \ref{table:X}).

The derived median [Fe/H] value is in excellent agreement with integrated [Fe/H] quantity derived from SFH studies  ($\left \langle\textrm{[Fe/H]}\right \rangle \sim -1.7$ dex, \citealp{Monelli2010}). 
Moreover, this value is in good agreement with the linear luminosity-stellar metallicity relation for LG dwarf galaxies reported by \cite{Kirby2013}, where the Cetus value lies below the relation but within the rms scatter. 
Looking at the metallicity spread, \cite{Leaman2013} have shown that the anti-correlation found by \cite{Kirby2011} between the mean metallicity of a dwarf galaxy and its intrinsic spread in [Fe/H] tends to saturate at high luminosities ($\gtrsim10^5 L_\odot$) (see also \citealp{Ho2015}). The calculated spread in [Fe/H] may be inflated by uncertainties on the individual measurements. 
In order to obtain the intrinsic spread, we applied Eq. 8 of \cite{Kirby2011}; this yielded $\sigma_{[Fe/H]} = 0.42\pm0.03$ dex, where the associated error is calculated as in the Appendix of \cite{Hargreaves1994}. The intrinsic spread is in agreement with the value calculated by simply subtracting in quadrature the mean error in metallicity of the sample from the standard deviation of the metallicity distribution function. Our $\sigma_{[Fe/H]}$ value is then compatible with the saturated trend followed by the other dwarf galaxies of similar or higher luminosities.
However, $\sigma_{[Fe/H]}$ is calculated on [Fe/H], which is a logarithmic quantity. Expressing [Fe/H] values in terms of linear metal fraction, $Z_i/Z_\odot = 10^{\textrm{[Fe/H]}_i}$, the flattened trend disappears, as shown in \cite{Leaman2012}.
Assuming $Z$ uncertainties as $\delta Z_i/Z_\odot = (Z_i/Z_\odot)\textrm{ln(10)} \delta \textrm{[Fe/H]}_i$, we got, in analogy to $\sigma_{[Fe/H]}$, the intrinsic $\sigma(Z/Z_\odot)^2=3.7\times10^{-4}$. This value is in very good agreement with the tight correlation between average linear metallicity $\bar{Z}$ and $\sigma(Z/Z_\odot)^2$ reported by \cite{Leaman2012}. 

\begin{figure}
        \centering
        \includegraphics[width=\hsize,height=.7\linewidth]{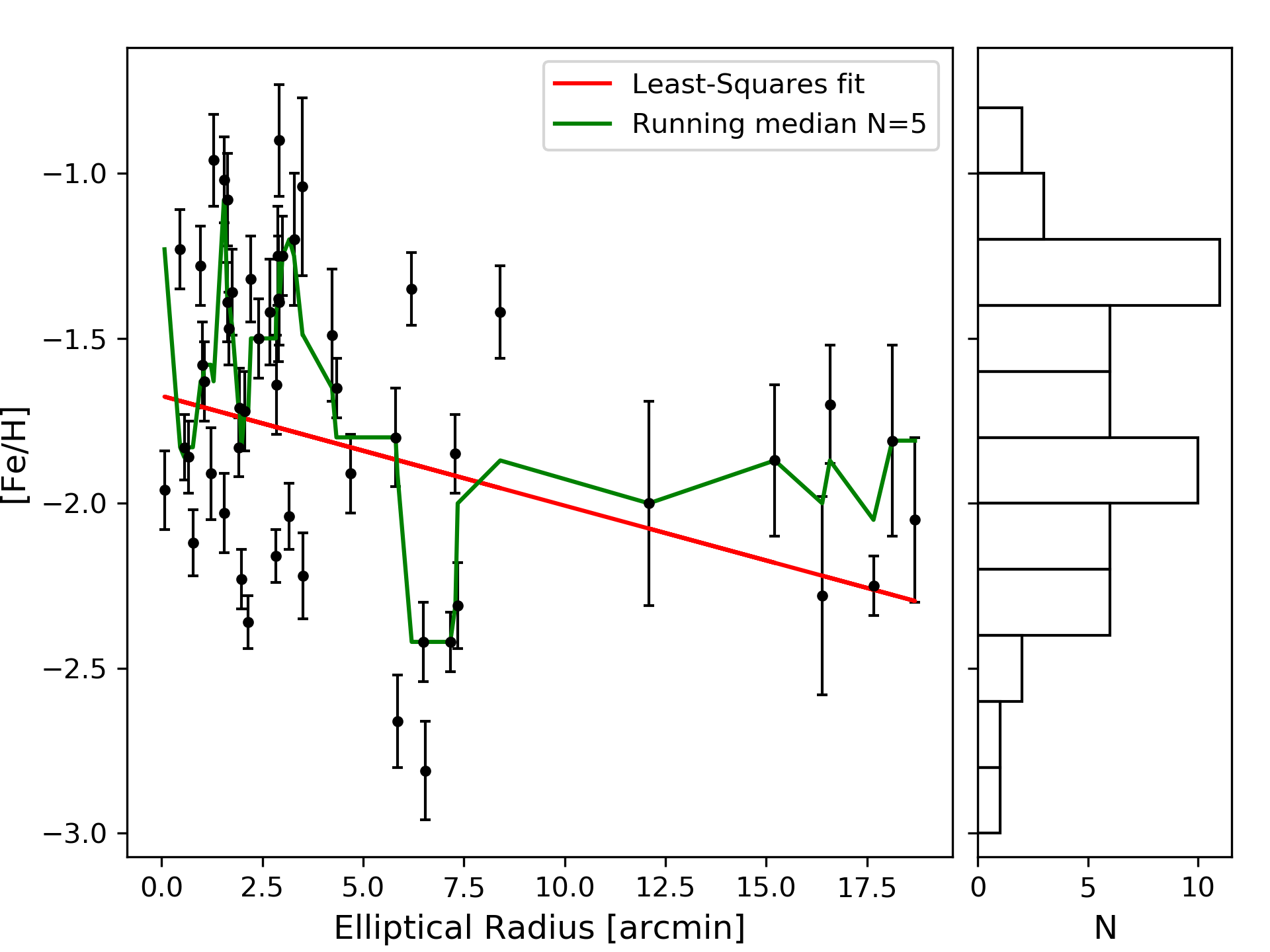}
        \caption{Individual metallicity measurement as a function of the elliptical radius. The red line is the least-squares linear fit to the data, while the green line represents the running median boxcar having a 5-point kernel size.}
        \label{Fig:feh2}
\end{figure}

We analyzed the spatial variation of the metallicity properties looking at the distribution of [Fe/H] as a function of elliptical radius: as shown in Fig.~\ref{Fig:feh2}, we observe a decreasing trend, that is, the metal-richer members look spatially concentrated toward the galactic center. An error-weighted linear least-square fit to the data confirmed the presence of a mild metallicity gradient of $m=-0.033 \pm 0.014$ dex/arcmin ($-0.15 \pm 0.06$ dex/kpc $=-0.09 \pm 0.04$ dex/$R_e$, with distance and $R_e$ values as reported in Table \ref{table:1}). The trend was also confirmed by a running-median: the gradient is almost constant inside $R_e$, it gets steeper up to $10'$ where it starts to flatten again.

The presence of a metallicity spread combined with the lack of a gradient within Cetus' half-light radius, and possible drop beyond that, is in excellent agreement with the analysis of deep photometric datasets: analyzing Cetus' SFH as a function of radius, \citealp{Hidalgo2013} detected no population gradient inside the galaxy half-light radius $R_e$. On the other hand, \cite{Monelli2012} have shown using photometry on a wider area (up to half of Cetus tidal radius $\sim15'$) that the metal poorer population on the RGB starts to dominate at radii grater than $\sim R_e$ and it is ubiquitous at all radii. 

When considered at face value, the slope of the best LSQ fit to the individual metallicities as a function of radius would suggest a much shallower metallicity gradient in Cetus than in the other LG dSphs, and make it more akin to what is seen for example in the WLM dIrr (see \citealp{Leaman2013}).
This could be seen as a contradiction with the possibility that angular momentum is one of the main parameters setting the strength of the metallicity gradient: in principle rotation can create a centrifugal barrier that prevents gas from efficiently funneling toward the galaxy center, producing therefore a smoother radial metallicity gradient and extended star formation (\citealp{Leaman2013, Schroyen2013}), while we find no evidence in support of rotation for Cetus and indeed its SFH was short \citep{Monelli2010}.

However, when looking at both simulations and observations in more detail, one finds that a simple linear fit is often not a complete description of the spatial variations of the metallicity properties. For example, in systems like Sculptor and VV~124, the mean metallicity remains fairly constant in the inner regions, and then declines and is followed by a plateau \citep[see Fig.~14 in][and references therein]{Schroyen2013}. This is exactly the kind of trend that we see in Cetus, when considering a running median of the metallicity as a function of radius.
For example, the slope of the declining region, expressed in dex/$R_e$, is -0.16 for VV~124, -0.26 for Sculptor (both values as in \citealp{Schroyen2013})\footnote{We note that  the metallicity values used for VV~124 in \citealp{Schroyen2013} were too high by about 0.4 dex, as reported in the errata by \citealp{Kirby2013Erratum}. However using the updated metallicity values we find that the slope of the declining region remains unchanged, as do our conclusions.} and $-0.17\pm0.08$ for Cetus. In this respect, the wide-area metallicity properties of Cetus appear to resemble those of similarly luminous, pressure-supported and old early- and transition-type dwarf galaxies. 
It is interesting to note that these three systems, all presenting similar metallicity gradients, span a range of environments: Sculptor is a MW satellite, VV~124 is extremely isolated ($D > 1$ Mpc), while Cetus is found in isolation at present, although it cannot be excluded that it might have experienced a pericentric passage around one of the large LG spirals. Furthermore, despite the complicated internal kinematics, a steep metallicity gradient is also seen in the Phoenix transition type, which is most likely at its first approach towards the MW (K17). All of the above might point to metallicity gradients being a consequence of internal mechanisms such as mass and angular momentum rather than environmental interactions with the larger LG galaxies.

\subsection{Two chemo-kinematically distinct populations?} \label{subsec:met-pop}
The occurrence of a metallicity gradient in Fig.~\ref{Fig:feh2} might also indicate the presence of sub-populations with different kinematics. We divided our dataset into a metal-poor (MP) and a metal-rich (MR) sample according to their [Fe/H] values being lower or greater than the median [Fe/H] (-1.71 dex), respectively.
We then ran the MultiNest code for both samples in order to estimate the associated velocity dispersion in the case of no rotation and obtained $\sigma_{v, MR} = 8.7_{-1.5}^{+1.9}$ km/s and $\sigma_{v, MP} = 13.8_{-2.3}^{+2.7}$ km/s. The two values are at 2-$\sigma$ from each other. Although this is a tentative result, it seems that the different spatial distributions of Cetus metal-richer and metal-poorer stars are reflected in different kinematic properties. This is along the same line as what
seen in some of the dSphs satellites of the MW, such as Sculptor, Fornax, Carina and Sextans, which have a spatially concentrated dynamically colder metal-rich stellar population together with a hotter and more extended metal-poorer one (e.g., \citealp{Tolstoy2004, Battaglia2006, Battaglia2008B,Koch2008, Battaglia2011,Amorisco+Evans2012B}).

As is the case for the majority of those LG dwarf galaxies where the evolved stellar population shows similar chemo-dynamical properties as seen here, it is difficult to ascertain whether this is due to the presence of chemo-dynamically distinct components or to a smooth variation of the spatial distribution of stellar populations of different mean metallicity. The SFH of Cetus and the other dwarf galaxies where this behavior has been detected does not show the presence of clear, separated bursts of star formation, which could be directly linked to two separate populations.
Unfortunately, in exclusively old systems such as Cetus, given that uncertainties in age determination increase at old ages, it is typically challenging or impossible to detect separate bursts based on their relative strength and separation.
So far the most clear case is that of Sculptor dSph, where the application of orbit-based dynamical modeling shows a distribution function that is bimodal in energy and angular momentum space for all of the best-fitting mass models explored, offering an independent and purely dynamically based confirmation of the existence of two physically distinct components (\citealp{Breddels+Helmi2014}; see also \citealp{Zhu2016}). 
The determination of the properties of chemo-dynamically distinct populations in dSphs is of strong interest not only to unravel the complex formation and evolution histories of these galaxies, but, importantly, also to shed light onto the dark-matter properties of these systems \citep{Battaglia2008, Walker+Penarrubia2011, Amorisco+Evans2012, Strigari2018}.

\section{Structural properties}
\label{sec:struct}
\begin{figure}
        \centering
        \includegraphics[width=\hsize]{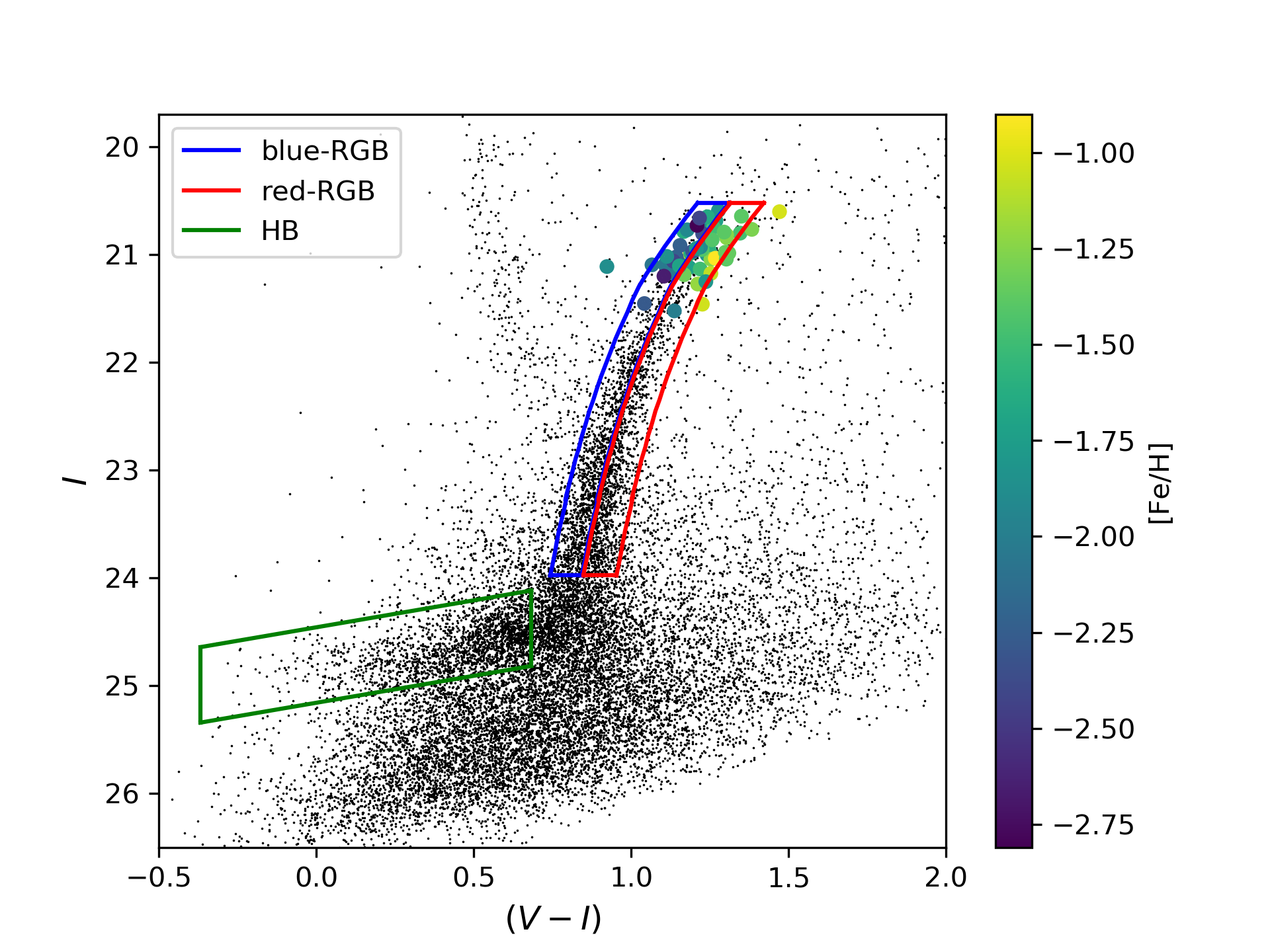}
        \caption{Color-magnitude diagram of the Cetus dSph from the Subaru/SuprimeCam photometric catalog. The FORS2 spectroscopic targets classified as probable members are overplotted, and color-coded according to their calculated [Fe/H]. Selection limits for the blue-RGB, red-RGB, and HB are marked as color boxes.}
        \label{FigCMD_sel}
 \end{figure}

 Given the observed metallicity gradient and chemo-kinematic hint of two distinct stellar population, we combined the spectroscopic information with the Subaru/SuprimeCam photometric data to gain insight into Cetus' structural properties. 
 We have already seen in the previous section that the MR sample of spectroscopically observed RGB stars is more spatially concentrated than the MP one (see Fig.~\ref{Fig:feh2}). Furthermore, as can be seen from the CMD reported in Fig.~\ref{FigCMD_sel}, the spectroscopic members show a clear correlation between their colors and metallicities, with the MP stars having bluer colors than the MR ones. This information guided us to split the overall RGB population into a red and blue part in order to check whether there is evidence of different structural properties. We also checked for the structural properties  of the entire RGB and HB populations, for completeness. The selection limits were defined as shown in Fig.~\ref{FigCMD_sel}, with the total RGB population corresponding to the sum of its blue and red parts.
 
 Following the \citet{Richardson2011} formalism, we employed a maximum likelihood approach to analyze the spatial distribution of the considered populations and get their structural parameters (for details see also \citealp{Cicuendez2018}). Assuming an exponential profile for the surface density distribution of the selected population, together with a constant contamination density, we had seven parameters to determine: the central surface density value ($\sigma_0$), the central coordinates ($\alpha_0,\delta_0$), the exponential scale length ($r_0$) measured on the semimajor axis, the position angle (P.A.), the ellipticity ($\epsilon$) and the constant contamination density ($\sigma_c$). We have numerically estimated the posterior distributions associated to these parameters using the \textit{emcee} code, already introduced in Sect.~\ref{subsec:kin-disp}. We set Gaussian priors for the central coordinates, the P.A. and ellipticity, all centered around the initial values with dispersion of $0.05^{\circ}$, $35^{\circ}$ and 0.5, respectively.  As initial values, we used those reported in Table \ref{table:1}. We also demanded positive values for the density parameters and the exponential scale length, while we fixed the ellipticity values to vary between 0 and 1. 
 The derived structural parameters obtained from the marginalized posterior distributions are reported in Table \ref{table:4}. 
 
\begin{figure}
        \centering
        \includegraphics[width=\hsize]{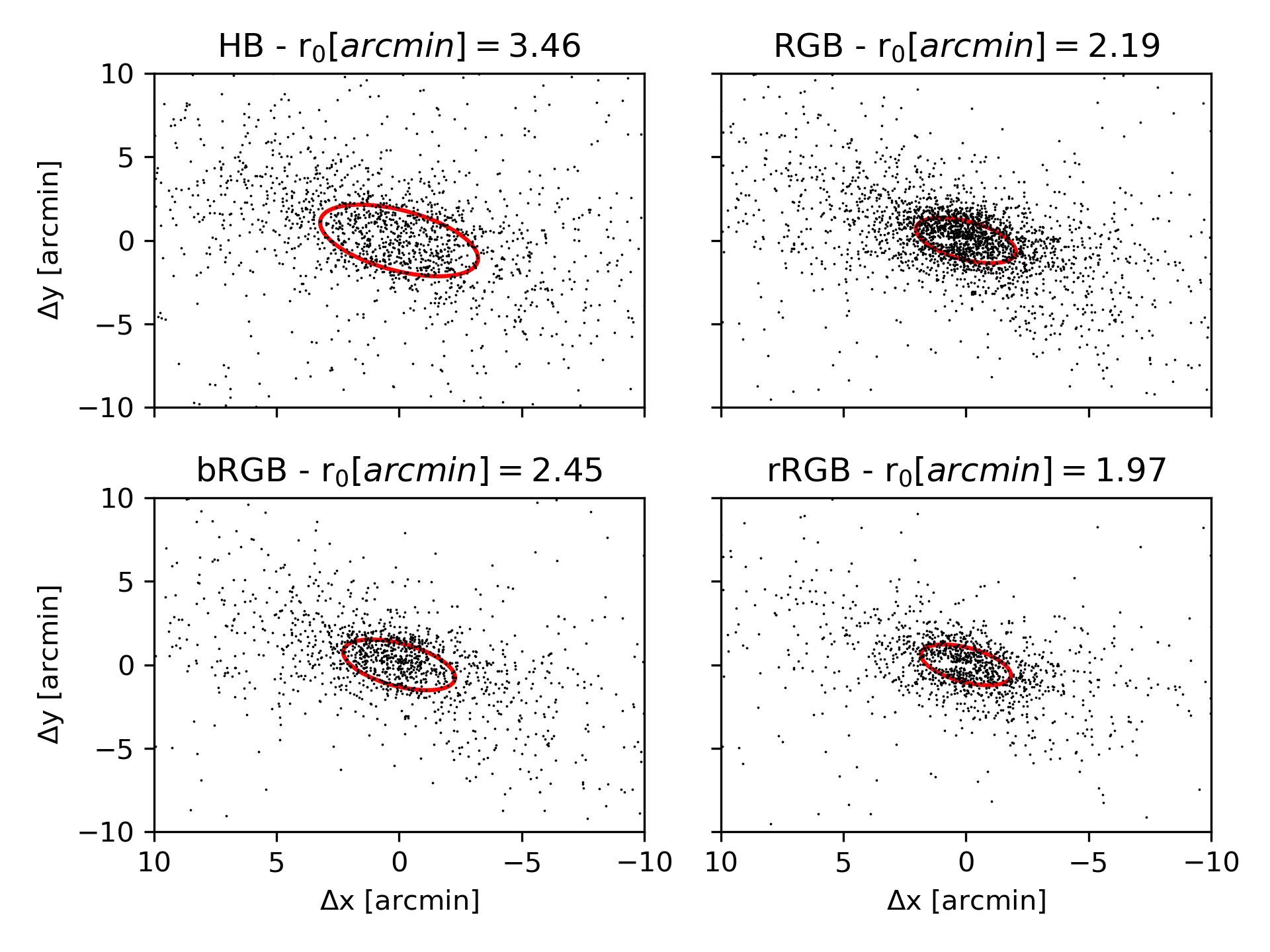}
        \caption{Spatial distribution of the selected HB, RGB, blue-RGB, and red-RGB populations. The ellipses have, as semimajor axis values, the calculated $r_0$ of the considered populations, together with their corresponding P.A. and ellipticity values. 
        North is toward the top of each panel, east is to the left. }
        \label{FigSpat_sel}
\end{figure}

 The spatial distribution corresponding to each selection can be seen in Fig.~\ref{FigSpat_sel}.  The different spatial extensions of the red and blue parts of the RGB can be distinguished, with the first one more centrally concentrated and less extended than the other. The HB population instead, although suffering from a higher level of contamination, is more extended and tends to resemble the blue RGB selection. This indicates, as expected, that the HB is dominated by an older population. The visual impression is confirmed by the determination of the structural parameters: we can see that the exponential radius tends to change within the selections, being smaller for the redder/more metal-rich selection on the RGB and larger for the bluer/more metal-poor one. It is however not obvious that there is a one-to-one correspondence between the bRGB and the HB selection, given the even less concentrated spatial distribution of the HB stars.

 We note that in the previous work by \citet{McConnachie+Irwin2006},  by fitting the total RGB population, the authors obtained a geometric averaged exponential radius of $1.59\arcmin \pm 0.05$. Our value of $2.19\arcmin \pm 0.05$ for the RGB fit, once changed from being measured on the semimajor axis to a geometric averaged radius, resulted in $1.53\arcmin \pm 0.05$, in very good agreement. On the other hand, the P.A. and ellipticity were almost equal for all our selections: roughly 65$^\circ$ and 0.5, respectively. In this case the ellipticity value of 0.33 reported by \citet{McConnachie+Irwin2006}  is significantly lower than our calculated value, while the P.A. of $63^\circ$ is compatible between the errors with our findings. 
 We attribute this discrepancy to the fact that our photometric dataset is deeper than that used by \citet{McConnachie+Irwin2006} in their study.

 Also interesting to note in Fig.~\ref{FigSpat_sel} is the elongated shape of the HB and the RGB along the major axis of the galaxy, a characteristic morphologically reminiscent of tidal tails. Even if the MW, or most likely M31, exerted tidal disturbance onto Cetus, it is difficult to explain how tidal features - if made by unbound stars - could still be visible at present. Assuming Cetus' motion is all in the radial direction moving away from M31 with a velocity of 46 km/s \citep{McConnachie2012}, the last pericentric passage around M31 could have occurred about 6~Gyr ago; this would correspond to $>200$ internal crossing times (defined as in \citealp{Penarrubia2009}), much in excess of the expectations for tidal tails to still be visible, as can be gathered in that same work. Curiously enough, elongations in the outer parts of the stellar component have also been detected in an extremely isolated dwarf galaxy like VV124 \citep{Kirby2012}, perhaps pointing to an explanation other than tidal disturbances.
 
\begin{table*}
        \caption{MCMC-Hammer output structural parameters for each selected population. Reported values represent the median of the corresponding marginalized posterior distributions, with $1\sigma$ errors set as the confidence intervals around the central value enclosing $68 \%$ of each distributions.}             
        \label{table:4}      
        \centering          
        \begin{tabular}{c|c|c|c|c|c|c}
                \hline
                Selections      & $\sigma_0$ & ($\alpha_0,\delta_0$) & $r_0$ & P.A. & $\epsilon$  & $\sigma_c$ \\ 
                & [stars/arcmin$^2$]& [deg] & [arcmin] & [deg] &  & [stars/arcmin$^2$]\\ 
                \hline
                blue-RGB & $74.6_{-4.0}^{+4.2}$ & $6.549 \pm 0.001$, $-11.038 \pm 0.001$& $2.45 \pm 0.08$ & $65.2 \pm 1.3$ & $0.53 \pm 0.02$ & $0.17 \pm 0.02$ \\
                red-RGB & $99.8_{-5.4}^{+5.6}$ & $6.546 \pm 0.001$, $-11.044 \pm 0.001$& $1.97_{-0.06}^{+0.07}$ & $65.2 \pm 1.4$ & $0.50 \pm 0.02$ & $0.14 \pm 0.02$ \\
                RGB &$159.5_{-6.4}^{+6.7}$ & $6.548 \pm 0.001$, $-11.041 \pm 0.001$& $2.19 \pm 0.05$ & $64.2 \pm 1.0$ & $0.51 \pm 0.01$ & $0.30 \pm 0.02$ \\
                HB & $47.1_{-2.8}^{+2.9}$ & $6.551 \pm 0.002$, $-11.037 \pm 0.001$& $3.46_{-0.15}^{+0.16}$ & $65.6 \pm 1.3$ & $0.56 \pm 0.02$ & $0.54\pm 0.04$ \\
                 \hline
        \end{tabular}
\end{table*}

\section{Tracing the Sagittarius stream in Cetus foreground} \label{sec:stream}
The CMD shown in the right part of Fig.~\ref{FigTarg} reveals the presence of an over-density of main sequence turn-off (MSTO) stars in the foreground to the Cetus dSph (see region at 19 $\lesssim$ I $\lesssim$ 23 and 0.4 $\lesssim$ (V-I) $\lesssim$ 0.7).

In order to test whether that region of the CMD is more populated than what would be expected in a smooth model of the MW, we generated a synthetic catalog in the direction of Cetus using the Besan\c{c}on model \citep{Robin2003}, encompassing the solid angle of the Subaru/SuprimeCam catalog ($0.25 \ deg^2$), a distance range up to 100 kpc and a magnitude error $\pm 0.05$ mag in both bands, with default values. We have found that the number counts predicted by the Besan\c{c}on model of the Galaxy
in that range of magnitude and colors are not sufficient to explain the MSTO feature:
indeed in the considered region we observe 220 stars, while the Besan\c{c}on model predicts only 132. 

Therefore it is evident that we are crossing some sort of structure placed in the foreground. The MSTO sources have a homogeneous spatial distribution on our f.o.v. and do not show any tight spatial clustering; therefore they appear to be part of a wider structure.  

In order to trace the upper part of the MSTO feature, we merged our catalog with the publicly available Sloan Digital Sky Survey (SDSS) DR13 Photometric catalog, which covers the Cetus area and probes a brighter magnitude range than the SuprimeCam dataset. Since the SDSS catalog is on a different photometric system, we searched for common targets and performed a rough photometric calibration to convert all magnitudes to the Johnson-Cousin system. This allowed us to set the upper part of the MSTO feature at I$\sim$19.5.

We find that the MSTO feature is well described by a set of isochrones (from \citealp{Girardi2000}) of old age ($10\leq$ t[Gyr] $\leq12.6$) that are metal-poor ($0.0004\leq$ Z $\leq0.001$), or intermediate-young age ($2\leq$ t[Gyr] $\leq5.6$) and metal-richer ($0.004\leq$ Z $\leq0.019$), occupying an increasing heliocentric distance range that goes from 25 to 40 kpc.

Searching in literature to possibly constrain these values, it appears that multiple MW streams cross close to the location on the sky of the Cetus dSph.
The Cetus Polar Stream \citep{Yanny2009, Newberg2009, Koposov2012, Yam2013} intersects the Sagittarius trailing tail at $(l,b)\sim (140^{\circ},70^{\circ})$, which was found by analyzing blue horizontal branch (BHB) stars in the Sloan data. Its distance was confined between $\sim$ 24 and 35 kpc, with most of its stars having metallicity of $-2.5 <$ [Fe/H] $< -2.0$. However, according to its position (e.g., as can be seen in Fig. 4 of \citet{Yam2013}) this stream is far enough from the Cetus dSph location and therefore should not be associated with our MSTO feature.

A narrow stellar stream was found by \citet{Koposov2014} in the ATLAS DR1 data, at least $12^{\circ}$ long and $\approx 0.25^{\circ}$ broad. Their optimal isochrones match was consistent with a metal-poor ([Fe/H]$< -2.1$) and old age (t $\sim$12.5 Gyr) stellar population positioned at a distance of $\sim$20 kpc. The stream great circle passes near the Cetus location (see \citet{Koposov2014}, their Fig.~1) and therefore it may be associated to our MSTO feature, but being a very narrow structure its contribution should also be negligible.
        
Cetus appears to project onto the southern Sagittarius stream, halfway from both its bright and faint arms (see \citet{Slater2013} and \citet{Belokurov2016}, their Fig.~1). \citet{Koposov2012} have shown that the two arms follow a similar distance gradient, going from $\sim$25 to 34 kpc across a sampled sky area of $30^{\circ}$, and present a spread in metallicity with [Fe/H] values between $-2.0$ and 0. At the location on the sky of Cetus, the Sagittarius arms are at a distance of $\sim$25 kpc and present stars with [Fe/H]$<-0.5$ \citep{Koposov2012}  and ages $> 5$ Gyr \citep{deBoer2015}. It is therefore likely that the MSTO feature we detect belongs to the Sagittarius stream. 

To further constrain the association of the MSTO feature to the Sagittarius stream, we  analyzed the three FORS2 MXU targets that fall on this area of the CMD. However, just one of them was found to be a stellar object with a visible CaT feature. The measured heliocentric velocity from the normalized stacked spectrum resulted in $ -70.25 \pm 8.1$ km/s. This velocity in the Galactic Standard of Rest (GSR) system is equivalent to $v_{GSR} \sim -13$ km/s. This is in agreement with the expectations from the \citet{Law+Majewski2010} model of the Sagittarius stream at this location and heliocentric distance range. 

We deem unlikely, though not completely excluded based on the comparison with isochrones, the hypothesis that the feature detected might be part of the smooth halo or be another feature other than Sagittarius. The Sagittarius stream dominates the faint MSTO features that appear when crossing a region of the sky around its orbital plane (\citealp{Deason2011, Belokurov2013, deBoer2015}), as is the case for the Cetus dSph.  It should also be noted that the outer parts of the Galactic halo (i.e., at Galactocentric distances > 15 kpc) show in general a lower metallicity compared to the inner parts, with a metallicity distribution function peaking around [Fe/H]$= -2.3$ dex (e.g., \citealp{Beers2012,AllendePrieto2014,FernandezAlvar2015}). Furthermore, the outer halo appears to be highly structured \citep{Janesh2016}, with the stellar streams being the main source of younger and more metal-rich stars with respect to the smooth halo component (see also \citealp{Huxor&Grebel2015}). This is also observed in M31 \citep{Ibata2014}. In our case, we suggest that the MSTO feature is most likely associated with the Sagittarius stream.

   \begin{figure}
        \centering
        \includegraphics[width=\hsize]{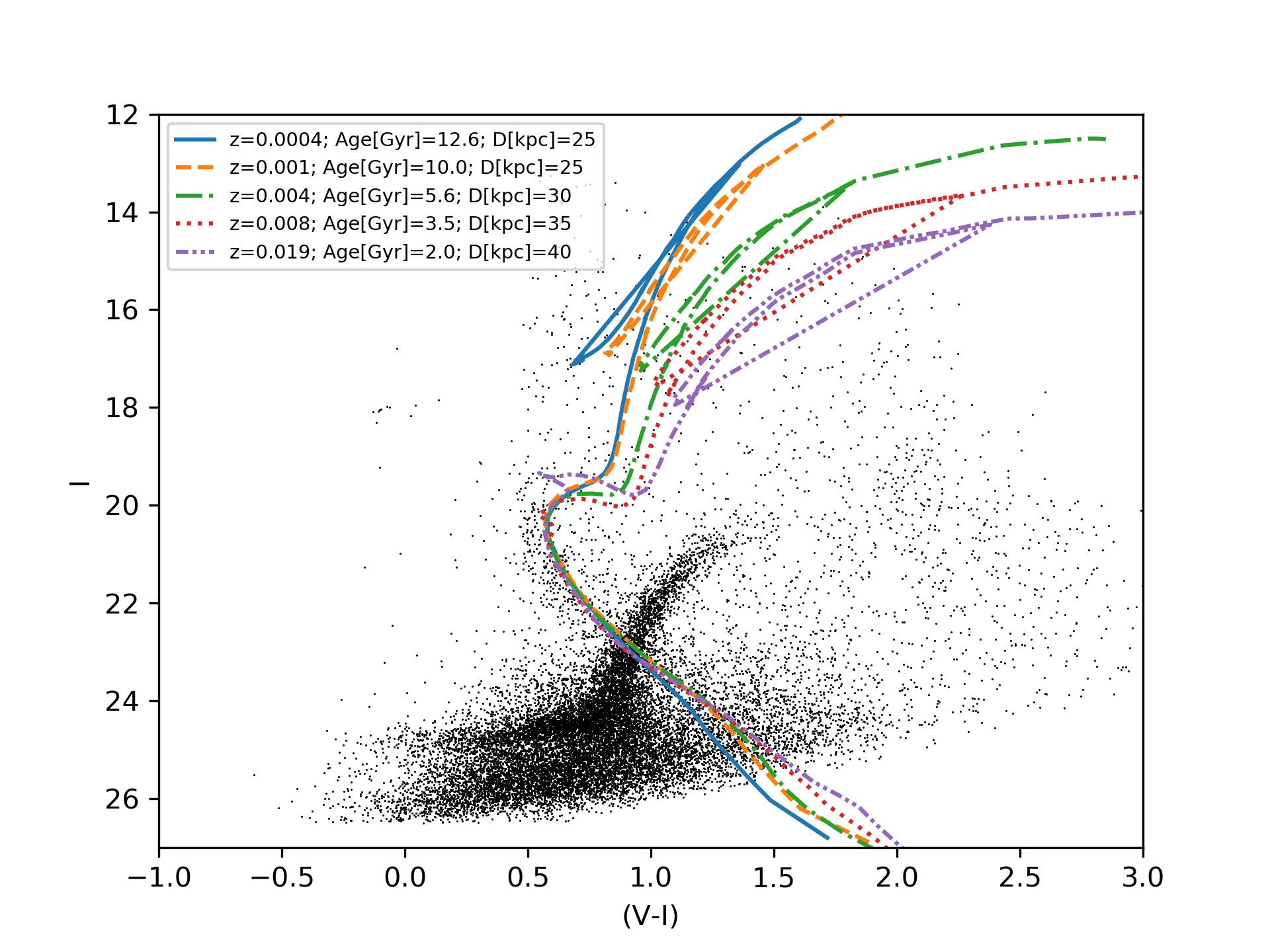}
        \caption{Isochrones overlaid on to the MSTO feature, probably belonging to the Sagittarius stream, on the color-magnitude diagram resulting from merging the photometric catalogs Subaru/SuprimeCam and the Sloan Digital Sky Survey (SDSS) DR13 in the Cetus dSph area. Extinction and reddening have been added to the isochrones (average extinction values calculated in the Cetus dSph area: $A_V= 0.076$, $A_I=0.043$, \citealp{S&F2011}). The figure legend reports the metallicity \textit{Z}, age, and distance of the corresponding isochrones.}
        \label{FigCMD_iso}
  \end{figure}
 
\section{Summary and Conclusions} \label{sec:summary}
In this paper we present an analysis carried out on VLT/FORS2 multi-object spectroscopic observations of 80 individual targets in the Cetus dwarf spheroidal galaxy, of which 54 are probable RGB member stars. We have been able to characterize from this sample the internal kinematics of the galaxy and present the first spectroscopic determination of its wide-area metallicity properties.

The Bayesian analysis of the internal kinematics shows that Cetus is a mainly pressure-supported system ($\bar{v}_{hel}=-78.9_{-1.6}^{+1.7}$ km/s, $\sigma_{v}=11.0_{-1.3}^{+1.6}$ km/s) with no significant signs of stellar rotation. Initially our results were found to be somewhat different from those reported in the previous studies of \cite{Lewis2007} and \cite{Kirby2014}. We attribute the difference in the systemic velocity to a small zero-point offset with respect to K14's study. The larger value that we obtained for the velocity dispersion with respect to K14 appears to be the result of the different spatial coverage of the two datasets combined with the presence of two chemo-kinematic stellar populations.
Moreover, the tests we conducted on mock datasets showed that the exclusion of  a significant rotation signal is robust, indicating that the presence of an eventual rotational component would be weak and not capable of producing the observed galaxy ellipticity.
This is in agreement with what has already been found by \cite{Kirby2014} and \cite{Wheeler2017}, although it contrasts with the \cite{Lewis2007} findings of a moderate rotation ($\sim8$ km/s) along the major axis. We can attribute this discrepancy to a probable artifact caused by the low S/N and nonhomogeneous distribution of \cite{Lewis2007} targets. 

From the analysis of Cetus metallicity properties we establish that the galaxy is predominantly a metal-poor system with a significant metallicity spread (median [Fe/H] $=-1.71$ dex, MAD $=0.49$ dex). The derived values are in very good agreement with integrated quantities from SFH studies (\citealp{Monelli2010, Hidalgo2013}). In addition, the derived median [Fe/H] value agrees, between the 1-$\sigma$ limits, with the $L_v - $[Fe/H] relation for LG dwarf galaxies reported by \cite{Kirby2013}. On the other hand, the intrinsic metallicity spread value ($\sigma_{[Fe/H]} = 0.42\pm0.03$ dex) does not follow the linear anti-correlation found by \cite{Kirby2011} between the intrinsic [Fe/H] spread and the luminosity of a dwarf system, resulting instead in agreement with the saturated trend found for other dwarf galaxies of similar or higher luminosities by \cite{Leaman2013} and \cite{Ho2015}. 

Looking at the radial distribution of the individual metallicity measurements, a LSQ linear fit revealed a mild metallicity gradient of $m=-0.033 \pm 0.014$ dex/arcmin. This value would set Cetus apart from other LG dSphs, where, in general, steeper gradients have been observed. However a simple linear fit can  often be an incomplete description of the radial metallicity distribution, as already seen in both simulations and observations (\citealp{Schroyen2013} and \citealp{Leaman2013}, respectively). By comparing the running-median of the metallicity as a function of radius, we see a complex radial stellar population profile, comparable to other pressure-supported dSphs of similar luminosity but inhabiting a range of environments.
This result seems to suggest that the formation of metallicity gradients in dwarf galaxies can be mainly driven by internal parameters like mass and angular momentum, rather than being the result of environmental interactions with the large hosts of the LG (see also \citealp{Kacharov2017}).

Albeit only at 2-$\sigma$ significance, we have also found evidence of two chemo-kinematically distinct stellar populations, with more metal-rich stars showing a lower dispersion velocity than the metal-poor ones ( $\sigma_{v, MR} = 8.7_{-1.5}^{+1.9}$ km/s and $\sigma_{v, MP} = 13.8_{-2.3}^{+2.7}$ km/s, respectively). Although tentative, this result is in line with what has already been found in other dSphs satellites of the MW (e.g., \citealp{Tolstoy2004, Battaglia2006, Battaglia2008B,Koch2008, Battaglia2011,Amorisco+Evans2012B}).

The presence of stellar population gradients is confirmed by our analysis of the photometric Subaru/SuprimeCam dataset, which reveals that the HB, blue (more-metal poor), and red (more metal-rich) portion of the RGB stars have decreasing values for the scale-length of the best-fitting exponential surface density profile. However, in our simple analysis, we do not find a one-to-one correspondence between the spatial distribution of the blue/metal-poor part of the RGB and the HB, as the latter has a significantly larger scale length than the former. This should be taken into account when adopting a spatial distribution for the tracer population in mass modeling studies that use multiple chemo-kinematic components. Intriguingly, features reminiscent of tidal tails are seen in the outer part of the system, although it is difficult to reconcile their presence with the timescales of a possible passage around M31 and the internal crossing times.

Finally, the ancillary photometric dataset used in this work from Subaru/SuprimeCam reveals the presence of a foreground population that most likely belongs to the Sagittarius stream, based  both on the analysis of the color-magnitude diagram and on the agreement of the radial velocity measurement for one of the observed spectroscopic targets with the Sagittarius stream orbit properties from the literature \citep{Law+Majewski2010}.

This study has not only increased our knowledge of the internal chemo-kinematic properties of the Cetus dSph, but has also added an extra piece to our understanding of the mechanisms that drive the evolution of such small systems.

\begin{acknowledgements}
We wish to thank the anonymous referee for the useful comments and the language editor Joshua Neve to have improved the readability of the paper.
We also thank Michele Bellazzini and Paolo Montegriffo for their useful help with the astrometry, Bo Milvang-Jensen for his useful discussion on the reduction process and J.Bermejo-Climent and L.Cicuendez for sharing their code to determine the galaxy's structural parameters. The authors are grateful to A.McConnachie and N.Arimoto for sharing the Subaru/SuprimeCam photometric data.
ST thankfully acknowledges the Spanish Ministry of Economy and Competitiveness (MINECO) for an FPI fellowship (BES-2015-074765).
GB gratefully acknowledges financial support by the Spanish Ministry of Economy and Competitiveness (MINECO) under the Ramon y Cajal Programme (RYC-2012-11537) and the grant AYA2014-56795-P. 
RL was supported by Sonderforschungsbereich SFB 881 ‘The Milky Way System’ (subproject A7) of the Deutsche Forschungsgemeinschaft (DFG), and acknowledges funding from the Natural Sciences and Engineering Research Council of Canada PDF award.

This research has made use of NASA’s Astrophysics Data System and extensive use of IRAF, Python and Scipy ecosystem.
\end{acknowledgements}

%
%

\bibliographystyle{bibtex/aa} 
\bibliography{bibtex/cetus.bib} 

\longtab{
        \begin{longtable}{rrlcccccl}
                \caption{\label{table:X} Properties of the observed VLT/FORS2 targets in Cetus dSph. Column (1) field (Cen = 1, NE = 2, SE = 3, NW = 4); (2) slit aperture: numbers < 30 indicate observed targets from chip-1, otherwise from chip-2 -- numbers counted from bottom to top of the CCD; (3) RA-Dec coordinated in J2000; (4) V band magnitude with error from Subaru/SuprimeCam photometric catalog (5) I band magnitude with error from Subaru/SuprimeCam photometric catalog (6) l.o.s. heliocentric velocity with velocity error; (6) metallicity with metallicity error; (8) S/N ratio in pxl$^{-1}$; (9) membership (Y= member; N= non-member). 3 stars had repeated measurements: targets 17, 20 and 21 from Cen field corresponding to targets 10, 8 and 6 from NE field, respectively; in this table we report the single measurements as well as the averaged values used during the analysis process.}\\
                \hline\hline
                Field & Slit &  Ra, Dec (J2000)& $V\pm\delta V$ & $I\pm\delta I$ & $v_{hel} \pm \delta v_{hel}$ & [Fe/H] $\pm \delta$ [Fe/H] & SNR & Memb. \\
                 &  &  [deg] & &        & [km/s]  & [dex] & [pxl$^{-1}$] &  \\
                \hline
                \endfirsthead
                \caption{continued.}\\
                \hline\hline
                Field & Slit &  Ra, Dec (J2000) & $V\pm\delta V$ & $I\pm\delta I$ & $v_{hel} \pm \delta v_{hel}$ & [Fe/H] $\pm \delta$ [Fe/H] & SNR & Memb. \\
                 &  & [deg]  & &        & [km/s]  & [dex] & [pxl$^{-1}$] &  \\
                \hline
                \endhead
                \hline
                \endfoot
                1  &   1  &  6.53688, -11.04748 &  22.245  $\pm$  0.007  &  21.004  $\pm$  0.004  &   -82.9  $\pm$   6.2  &  -1.83  $\pm$  0.10  &  37.8  &  Y  \\ 
                1  &   2  &  6.54593, -11.06429 &  22.149  $\pm$  0.006  &  20.803  $\pm$  0.004  &   -91.2  $\pm$   6.3  &  -1.47  $\pm$  0.11  &  28.9  &  Y  \\ 
                1  &   3  &  6.53830, -11.03623 &  22.151  $\pm$  0.006  &  20.768  $\pm$  0.004  &   -77.4  $\pm$  10.1  &  -1.28  $\pm$  0.12  &  12.6  &  Y  \\
                1  &   4  &  6.54504, -11.04530 &  21.943  $\pm$  0.005  &  20.711  $\pm$  0.003  &   -77.8  $\pm$   4.6  &  -1.96  $\pm$  0.12  &  47.7  &  Y  \\
                1  &   5  &  6.54512, -11.03542 &  22.222  $\pm$  0.007  &  21.063  $\pm$  0.004  &   -75.7  $\pm$   7.2  &  -2.12  $\pm$  0.10  &  25.4  &  Y  \\
                1  &   6  &  6.54664, -11.03221 &  22.216  $\pm$  0.007  &  20.967  $\pm$  0.004  &   -69.4  $\pm$   4.3  &  -1.58  $\pm$  0.13  &  25.2  &  Y  \\
                1  &   7  &  6.55304, -11.04146 &  22.123  $\pm$  0.006  &  20.820  $\pm$  0.004  &   -78.1  $\pm$   3.9  &  -1.23  $\pm$  0.12  &  27.8  &  Y  \\
                1  &   8  &  6.55643, -11.04140 &  22.247  $\pm$  0.007  &  21.126  $\pm$  0.005  &   -76.8  $\pm$   4.4  &  -1.86  $\pm$  0.11  &  27.0  &  Y  \\
                1  &   9  &  6.56251, -11.04938 &  22.179  $\pm$  0.006  &  20.992  $\pm$  0.004  &   -85.6  $\pm$   4.1  &  -0.96  $\pm$  0.14  &  28.6  &  Y  \\
                1  &  10  &  6.56207, -11.03662 &  21.955  $\pm$  0.006  &  20.788  $\pm$  0.004  &   -76.9  $\pm$   3.1  &  -1.63  $\pm$  0.12  &  32.6  &  Y  \\
                1  &  11  &  6.56366, -11.03405 &  21.950  $\pm$  0.005  &  20.772  $\pm$  0.004  &   -78.3  $\pm$   5.7  &  -1.91  $\pm$  0.14  &  31.2  &  Y  \\
                1  &  12  &  6.56208, -11.02086 &  21.957  $\pm$  0.006  &  20.688  $\pm$  0.003  &   -91.4  $\pm$   3.6  &  -1.71  $\pm$  0.12  &  32.4  &  Y  \\
                1  &  13  &  6.57068, -11.03632 &  22.202  $\pm$  0.007  &  21.095  $\pm$  0.004  &   -61.3  $\pm$   4.6  &  -2.03  $\pm$  0.12  &  22.3  &  Y  \\
                1  &  14  &  6.56771, -11.02073 &  22.305  $\pm$  0.007  &  21.121  $\pm$  0.005  &   -88.8  $\pm$   6.3  &  -1.72  $\pm$  0.12  &  21.2  &  Y  \\
                1  &  15  &  6.57265, -11.02699 &  21.870  $\pm$  0.005  &  20.593  $\pm$  0.003  &   -84.5  $\pm$   5.2  &  -1.83  $\pm$  0.09  &  35.3  &  Y  \\
                1  &  16  &  6.59213, -11.06961 &  22.355  $\pm$  0.007  &  21.136  $\pm$  0.005  &   -82.7  $\pm$   6.5  &  -1.49  $\pm$  0.20  &  17.8  &  Y  \\
                
                 & rep.  &  6.57904, -11.02699 &  22.357  $\pm$  0.007  &  21.189  $\pm$  0.005  &  -91.4 $\pm$  3.3  &  -1.32  $\pm$   0.13 &  \dots  &  Y  \\
                 1 &  17  &   &    &    &   -91.2  $\pm$   3.7  &  -1.29  $\pm$  0.17  &  24.8  &   \\
                 2 &  10  &   &    &    &   -92.3  $\pm$   7.3  &  -1.37  $\pm$  0.19  &  22.1  &   \\
                
                1  &  18  &  6.58434, -11.03270 &  22.010  $\pm$  0.006  &  20.749  $\pm$  0.004  &   -89.3  $\pm$   4.0  &  -1.50  $\pm$  0.12  &  30.7  &  Y  \\
                1  &  19  &  6.57914, -11.01265 &  22.298  $\pm$  0.007  &  21.045  $\pm$  0.004  &   -92.2  $\pm$   4.7  &  -1.25  $\pm$  0.15  &  25.3  &  Y  \\
                
                  &  rep.  &  6.58899, -11.02351 &  22.166  $\pm$  0.006  &  20.962  $\pm$  0.004  &   -83.9 $\pm$  4.1  &  -2.16  $\pm$  0.08  &  \dots  &  Y  \\
                1  &  20  &   &    &    &    -97.3  $\pm$   5.6  &  -2.22  $\pm$  0.11  &  30.6  &    \\
                2  &    8   &   &    &    &    -69.1  $\pm$   5.9  &  -2.07  $\pm$  0.13  &  29.5  &    \\
                
                  &  rep. &  6.59423, -11.02139 &  22.313  $\pm$  0.007  &  21.178  $\pm$  0.005  &  -66.8         $\pm$ 3.8 &   -2.04   $\pm$  0.10  &  \dots  &  Y  \\
                1  &  21  &   &    &    &   -74.3  $\pm$   8.6  &  -2.07  $\pm$  0.13  &  25.7  &    \\
                2  &   6  &   &    &    &   -64.9  $\pm$   4.4  &  -1.99  $\pm$  0.18  &  20.8  &    \\
                
                1  &  31  &  6.48997, -11.06058 &  22.688  $\pm$  0.009  &  21.462  $\pm$  0.006  &   -66.5  $\pm$   4.2  &  -1.04  $\pm$  0.27  &  19.3  &  Y  \\
                1  &  32  &  6.49335, -11.05962 &  22.486  $\pm$  0.008  &  21.275  $\pm$  0.005  &   -65.3  $\pm$   4.1  &  -1.20  $\pm$  0.20  &  22.6  &  Y  \\
                1  &  33  &  6.49151, -11.04743 &  22.167  $\pm$  0.006  &  21.024  $\pm$  0.004  &   -67.0  $\pm$   5.1  &  -2.22  $\pm$  0.13  &  26.7  &  Y  \\
                1  &  34  &  6.49907, -11.05631 &  22.303  $\pm$  0.007  &  21.036  $\pm$  0.004  &   -75.5  $\pm$   4.0  &  -0.90  $\pm$  0.17  &  27.4  &  Y  \\
                1  &  35  &  6.49919, -11.04752 &  22.148  $\pm$  0.006  &  20.845  $\pm$  0.004  &   -73.5  $\pm$   4.0  &  -1.25  $\pm$  0.12  &  29.2  &  Y  \\
                1  &  36  &  6.51416, -11.07648 &  22.178  $\pm$  0.006  &  20.989  $\pm$  0.004  &  -100.3  $\pm$   7.0  &  -1.64  $\pm$  0.15  &  22.5  &  Y  \\
                1  &  37  &  6.50501, -11.04473 &  22.180  $\pm$  0.006  &  20.955  $\pm$  0.004  &   -71.5  $\pm$   6.2  &  -1.42  $\pm$  0.16  &  27.3  &  Y  \\
                1  &  38  &  6.51469, -11.06296 &  22.036  $\pm$  0.006  &  20.809  $\pm$  0.004  &   -67.5  $\pm$   5.8  &  -2.36  $\pm$  0.08  &  30.2  &  Y  \\
                1  &  39  &  6.52432, -11.08034 &  22.346  $\pm$  0.007  &  21.044  $\pm$  0.004  &   -80.8  $\pm$   5.0  &  -1.38  $\pm$  0.19  &  20.5  &  Y  \\
                1  &  40  &  6.51815, -11.05490 &  22.298  $\pm$  0.007  &  20.988  $\pm$  0.004  &   -73.0  $\pm$   4.4  &  -1.36  $\pm$  0.13  &  26.5  &  Y  \\
                1  &  41  &  6.52159, -11.05413 &  22.075  $\pm$  0.006  &  20.604  $\pm$  0.003  &   -82.7  $\pm$   4.4  &  -1.02  $\pm$  0.13  &  33.5  &  Y  \\
                1  &  42  &  6.52743, -11.06280 &  21.996  $\pm$  0.006  &  20.646  $\pm$  0.003  &   -99.7  $\pm$   4.2  &  -1.39  $\pm$  0.12  &  38.9  &  Y  \\
                1  &  43  &  6.52120, -11.03572 &  22.072  $\pm$  0.006  &  20.917  $\pm$  0.004  &   -66.3  $\pm$   5.9  &  -2.23  $\pm$  0.09  &  29.9  &  Y  \\
                1  &  44  &  6.52532, -11.03752 &  22.430  $\pm$  0.008  &  21.177  $\pm$  0.005  &   -80.0  $\pm$   4.1  &  -1.08  $\pm$  0.14  &  28.4  &  Y  \\
                2  &   1  &  6.62692, -10.98979 &  22.306  $\pm$  0.007  &  21.202  $\pm$  0.005  &   -70.7  $\pm$   5.3  &  -2.66  $\pm$  0.14  &  32.2  &  Y  \\
                2  &   2  &  6.62047, -11.02169 &  22.151  $\pm$  0.006  &  20.931  $\pm$  0.004  &   -84.6  $\pm$   4.6  &  -1.91  $\pm$  0.12  &  29.5  &  Y  \\
                2  &   3  &  6.62018, -10.97139 &  21.940  $\pm$  0.005  &  20.731  $\pm$  0.004  &   -89.1  $\pm$   5.5  &  -2.81  $\pm$  0.15  &  43.1  &  Y  \\
                2  &   4  &  6.61073, -11.01024 &  21.893  $\pm$  0.005  &  20.651  $\pm$  0.003  &   -85.9  $\pm$   3.2  &  -1.65  $\pm$  0.09  &  33.0  &  Y  \\
                2  &   5  &  6.60175, -10.96567 &  22.197  $\pm$  0.006  &  21.063  $\pm$  0.004  &   -80.9  $\pm$   5.6  &  -2.42  $\pm$  0.12  &  28.3  &  Y  \\
                2  &   7  &  6.59280, -11.01003 &  22.057  $\pm$  0.006  &  20.866  $\pm$  0.004  &  -350.8  $\pm$   7.1  &      /  &  34.1  &  N\tablefootmark{b}   \\
                2  &   9  &  6.58461, -11.01562 &  22.088  $\pm$  0.006  &  20.793  $\pm$  0.004  &   -73.8  $\pm$   6.4  &  -1.39  $\pm$  0.13  &  33.3  &  Y  \\
                2  &  11  &  6.57328, -11.01102 &  21.771  $\pm$  0.005  &  20.59   $\pm$  0.003  &     9.1  $\pm$   5.9  &      /  &  41.4  &  N\tablefootmark{b}   \\ 
                2  &  31  &  6.66900, -10.97425 &  22.122  $\pm$  0.006  &  20.865  $\pm$  0.004  &   -74.6  $\pm$   5.7  &  -1.42  $\pm$  0.14  &  32.8  &  Y  \\ 
                2  &  32  &  6.66218, -11.014      &  22.076  $\pm$  0.006  &  20.739  $\pm$  0.004  &   /   &      /   &   4.3  &  N\tablefootmark{$\star$}   \\
                2  &  33  &  6.66026, -10.99978 &  22.132  $\pm$  0.006  &  21.019  $\pm$  0.004  &   -59.1  $\pm$   5.1  &  -1.85  $\pm$  0.12  &  31.9  &  Y  \\
                2  &  34  &  6.65242, -10.98151 &  22.167  $\pm$  0.006  &  20.972  $\pm$  0.004  &   -59.4  $\pm$   6.5  &  -2.31  $\pm$  0.13  &  33.0  &  Y  \\
                2  &  35  &  6.64503, -11.01836 &  22.279  $\pm$  0.007  &  20.980  $\pm$  0.004  &   -55.4  $\pm$  11.6  &  -1.35  $\pm$  0.11  &  24.5  &  Y  \\
                2  &  36  &  6.64520, -10.97766 &  21.880  $\pm$  0.005  &  20.663  $\pm$  0.003  &   -86.4  $\pm$   4.7  &  -2.42  $\pm$  0.09  &  40.0  &  Y  \\
                2  &  37  &  6.63863, -11.02350 &  22.152  $\pm$  0.006  &  20.945  $\pm$  0.004  &   -76.6  $\pm$   5.7  &  -1.80  $\pm$  0.15  &  21.8  &  Y  \\
                3  &   1  &  6.69005, -11.18939 &  22.491  $\pm$  0.008  &  21.445  $\pm$  0.006  &  -269.2  $\pm$   8.6  &      /  &  14.3  &  N\tablefootmark{b}   \\
                3  &   2  &  6.67327, -11.16669 &  22.036  $\pm$  0.006  &  21.113  $\pm$  0.005  &  -113.1  $\pm$   7.1  &  -1.87  $\pm$  0.23  &  20.8  &  Y  \\
                3  &   3  &  6.67343, -11.15788 &  22.406  $\pm$  0.007  &  20.663  $\pm$  0.003  &   -15.3  $\pm$   4.9  &  -1.79  $\pm$  0.12  &  34.0  &  N\tablefootmark{c}  \\
                3  &   4  &  6.65313, -11.13756 &  21.444  $\pm$  0.004  &  20.707  $\pm$  0.003  &   -99.8  $\pm$  59.9  &     /  &  17.0  &  N\tablefootmark{a}   \\
                3  &   5  &  6.70653, -11.16299 &  20.980  $\pm$  0.003  &  20.571  $\pm$  0.003  & -7223.6  $\pm$  99.9  &     /  &   7.5  &  N\tablefootmark{a}   \\ 
                3  &   6  &  6.66050, -11.13212 &  21.982  $\pm$  0.006  &  21.386  $\pm$  0.005  &   -70.3  $\pm$    8.1  &     /  &  13.3  &  N\tablefootmark{a, $\dagger $}   \\
                3  &   7  &  6.71027, -11.15561 &  23.332  $\pm$  0.015  &  20.956  $\pm$  0.004  & -3126.8  $\pm$  25.1  &      /  &  16.5  &  N\tablefootmark{a}   \\ 
                3  &   8  &  6.67261, -11.12727 &  23.622  $\pm$  0.019  &  21.285  $\pm$  0.005  &   989.8  $\pm$  33.3  &      /  &  10.6  &  N\tablefootmark{a}   \\ 
                3  &   9  &  6.67263, -11.12138 &  22.661  $\pm$  0.009  &  21.524  $\pm$  0.006  &   -46.3  $\pm$   8.0  &  -2.00  $\pm$  0.31  &  13.6  &  Y  \\
                3  &  10  &  6.70292, -11.12720 &  22.492  $\pm$  0.008  &  21.370  $\pm$  0.005  &  -125.5  $\pm$   9.7  &  -1.67  $\pm$  0.27  &  15.1  &  N\tablefootmark{c}   \\
                3  &  31  &  6.65333, -11.22907 &  23.345  $\pm$  0.015  &  20.770   $\pm$  0.004  &   -61.5  $\pm$   8.4  &      /  &  23.5  &  N\tablefootmark{a}   \\ 
                3  &  32  &  6.65380, -11.22331 &  21.341  $\pm$  0.004  &  20.080   $\pm$  0.002  &  3713.0  $\pm$  30.3  &      /  &  32.1  &  N\tablefootmark{a}   \\ 
                3  &  33  &  6.64821, -11.21443 &  22.561  $\pm$  0.008  &  20.494  $\pm$  0.003  &   -66.7  $\pm$   5.7  &  -2.25  $\pm$  0.09  &  32.4  &  Y  \\
                3  &  34  &  6.67011, -11.21540 &  22.163  $\pm$  0.006  &  21.097  $\pm$  0.004  &   -94.5  $\pm$   6.9  &  -2.05  $\pm$  0.25  &  17.2  &  Y  \\
                3  &  35  &  6.62680, -11.184   &  22.554  $\pm$  0.008  &  21.334  $\pm$  0.005  &    55.9  $\pm$  10.1  &     /  &  13.6  &  N\tablefootmark{b}   \\ 
                3  &  36  &  6.63913, -11.18699 &  23.612  $\pm$  0.019  &  20.930   $\pm$  0.004  &   -31.4  $\pm$  21.5  &      /  &  21.2  &  N\tablefootmark{a}   \\ 
                3  &  37  &  6.66534, -11.19491 &  22.161  $\pm$  0.006  &  21.038  $\pm$  0.004  &    91.6  $\pm$   5.9  &      /  &  15.8  &  N\tablefootmark{b}   \\
                3  &  38  &  6.68470, -11.19680 &  22.382  $\pm$  0.007  &  21.042  $\pm$  0.004  &    -9.6  $\pm$   6.8  &  -2.04  $\pm$  0.26  &  18.4  &  N\tablefootmark{c}  \\
                4  &   1  &  6.37661, -10.93484 &  22.260  $\pm$  0.007  &  21.107  $\pm$  0.004  &   -80.6  $\pm$   5.3  &  -1.70  $\pm$  0.18  &  23.5  &  Y  \\
                4  &   2  &  6.34739, -10.91718 &  22.416  $\pm$  0.008  &  20.589  $\pm$  0.003  &    13.4  $\pm$   3.5  &       /  &  25.7  &  N\tablefootmark{b}   \\ 
                4  &   3  &  6.35494, -10.90558 &  23.172  $\pm$  0.014  &  20.975  $\pm$  0.004  &    -9.1  $\pm$   8.3  &      /  &  17.5  &  N\tablefootmark{a}   \\
                4  &   4  &  6.40200, -10.91746 &  22.498  $\pm$  0.008  &  21.456  $\pm$  0.006  &   -59.9  $\pm$  19.4  &  -2.28  $\pm$  0.30  &   8.0  &  Y  \\
                4  &   5  &  6.36032, -10.89001 &  23.304  $\pm$  0.015  &  21.804  $\pm$  0.007  &    33.7  $\pm$  16.7  &      /  &   9.1  &  N\tablefootmark{b}   \\
                4  &   6  &  6.38581, -10.89334 &  23.582  $\pm$  0.018  &  22.265  $\pm$  0.010   &   -46.4  $\pm$  19.1  &      /  &   6.3  &  N\tablefootmark{b, $\ddagger$}   \\
                4  &   7  &  6.34763, -10.8755  &  22.505  $\pm$  0.008  &  21.262  $\pm$  0.005  &     9.6  $\pm$  13.9  &     /  &  11.7  &  N\tablefootmark{b}   \\ 
                4  &  31  &  6.31670, -10.96733 &  22.207  $\pm$  0.007  &  20.696  $\pm$  0.003  &    -7.9  $\pm$   4.0  &  -1.76  $\pm$  0.19  &  17.4  &  N\tablefootmark{c}   \\
                4  &  32  &  6.32031, -10.95838 &  22.132  $\pm$  0.006  &  20.399  $\pm$  0.003  &     6.1  $\pm$   3.9  &     /  &  25.7  &  N\tablefootmark{b}   \\ 
                4  &  33  &  6.32984, -10.95438 &  22.490  $\pm$  0.008  &  21.253  $\pm$  0.005  &  -114.4  $\pm$   8.3  &  -1.81  $\pm$  0.29  &  15.9  &  Y  \\
                4  &  34  &  6.35298, -10.95691 &  22.658  $\pm$  0.009  &  20.292  $\pm$  0.003  &     3.6  $\pm$   7.9  &    /  &  27.7  &  N\tablefootmark{b}   \\ 
                4  &  35  &  6.38168, -10.95799 &  23.458  $\pm$  0.017  &  21.419  $\pm$  0.006  & -2426.4  $\pm$  28.5  &      /  &  10.7  &  N\tablefootmark{a}   \\    \hline    
        \end{longtable}
        \tablefoot{Stars marked as nonmembers were excluded according to the following criteria (see Sect.~\ref{sec:membership} for full description):
        \tablefoottext{$\star$}{target excluded because without a reliable spectral extraction};   
        \tablefoottext{a}{based on their magnitudes and colors:
                ($\dagger $) target belong to the MSTO feature described in Sect.~\ref{sec:stream}};
        \tablefoottext{b}{based on their kinematics:
                ($\ddagger$) target excluded because without a reliable metallicity estimation};
        \tablefoottext{c}{based on the iterative kinematic selection}.}
}

\begin{appendix}
\section{MultiNest mock tests - Output tables}
In this section we report tables with the output values from MultiNest mock tests. There are six tables, depending on whether we created the input databases according to the solid-body or the flat rotation velocity law, and on which catalog (ours, K14, L07) we used to obtain the target spatial positions and error distribution information. 
In each table, input databases are indicated according to their \textit{n} and $\theta$ values ($n=\left \{ 2, 1, 0.5, 0.25, 0 \right \}$ and $\theta = \theta_C + \left \{ 0^\circ, 45^\circ, 90^\circ \right \}$, where $ \theta_C=63^\circ$).
Note that we have simulated an additional case with $n=0.75$ for input databases created according to the constant rotation law, in order to have a smoother transition between cases $n=1$ and 0.5. This allowed to compensate for the fact that constant rotation resulted more difficult to detect with respect to solid-body rotation, as explained in Sect.~5.4. We got similar result to the $n=0.5$ cases created according to the solid-body rotation law, with strong to moderate evidences of rotation, depending on the axis of the input constant velocity.

\begin{sidewaystable}
        \centering
        \caption{Output results for catalogs created according to the solid-body rotation law and following our targets spatial positions and error distribution. Columns represent: \textit{n} and $\theta$ input values; output parameters for the linear rotation model (systemic velocity, velocity dispersion, gradient and position angle); output parameters for the flat rotation model (systemic velocity, velocity dispersion, constant rotation velocity and position angle); output parameters for the dispersion-only model (systemic velocity and velocity dispersion); Bayes factor for the linear versus the flat rotation model; Bayes factor for the favored rotation model against the dispersion-only one.} 
        \label{targs1}
        \begin{tabular}{cr|rrrr|rrrr|rr|rr}
                \hline
                n & $\theta$ & V$_{sys}$ & $\sigma_v$ & \textit{k} & $\theta_k$ & V$_{sys}$ & $\sigma_v$  & V$_c$ & $\theta_{V_c}$ & V$_{sys}$ & $\sigma_v$ &  $\textrm{ln}B_{lin,flat}$ & $\textrm{ln}B_{rot,disp}$ \\ 
                 & [$^\circ$] & [km/s]& [km/s] & [km/s/$'$] & [$^\circ$]& [km/s]&  [km/s] & [km/s] &  [$^\circ$]& [km/s]& [km/s]&  & \\ 
                \hline 
                & 63 & $-0.05_{-1.66}^{+1.66}$ & $10.30_{-1.20}^{+1.39}$ & $7.39_{-0.49}^{+0.50}$ & $63.49_{-2.42}^{+2.58}$ & $4.46_{-2.62}^{+2.63}$ & $18.45_{-1.44}^{+1.05}$ & $19.01_{-1.45}^{+0.72}$ & $80.22_{-10.74}^{+10.62}$ & $7.70_{-2.77}^{+2.77}$ & $19.70_{-0.46}^{+0.22}$ & $28.61_{-4.31}^{+6.57}$ & $51.61_{-7.26}^{+6.27}$ \\
                2    & 108 & $-0.06_{-1.62}^{+1.63}$ & $9.78_{-1.19}^{+1.35}$ & $7.40_{-0.33}^{+0.34}$ & $108.57_{-3.52}^{+3.71}$ & $5.06_{-2.85}^{+2.85}$ & $19.82_{-0.29}^{+0.13}$ & $19.50_{-0.81}^{+0.37}$ & $119.24_{-8.26}^{+8.01}$ & $6.45_{-2.81}^{+2.81}$ & $19.90_{-0.16}^{+0.07}$ & $77.48_{-6.62}^{+8.16}$ & $108.90_{-6.95}^{+8.60}$ \\
                & 153 & $-0.31_{-1.68}^{+1.68}$ & $10.31_{-1.22}^{+1.42}$ & $7.43_{-0.32}^{+0.33}$ & $153.24_{-3.92}^{+3.78}$ & $1.76_{-2.86}^{+2.86}$ & $19.86_{-0.23}^{+0.10}$ & $19.49_{-0.83}^{+0.38}$ & $147.64_{-7.78}^{+7.67}$ & $1.03_{-2.81}^{+2.81}$ & $19.91_{-0.15}^{+0.07}$ & $87.62_{-6.39}^{+8.90}$ & $116.09_{-7.09}^{+10.96}$ \\ \rowcolor{Gray}
                & 63 & $0.06_{-1.64}^{+1.65}$ & $10.03_{-1.18}^{+1.38}$ & $3.67_{-0.49}^{+0.49}$ & $63.50_{-4.86}^{+5.25}$ & $1.78_{-1.97}^{+1.97}$ & $12.56_{-1.40}^{+1.65}$ & $12.84_{-2.45}^{+2.47}$ & $81.78_{-15.83}^{+14.47}$ & $3.99_{-2.32}^{+2.32}$ & $16.12_{-1.59}^{+1.76}$ & $8.31_{-4.72}^{+4.50}$ & $17.18_{-4.01}^{+5.35}$ \\    \rowcolor{Gray}
                1    & 108 & $-0.07_{-1.67}^{+1.67}$ & $10.18_{-1.20}^{+1.40}$ & $3.72_{-0.34}^{+0.35}$ & $107.90_{-7.04}^{+7.74}$ & $2.37_{-2.44}^{+2.43}$ & $16.70_{-1.66}^{+1.74}$ & $18.12_{-2.44}^{+1.36}$ & $125.76_{-10.18}^{+9.36}$ & $3.15_{-2.68}^{+2.68}$ & $19.18_{-1.03}^{+0.59}$ & $21.39_{-3.94}^{+4.98}$ & $33.63_{-5.26}^{+4.96}$ \\ \rowcolor{Gray}
                & 153 & $0.37_{-1.65}^{+1.65}$ & $10.23_{-1.19}^{+1.39}$ & $3.70_{-0.31}^{+0.32}$ & $152.57_{-7.92}^{+7.37}$ & $1.78_{-2.54}^{+2.56}$ & $17.41_{-1.67}^{+1.56}$ & $18.47_{-2.22}^{+1.14}$ & $149.93_{-9.24}^{+8.95}$ & $0.81_{-2.71}^{+2.70}$ & $19.28_{-0.94}^{+0.53}$ & $25.71_{-4.50}^{+4.85}$ & $36.81_{-5.88}^{+4.10}$ \\ 
                & 63 & $0.13_{-1.62}^{+1.63}$ & $9.82_{-1.20}^{+1.38}$ & $1.77_{-0.49}^{+0.49}$ & $63.13_{-9.49}^{+11.07}$ & $1.09_{-1.68}^{+1.71}$ & $10.56_{-1.24}^{+1.43}$ & $6.27_{-2.26}^{+2.23}$ & $77.75_{-26.22}^{+26.67}$ & $2.00_{-1.76}^{+1.75}$ & $11.70_{-1.29}^{+1.51}$ & $1.50_{-2.21}^{+2.34}$ & $3.59_{-2.61}^{+3.83}$ \\
                0.5  & 108 & $0.45_{-1.65}^{+1.65}$ & $10.07_{-1.19}^{+1.39}$ & $1.91_{-0.33}^{+0.35}$ & $106.72_{-12.87}^{+15.00}$ & $1.64_{-1.87}^{+1.85}$ & $11.97_{-1.36}^{+1.58}$ & $10.87_{-3.08}^{+3.07}$ & $128.99_{-16.25}^{+13.37}$ & $2.15_{-2.01}^{+2.01}$ & $13.72_{-1.47}^{+1.72}$ & $6.55_{-3.75}^{+2.52}$ & $10.84_{-5.28}^{+3.20}$ \\
                & 153 & $-0.00_{-1.66}^{+1.66}$ & $10.19_{-1.19}^{+1.38}$ & $1.88_{-0.32}^{+0.33}$ & $153.36_{-14.71}^{+13.31}$ & $1.00_{-1.89}^{+1.88}$ & $12.16_{-1.38}^{+1.61}$ & $12.37_{-3.35}^{+3.21}$ & $153.87_{-10.84}^{+10.81}$ & $0.41_{-2.04}^{+2.04}$ & $13.85_{-1.53}^{+1.75}$ & $7.27_{-3.02}^{+3.69}$ & $11.52_{-4.31}^{+5.05}$ \\    \rowcolor{Gray}
                & 63 & $0.30_{-1.62}^{+1.65}$ & $10.16_{-1.19}^{+1.37}$ & $0.85_{-0.47}^{+0.48}$ & $62.17_{-20.16}^{+27.25}$ & $0.66_{-1.64}^{+1.63}$ & $10.15_{-1.19}^{+1.38}$ & $3.46_{-2.49}^{+2.42}$ & $83.25_{-39.49}^{+38.81}$ & $1.11_{-1.61}^{+1.62}$ & $10.43_{-1.20}^{+1.39}$ & $-0.84_{-0.81}^{+1.36}$ & $-0.38_{-1.20}^{+2.02}$ \\    \rowcolor{Gray}
                0.25 & 108 & $-0.03_{-1.65}^{+1.64}$ & $10.08_{-1.20}^{+1.38}$ & $0.96_{-0.36}^{+0.37}$ & $107.41_{-25.53}^{+28.25}$ & $0.54_{-1.68}^{+1.67}$ & $10.52_{-1.24}^{+1.43}$ & $5.09_{-2.83}^{+2.79}$ & $131.35_{-31.08}^{+24.94}$ & $0.79_{-1.68}^{+1.68}$ & $10.92_{-1.25}^{+1.43}$ & $0.55_{-1.43}^{+1.81}$ & $1.15_{-1.77}^{+2.67}$ \\     \rowcolor{Gray}
                & 153 & $-0.28_{-1.63}^{+1.63}$ & $9.96_{-1.18}^{+1.37}$ & $0.86_{-0.35}^{+0.35}$ & $134.88_{-28.97}^{+26.60}$ & $0.44_{-1.67}^{+1.67}$ & $10.40_{-1.23}^{+1.43}$ & $5.33_{-3.02}^{+3.02}$ & $148.73_{-26.00}^{+24.15}$ & $0.13_{-1.67}^{+1.67}$ & $10.85_{-1.26}^{+1.45}$ & $0.82_{-1.62}^{+2.28}$ & $1.18_{-1.81}^{+2.46}$ \\ 
                & 63 & $-0.22_{-1.62}^{+1.62}$ & $10.06_{-1.17}^{+1.38}$ & $-0.05_{-0.42}^{+0.41}$ & $60.57_{-48.23}^{+49.54}$ & $-0.20_{-1.59}^{+1.61}$ & $10.11_{-1.19}^{+1.38}$ & $0.58_{-2.48}^{+2.52}$ & $118.00_{-50.83}^{+49.74}$ & $-0.01_{-1.55}^{+1.57}$ & $10.10_{-1.18}^{+1.34}$ & $-1.21_{-0.44}^{+0.57}$ & $-1.58_{-0.27}^{+0.95}$ \\
                0    & 108 & $-0.01_{-1.66}^{+1.65}$ & $10.19_{-1.20}^{+1.37}$ & $0.01_{-0.41}^{+0.41}$ & $65.22_{-47.89}^{+48.95}$ & $0.06_{-1.64}^{+1.62}$ & $10.21_{-1.20}^{+1.38}$ & $0.93_{-2.60}^{+2.56}$ & $111.43_{-51.31}^{+51.99}$ & $0.10_{-1.60}^{+1.59}$ & $10.22_{-1.18}^{+1.36}$ & $-1.13_{-0.60}^{+0.48}$ & $-1.56_{-0.29}^{+0.78}$ \\
                & 153 & $-0.11_{-1.60}^{+1.60}$ & $9.87_{-1.18}^{+1.36}$ & $0.11_{-0.41}^{+0.42}$ & $62.45_{-48.85}^{+48.21}$ & $-0.03_{-1.59}^{+1.58}$ & $9.97_{-1.17}^{+1.36}$ & $0.74_{-2.48}^{+2.47}$ & $96.89_{-52.80}^{+51.96}$ & $0.10_{-1.54}^{+1.56}$ & $9.99_{-1.17}^{+1.34}$ & $-1.07_{-0.44}^{+0.56}$ & $-1.62_{-0.28}^{+0.92}$ \\\hline
        \end{tabular}
\end{sidewaystable}

\begin{sidewaystable}[] 
        \centering
        \caption{Output results for catalogs created according to the solid-body rotation law and following K14 spatial positions and error distribution. Columns as in Table \ref{targs1}.} 
        \label{kirby1}
        \begin{tabular}{cr|rrrr|rrrr|rr|rr}
        \hline
        n & $\theta$ & V$_{sys}$ & $\sigma_v$ & \textit{k} & $\theta_k$ & V$_{sys}$ & $\sigma_v$  & V$_c$ & $\theta_{V_c}$ & V$_{sys}$ & $\sigma_v$ &  $\textrm{ln}B_{lin,flat}$ & $\textrm{ln}B_{rot,disp}$ \\ 
        & [$^\circ$] & [km/s]& [km/s] & [km/s/$'$] & [$^\circ$]& [km/s]&  [km/s] & [km/s] &  [$^\circ$]& [km/s]& [km/s]&  & \\ 
        \hline 
                & 63 & $0.07_{-1.07}^{+1.07}$ & $9.82_{-0.79}^{+0.87}$ & $7.52_{-0.32}^{+0.33}$ & $63.17_{-4.22}^{+4.44}$ & $-0.05_{-1.80}^{+1.80}$ & $18.19_{-1.17}^{+1.07}$ & $19.66_{-0.55}^{+0.26}$ & $71.11_{-6.67}^{+6.71}$ & $-1.31_{-1.94}^{+1.95}$ & $19.87_{-0.20}^{+0.10}$ & $60.78_{-7.34}^{+7.15}$ & $125.08_{-8.88}^{+9.07}$ \\
                2    & 108 & $-0.20_{-1.07}^{+1.06}$ & $9.80_{-0.79}^{+0.88}$ & $7.38_{-0.36}^{+0.37}$ & $107.44_{-4.35}^{+4.10}$ & $0.15_{-1.75}^{+1.76}$ & $17.78_{-1.16}^{+1.17}$ & $19.57_{-0.67}^{+0.32}$ & $94.28_{-6.93}^{+6.76}$ & $-1.61_{-1.94}^{+1.93}$ & $19.85_{-0.25}^{+0.11}$ & $56.78_{-5.51}^{+7.04}$ & $113.26_{-8.29}^{+10.26}$ \\
                & 153 & $-0.06_{-1.07}^{+1.06}$ & $9.73_{-0.78}^{+0.86}$ & $7.38_{-0.57}^{+0.57}$ & $152.87_{-2.61}^{+2.47}$ & $1.04_{-1.36}^{+1.37}$ & $13.26_{-0.97}^{+1.07}$ & $18.70_{-1.57}^{+0.93}$ & $144.27_{-5.48}^{+5.18}$ & $-0.71_{-1.77}^{+1.76}$ & $17.90_{-1.17}^{+1.16}$ & $28.26_{-8.30}^{+5.33}$ & $54.71_{-9.38}^{+7.54}$ \\ \rowcolor{Gray}
                & 63 & $-0.15_{-1.07}^{+1.07}$ & $9.93_{-0.79}^{+0.88}$ & $3.77_{-0.32}^{+0.33}$ & $63.53_{-8.45}^{+9.13}$ & $-0.03_{-1.27}^{+1.27}$ & $12.05_{-0.91}^{+1.01}$ & $14.13_{-1.55}^{+1.56}$ & $68.02_{-9.21}^{+9.45}$ & $-0.93_{-1.67}^{+1.65}$ & $16.66_{-1.16}^{+1.26}$ & $17.27_{-5.16}^{+5.25}$ & $45.94_{-6.39}^{+7.73}$ \\ \rowcolor{Gray}
                1    & 108 & $0.04_{-1.06}^{+1.06}$ & $9.71_{-0.78}^{+0.86}$ & $3.74_{-0.35}^{+0.37}$ & $108.29_{-8.82}^{+7.82}$ & $0.54_{-1.26}^{+1.26}$ & $11.91_{-0.89}^{+1.00}$ & $13.87_{-1.61}^{+1.65}$ & $99.85_{-9.41}^{+8.83}$ & $-0.80_{-1.59}^{+1.61}$ & $15.97_{-1.13}^{+1.24}$ & $17.65_{-6.85}^{+4.93}$ & $43.60_{-6.43}^{+5.46}$ \\ \rowcolor{Gray}
                & 153 & $0.20_{-1.08}^{+1.07}$ & $9.97_{-0.79}^{+0.88}$ & $3.66_{-0.57}^{+0.58}$ & $152.43_{-5.53}^{+4.87}$ & $0.84_{-1.17}^{+1.17}$ & $10.90_{-0.84}^{+0.94}$ & $10.53_{-2.05}^{+2.07}$ & $145.37_{-9.23}^{+8.21}$ & $-0.25_{-1.30}^{+1.31}$ & $12.59_{-0.93}^{+1.03}$ & $6.71_{-3.33}^{+3.88}$ & $16.79_{-4.70}^{+5.23}$ \\ 
                & 63 & $-0.11_{-1.06}^{+1.06}$ & $9.84_{-0.79}^{+0.87}$ & $1.94_{-0.33}^{+0.35}$ & $63.53_{-15.70}^{+17.51}$ & $0.02_{-1.13}^{+1.13}$ & $10.32_{-0.81}^{+0.91}$ & $7.18_{-1.42}^{+1.47}$ & $69.80_{-16.00}^{+16.72}$ & $-0.58_{-1.22}^{+1.23}$ & $11.76_{-0.90}^{+0.99}$ & $4.26_{-2.98}^{+2.87}$ & $13.09_{-5.14}^{+6.15}$ \\
                0.5  & 108 & $-0.03_{-1.09}^{+1.08}$ & $10.09_{-0.80}^{+0.88}$ & $1.89_{-0.34}^{+0.36}$ & $102.43_{-17.96}^{+15.57}$ & $0.24_{-1.15}^{+1.15}$ & $10.56_{-0.83}^{+0.92}$ & $6.94_{-1.50}^{+1.52}$ & $97.06_{-17.11}^{+15.81}$ & $-0.51_{-1.24}^{+1.25}$ & $11.95_{-0.89}^{+0.99}$ & $4.35_{-3.05}^{+3.78}$ & $12.05_{-4.25}^{+5.89}$ \\
                & 153 & $-0.05_{-1.07}^{+1.07}$ & $9.89_{-0.79}^{+0.89}$ & $1.73_{-0.57}^{+0.59}$ & $151.66_{-14.25}^{+10.21}$ & $0.22_{-1.11}^{+1.11}$ & $10.18_{-0.80}^{+0.89}$ & $4.88_{-1.96}^{+1.97}$ & $143.57_{-20.41}^{+16.62}$ & $-0.35_{-1.12}^{+1.12}$ & $10.54_{-0.82}^{+0.91}$ & $1.06_{-1.77}^{+1.88}$ & $2.31_{-2.35}^{+3.33}$ \\ \rowcolor{Gray}
                & 63 & $0.08_{-1.07}^{+1.07}$ & $9.86_{-0.79}^{+0.88}$ & $1.00_{-0.36}^{+0.38}$ & $67.41_{-29.53}^{+31.25}$ & $0.07_{-1.10}^{+1.09}$ & $10.04_{-0.80}^{+0.89}$ & $3.52_{-1.56}^{+1.54}$ & $75.19_{-30.79}^{+31.84}$ & $-0.20_{-1.11}^{+1.10}$ & $10.42_{-0.81}^{+0.90}$ & $0.37_{-1.39}^{+1.54}$ & $1.49_{-2.62}^{+3.16}$ \\ \rowcolor{Gray}
                0.25 & 108 & $0.00_{-1.06}^{+1.06}$ & $9.81_{-0.78}^{+0.86}$ & $0.95_{-0.37}^{+0.39}$ & $105.68_{-33.76}^{+28.36}$ & $0.10_{-1.09}^{+1.09}$ & $9.86_{-0.80}^{+0.88}$ & $3.38_{-1.60}^{+1.55}$ & $99.11_{-33.28}^{+31.57}$ & $-0.19_{-1.09}^{+1.10}$ & $10.17_{-0.80}^{+0.89}$ & $0.36_{-1.37}^{+1.83}$ & $1.26_{-2.33}^{+3.01}$ \\ \rowcolor{Gray}
                & 153 & $-0.13_{-1.07}^{+1.05}$ & $9.78_{-0.79}^{+0.87}$ & $0.66_{-0.49}^{+0.56}$ & $150.20_{-35.86}^{+29.11}$ & $0.00_{-1.07}^{+1.09}$ & $9.82_{-0.78}^{+0.88}$ & $2.20_{-1.81}^{+1.88}$ & $139.01_{-37.16}^{+33.35}$ & $-0.27_{-1.07}^{+1.08}$ & $10.01_{-0.79}^{+0.88}$ & $-0.50_{-0.73}^{+1.05}$ & $-0.95_{-1.07}^{+2.42}$ \\ 
                & 63 & $-0.14_{-1.05}^{+1.05}$ & $9.78_{-0.78}^{+0.86}$ & $0.15_{-0.45}^{+0.46}$ & $113.24_{-46.38}^{+46.63}$ & $-0.17_{-1.07}^{+1.07}$ & $9.77_{-0.78}^{+0.86}$ & $0.51_{-1.69}^{+1.70}$ & $120.21_{-50.15}^{+49.60}$ & $-0.17_{-1.05}^{+1.05}$ & $9.76_{-0.77}^{+0.86}$ & $-0.68_{-0.32}^{+0.46}$ & $-1.98_{-0.26}^{+1.07}$ \\
                0    & 108 & $-0.21_{-1.06}^{+1.07}$ & $9.80_{-0.78}^{+0.87}$ & $0.15_{-0.47}^{+0.47}$ & $107.47_{-44.98}^{+45.13}$ & $-0.19_{-1.08}^{+1.07}$ & $9.83_{-0.78}^{+0.88}$ & $-0.31_{-1.69}^{+1.71}$ & $104.21_{-49.65}^{+49.05}$ & $-0.25_{-1.06}^{+1.07}$ & $9.85_{-0.77}^{+0.87}$ & $-0.70_{-0.54}^{+0.46}$ & $-1.79_{-0.43}^{+0.89}$ \\
                & 153 & $0.16_{-1.07}^{+1.06}$ & $9.88_{-0.78}^{+0.86}$ & $-0.15_{-0.44}^{+0.45}$ & $96.35_{-48.69}^{+48.46}$ & $0.13_{-1.08}^{+1.07}$ & $9.90_{-0.79}^{+0.86}$ & $-0.55_{-1.70}^{+1.68}$ & $98.83_{-49.92}^{+51.63}$ & $0.15_{-1.06}^{+1.06}$ & $9.86_{-0.78}^{+0.85}$ & $-0.65_{-0.51}^{+0.48}$ & $-1.93_{-0.30}^{+0.77}$ \\\hline
        \end{tabular}
\end{sidewaystable}

\begin{sidewaystable}[] 
        \centering
        \caption{Output results for catalogs created according to the solid-body rotation law and following L07 spatial positions and error distribution. Columns as in Table \ref{targs1}.} 
        \label{lewis1}
        \begin{tabular}{cr|rrrr|rrrr|rr|rr}
        \hline
        n & $\theta$ & V$_{sys}$ & $\sigma_v$ & \textit{k} & $\theta_k$ & V$_{sys}$ & $\sigma_v$  & V$_c$ & $\theta_{V_c}$ & V$_{sys}$ & $\sigma_v$ &  $\textrm{ln}B_{lin,flat}$ & $\textrm{ln}B_{rot,disp}$ \\ 
        & [$^\circ$] & [km/s]& [km/s] & [km/s/$'$] & [$^\circ$]& [km/s]&  [km/s] & [km/s] &  [$^\circ$]& [km/s]& [km/s]&  & \\ 
        \hline 
                & 63 & $-0.27_{-1.98}^{+1.94}$ & $10.35_{-1.42}^{+1.62}$ & $7.39_{-0.86}^{+0.92}$ & $63.89_{-8.97}^{+10.78}$ & $0.52_{-2.46}^{+2.40}$ & $15.26_{-1.70}^{+1.87}$ & $18.68_{-1.79}^{+0.97}$ & $79.07_{-10.79}^{+11.91}$ & $-5.98_{-2.73}^{+2.74}$ & $19.26_{-0.95}^{+0.54}$ & $15.94_{-4.68}^{+4.88}$ & $33.67_{-6.51}^{+6.74}$ \\
                2    & 108 & $-0.32_{-1.95}^{+1.97}$ & $10.14_{-1.42}^{+1.61}$ & $7.52_{-0.59}^{+0.63}$ & $106.93_{-11.58}^{+10.98}$ & $1.98_{-2.65}^{+2.61}$ & $16.79_{-1.69}^{+1.75}$ & $19.11_{-1.32}^{+0.65}$ & $99.01_{-11.17}^{+11.31}$ & $-3.54_{-2.80}^{+2.79}$ & $19.58_{-0.61}^{+0.31}$ & $22.13_{-5.52}^{+4.13}$ & $44.49_{-5.69}^{+5.87}$ \\
                & 153 & $0.38_{-1.93}^{+1.94}$ & $10.23_{-1.41}^{+1.60}$ & $7.25_{-1.16}^{+1.18}$ & $150.39_{-9.48}^{+6.90}$ & $3.30_{-2.42}^{+2.43}$ & $13.41_{-1.59}^{+1.82}$ & $17.31_{-2.54}^{+1.83}$ & $129.57_{-13.16}^{+11.36}$ & $1.07_{-2.60}^{+2.62}$ & $18.15_{-1.59}^{+1.21}$ & $9.65_{-3.63}^{+4.07}$ & $22.67_{-4.31}^{+4.44}$ \\ \rowcolor{Gray}
                & 63 & $-0.59_{-1.91}^{+1.91}$ & $9.78_{-1.42}^{+1.61}$ & $3.62_{-0.70}^{+0.90}$ & $71.71_{-18.14}^{+23.21}$ & $1.03_{-2.21}^{+2.18}$ & $11.16_{-1.48}^{+1.70}$ & $12.28_{-2.61}^{+2.66}$ & $76.93_{-17.01}^{+19.22}$ & $-3.23_{-2.19}^{+2.20}$ & $14.40_{-1.63}^{+1.81}$ & $3.84_{-2.81}^{+2.08}$ & $11.43_{-4.10}^{+4.36}$ \\ \rowcolor{Gray}
                1    & 108 & $0.05_{-1.94}^{+1.95}$ & $10.05_{-1.40}^{+1.58}$ & $4.04_{-0.62}^{+0.70}$ & $103.79_{-20.69}^{+19.60}$ & $2.16_{-2.24}^{+2.21}$ & $11.86_{-1.49}^{+1.70}$ & $13.96_{-2.49}^{+2.44}$ & $96.32_{-16.86}^{+16.96}$ & $-1.92_{-2.35}^{+2.39}$ & $15.76_{-1.68}^{+1.82}$ & $5.15_{-3.59}^{+2.64}$ & $15.39_{-4.18}^{+4.22}$ \\ \rowcolor{Gray}
                & 153 & $-0.07_{-1.98}^{+1.99}$ & $10.55_{-1.42}^{+1.62}$ & $3.33_{-1.08}^{+1.31}$ & $151.78_{-23.86}^{+14.03}$ & $1.30_{-2.16}^{+2.17}$ & $11.15_{-1.45}^{+1.64}$ & $9.36_{-2.91}^{+2.98}$ & $135.13_{-23.61}^{+19.67}$ & $0.38_{-2.02}^{+2.01}$ & $12.74_{-1.52}^{+1.73}$ & $1.58_{-2.09}^{+2.26}$ & $5.75_{-3.08}^{+3.19}$ \\
                & 63 & $-0.45_{-1.97}^{+1.94}$ & $10.41_{-1.42}^{+1.61}$ & $2.13_{-0.83}^{+0.99}$ & $67.67_{-28.92}^{+36.66}$ & $0.19_{-2.14}^{+2.17}$ & $11.05_{-1.46}^{+1.66}$ & $6.77_{-2.95}^{+3.03}$ & $69.96_{-28.29}^{+33.74}$ & $-1.64_{-1.92}^{+1.94}$ & $11.93_{-1.48}^{+1.68}$ & $0.75_{-1.29}^{+1.91}$ & $2.28_{-2.52}^{+3.30}$ \\
                0.5  & 108 & $-0.03_{-1.91}^{+1.90}$ & $10.15_{-1.40}^{+1.58}$ & $2.08_{-0.71}^{+0.82}$ & $105.10_{-36.76}^{+33.63}$ & $0.93_{-2.08}^{+2.09}$ & $10.58_{-1.43}^{+1.63}$ & $7.08_{-2.63}^{+2.66}$ & $97.74_{-30.44}^{+30.37}$ & $-0.88_{-1.89}^{+1.90}$ & $11.72_{-1.46}^{+1.67}$ & $0.51_{-1.50}^{+1.56}$ & $2.92_{-2.47}^{+3.10}$ \\
                & 153 & $-0.35_{-1.87}^{+1.89}$ & $9.95_{-1.39}^{+1.58}$ & $1.53_{-1.03}^{+1.18}$ & $152.37_{-40.14}^{+29.29}$ & $0.35_{-2.00}^{+2.03}$ & $10.13_{-1.41}^{+1.60}$ & $5.14_{-3.06}^{+3.02}$ & $134.21_{-39.53}^{+35.79}$ & $-0.20_{-1.79}^{+1.81}$ & $10.72_{-1.42}^{+1.60}$ & $-0.14_{-0.88}^{+1.27}$ & $0.40_{-1.44}^{+2.27}$ \\ \rowcolor{Gray}
                & 63 & $0.22_{-1.84}^{+1.86}$ & $9.79_{-1.39}^{+1.57}$ & $1.15_{-0.97}^{+1.03}$ & $52.75_{-42.67}^{+42.18}$ & $0.56_{-1.95}^{+1.97}$ & $9.84_{-1.39}^{+1.57}$ & $4.14_{-3.08}^{+3.05}$ & $64.54_{-40.58}^{+41.29}$ & $-0.65_{-1.70}^{+1.71}$ & $10.03_{-1.39}^{+1.58}$ & $-0.48_{-0.66}^{+0.78}$ & $-0.62_{-0.97}^{+2.38}$ \\ \rowcolor{Gray}
                0.25 & 108 & $-0.20_{-1.91}^{+1.90}$ & $10.27_{-1.41}^{+1.59}$ & $1.36_{-0.96}^{+0.99}$ & $85.78_{-41.55}^{+43.86}$ & $0.25_{-2.01}^{+2.02}$ & $10.20_{-1.41}^{+1.60}$ & $4.03_{-3.02}^{+2.93}$ & $96.09_{-42.66}^{+41.20}$ & $-0.71_{-1.80}^{+1.79}$ & $10.73_{-1.39}^{+1.60}$ & $-0.36_{-0.77}^{+1.51}$ & $-0.00_{-1.39}^{+2.67}$ \\ \rowcolor{Gray}
                & 153 & $-0.08_{-1.91}^{+1.93}$ & $10.34_{-1.42}^{+1.61}$ & $0.66_{-1.09}^{+1.07}$ & $130.96_{-45.67}^{+42.77}$ & $0.12_{-2.02}^{+2.01}$ & $10.33_{-1.42}^{+1.60}$ & $2.46_{-3.20}^{+3.12}$ & $92.82_{-45.84}^{+46.12}$ & $-0.22_{-1.76}^{+1.78}$ & $10.57_{-1.41}^{+1.59}$ & $-0.52_{-0.49}^{+0.65}$ & $-0.95_{-0.65}^{+1.41}$ \\
                & 63 & $0.27_{-1.81}^{+1.80}$ & $9.62_{-1.38}^{+1.54}$ & $-0.12_{-0.98}^{+0.97}$ & $53.35_{-45.40}^{+44.01}$ & $0.23_{-1.91}^{+1.91}$ & $9.64_{-1.37}^{+1.56}$ & $-0.01_{-3.04}^{+3.08}$ & $63.54_{-47.64}^{+47.70}$ & $0.17_{-1.68}^{+1.68}$ & $9.66_{-1.36}^{+1.53}$ & $-0.59_{-0.52}^{+0.30}$ & $-1.41_{-0.27}^{+0.82}$ \\
                0    & 108 & $0.09_{-1.85}^{+1.85}$ & $10.20_{-1.40}^{+1.56}$ & $0.05_{-0.99}^{+0.99}$ & $86.01_{-44.89}^{+45.79}$ & $-0.05_{-1.96}^{+1.97}$ & $10.16_{-1.40}^{+1.58}$ & $-0.04_{-3.19}^{+3.11}$ & $71.75_{-48.52}^{+48.85}$ & $-0.07_{-1.74}^{+1.75}$ & $10.31_{-1.38}^{+1.56}$ & $-0.59_{-0.38}^{+0.40}$ & $-1.30_{-0.35}^{+0.90}$ \\
                & 153 & $-0.28_{-1.80}^{+1.80}$ & $9.71_{-1.39}^{+1.56}$ & $0.09_{-1.01}^{+1.04}$ & $24.71_{-42.34}^{+45.87}$ & $-0.16_{-1.91}^{+1.89}$ & $9.72_{-1.39}^{+1.57}$ & $-0.49_{-3.02}^{+3.03}$ & $67.49_{-48.83}^{+48.11}$ & $-0.09_{-1.68}^{+1.68}$ & $9.75_{-1.35}^{+1.55}$ & $-0.55_{-0.28}^{+0.46}$ & $-1.43_{-0.29}^{+0.82}$ \\\hline
        \end{tabular}
\end{sidewaystable}

\begin{sidewaystable}
        \centering
        \caption{Output results for catalogs created according to the constant rotation law and following our targets spatial positions and error distribution. Columns as in Table \ref{targs1}.} 
        \label{targs2}
        \begin{tabular}{cr|rrrr|rrrr|rr|rr}
        \hline
        n & $\theta$ & V$_{sys}$ & $\sigma_v$ & \textit{k} & $\theta_k$ & V$_{sys}$ & $\sigma_v$  & V$_c$ & $\theta_{V_c}$ & V$_{sys}$ & $\sigma_v$ &  $\textrm{ln}B_{lin,flat}$ & $\textrm{ln}B_{rot,disp}$ \\ 
        & [$^\circ$] & [km/s]& [km/s] & [km/s/$'$] & [$^\circ$]& [km/s]&  [km/s] & [km/s] &  [$^\circ$]& [km/s]& [km/s]&  & \\ 
        \hline 
                & 63 & $0.03_{-2.20}^{+2.18}$ & $14.13_{-1.50}^{+1.74}$ & $3.96_{-0.68}^{+0.67}$ & $54.16_{-5.55}^{+5.81}$ & $-0.19_{-1.61}^{+1.60}$ & $9.83_{-1.19}^{+1.37}$ & $20.11_{-1.92}^{+1.93}$ & $64.12_{-8.29}^{+8.35}$ & $4.06_{-2.74}^{+2.73}$ & $19.17_{-1.90}^{+2.21}$ & $-11.12_{-5.02}^{+4.22}$ & $23.12_{-4.34}^{+5.93}$ \\
                2    & 108 & $-0.70_{-2.14}^{+2.14}$ & $13.79_{-1.49}^{+1.72}$ & $2.93_{-0.58}^{+0.60}$ & $74.48_{-8.39}^{+10.45}$ & $0.18_{-1.65}^{+1.65}$ & $10.12_{-1.20}^{+1.40}$ & $20.05_{-2.36}^{+2.39}$ & $106.73_{-7.73}^{+7.22}$ & $2.25_{-2.60}^{+2.62}$ & $18.20_{-1.80}^{+2.15}$ & $-9.39_{-2.78}^{+4.64}$ & $20.21_{-5.18}^{+4.77}$ \\
                & 153 & $-1.18_{-2.03}^{+2.03}$ & $12.87_{-1.42}^{+1.65}$ & $1.49_{-0.39}^{+0.40}$ & $149.33_{-22.77}^{+20.20}$ & $0.20_{-1.61}^{+1.64}$ & $10.10_{-1.20}^{+1.39}$ & $20.37_{-2.94}^{+2.94}$ & $154.16_{-5.67}^{+5.42}$ & $-0.94_{-2.19}^{+2.19}$ & $14.94_{-1.57}^{+1.83}$ & $-6.23_{-3.59}^{+4.03}$ & $12.48_{-4.68}^{+4.21}$ \\ \rowcolor{Gray}
                & 63 & $0.46_{-1.78}^{+1.78}$ & $11.08_{-1.26}^{+1.47}$ & $1.91_{-0.56}^{+0.55}$ & $54.87_{-9.58}^{+10.85}$ & $0.38_{-1.63}^{+1.63}$ & $9.98_{-1.18}^{+1.40}$ & $9.92_{-2.00}^{+2.01}$ & $62.58_{-15.71}^{+16.17}$ & $2.48_{-1.89}^{+1.88}$ & $12.67_{-1.35}^{+1.59}$ & $-2.12_{-2.97}^{+2.66}$ & $6.35_{-2.65}^{+3.78}$ \\ \rowcolor{Gray}
                1    & 108 & $-0.33_{-1.76}^{+1.76}$ & $11.00_{-1.28}^{+1.48}$ & $1.46_{-0.48}^{+0.50}$ & $76.49_{-13.58}^{+18.54}$ & $0.18_{-1.62}^{+1.62}$ & $9.95_{-1.19}^{+1.36}$ & $10.32_{-2.27}^{+2.34}$ & $106.86_{-15.56}^{+13.35}$ & $1.26_{-1.90}^{+1.90}$ & $12.61_{-1.36}^{+1.56}$ & $-1.79_{-2.51}^{+2.17}$ & $5.71_{-3.01}^{+4.68}$ \\ \rowcolor{Gray}
                & 153 & $-0.44_{-1.74}^{+1.74}$ & $11.01_{-1.27}^{+1.47}$ & $0.69_{-0.40}^{+0.39}$ & $131.11_{-36.53}^{+34.58}$ & $0.34_{-1.65}^{+1.66}$ & $10.13_{-1.22}^{+1.41}$ & $9.57_{-3.08}^{+3.03}$ & $154.88_{-12.54}^{+11.75}$ & $-0.39_{-1.75}^{+1.76}$ & $11.47_{-1.29}^{+1.50}$ & $0.10_{-2.38}^{+1.74}$ & $1.51_{-1.53}^{+3.10}$ \\
                & 63 & $0.42_{-1.76}^{+1.78}$ & $10.96_{-1.25}^{+1.45}$ & $1.21_{-0.56}^{+0.56}$ & $56.17_{-15.85}^{+18.91}$ & $0.24_{-1.65}^{+1.65}$ & $10.16_{-1.21}^{+1.40}$ & $7.70_{-2.15}^{+2.13}$ & $63.86_{-21.88}^{+22.23}$ & $1.80_{-1.78}^{+1.80}$ & $11.87_{-1.29}^{+1.51}$ & $-1.16_{-2.20}^{+2.11}$ & $2.46_{-2.35}^{+3.29}$ \\
                0.75 & 108 & $-0.21_{-1.70}^{+1.71}$ & $10.63_{-1.23}^{+1.43}$ & $0.90_{-0.46}^{+0.48}$ & $78.39_{-23.46}^{+30.01}$ & $0.06_{-1.63}^{+1.63}$ & $10.07_{-1.19}^{+1.37}$ & $7.36_{-2.35}^{+2.36}$ & $107.47_{-21.20}^{+17.70}$ & $0.57_{-1.72}^{+1.72}$ & $11.19_{-1.26}^{+1.45}$ & $-0.61_{-2.54}^{+1.72}$ & $1.82_{-2.28}^{+2.81}$ \\
                & 153 & $-0.82_{-1.66}^{+1.67}$ & $10.24_{-1.22}^{+1.41}$ & $0.52_{-0.40}^{+0.40}$ & $118.57_{-40.12}^{+37.14}$ & $-0.33_{-1.62}^{+1.62}$ & $10.11_{-1.20}^{+1.39}$ & $6.29_{-2.96}^{+2.92}$ & $150.79_{-20.46}^{+18.26}$ & $-0.53_{-1.66}^{+1.65}$ & $10.70_{-1.24}^{+1.43}$ & $0.69_{-2.18}^{+1.14}$ & $0.36_{-1.37}^{+1.98}$ \\ \rowcolor{Gray}
                & 63 & $0.14_{-1.71}^{+1.72}$ & $10.44_{-1.24}^{+1.43}$ & $0.94_{-0.52}^{+0.52}$ & $54.74_{-20.85}^{+23.54}$ & $0.11_{-1.65}^{+1.64}$ & $10.16_{-1.22}^{+1.40}$ & $5.80_{-2.33}^{+2.30}$ & $58.91_{-26.14}^{+28.16}$ & $1.02_{-1.69}^{+1.68}$ & $11.00_{-1.26}^{+1.46}$ & $-0.31_{-2.32}^{+1.14}$ & $0.87_{-2.17}^{+2.45}$ \\ \rowcolor{Gray}
                0.5  & 108 & $0.10_{-1.67}^{+1.67}$ & $10.38_{-1.22}^{+1.40}$ & $0.61_{-0.43}^{+0.45}$ & $73.86_{-33.85}^{+40.52}$ & $0.30_{-1.64}^{+1.65}$ & $10.17_{-1.21}^{+1.39}$ & $5.03_{-2.50}^{+2.45}$ & $104.56_{-32.44}^{+28.53}$ & $0.66_{-1.65}^{+1.65}$ & $10.62_{-1.23}^{+1.41}$ & $-0.01_{-2.23}^{+1.17}$ & $-0.22_{-1.40}^{+2.25}$ \\ \rowcolor{Gray}
                & 153 & $-0.59_{-1.64}^{+1.65}$ & $10.35_{-1.21}^{+1.40}$ & $0.29_{-0.41}^{+0.40}$ & $84.87_{-46.58}^{+47.83}$ & $-0.24_{-1.64}^{+1.61}$ & $10.07_{-1.20}^{+1.38}$ & $3.38_{-2.82}^{+2.86}$ & $149.17_{-33.81}^{+31.77}$ & $-0.42_{-1.60}^{+1.60}$ & $10.37_{-1.20}^{+1.38}$ & $0.17_{-1.57}^{+0.99}$ & $-0.93_{-1.02}^{+1.69}$ \\
                & 63 & $0.16_{-1.64}^{+1.64}$ & $10.08_{-1.20}^{+1.37}$ & $0.33_{-0.43}^{+0.42}$ & $61.63_{-47.40}^{+47.22}$ & $0.04_{-1.62}^{+1.63}$ & $10.00_{-1.19}^{+1.38}$ & $2.61_{-2.52}^{+2.45}$ & $67.88_{-46.56}^{+48.39}$ & $0.31_{-1.61}^{+1.57}$ & $10.23_{-1.19}^{+1.36}$ & $-0.50_{-0.52}^{+0.49}$ & $-1.53_{-0.61}^{+1.37}$ \\
                0.25 & 108 & $0.32_{-1.62}^{+1.62}$ & $10.04_{-1.19}^{+1.38}$ & $0.31_{-0.42}^{+0.43}$ & $64.26_{-45.77}^{+47.53}$ & $0.33_{-1.60}^{+1.61}$ & $9.91_{-1.20}^{+1.37}$ & $2.62_{-2.55}^{+2.51}$ & $109.65_{-45.84}^{+43.49}$ & $0.45_{-1.57}^{+1.58}$ & $10.11_{-1.18}^{+1.37}$ & $-0.47_{-1.30}^{+0.45}$ & $-1.50_{-0.60}^{+1.41}$ \\
                & 153 & $-0.67_{-1.61}^{+1.60}$ & $10.04_{-1.19}^{+1.36}$ & $0.21_{-0.41}^{+0.40}$ & $69.97_{-50.12}^{+48.15}$ & $-0.48_{-1.61}^{+1.61}$ & $9.90_{-1.20}^{+1.36}$ & $2.11_{-2.65}^{+2.66}$ & $127.60_{-45.92}^{+44.66}$ & $-0.70_{-1.56}^{+1.56}$ & $10.03_{-1.18}^{+1.36}$ & $-0.49_{-0.88}^{+0.62}$ & $-1.65_{-0.51}^{+1.68}$ \\\hline
        \end{tabular}
\end{sidewaystable}

\begin{sidewaystable}[] 
        \centering
        \caption{Output results for catalogs created according to the constant rotation law and following K14 spatial positions and error distribution. Columns as in Table \ref{targs1}.} 
        \label{kirby2}
        \begin{tabular}{cr|rrrr|rrrr|rr|rr}
        \hline
        n & $\theta$ & V$_{sys}$ & $\sigma_v$ & \textit{k} & $\theta_k$ & V$_{sys}$ & $\sigma_v$  & V$_c$ & $\theta_{V_c}$ & V$_{sys}$ & $\sigma_v$ &  $\textrm{ln}B_{lin,flat}$ & $\textrm{ln}B_{rot,disp}$ \\ 
        & [$^\circ$] & [km/s]& [km/s] & [km/s/$'$] & [$^\circ$]& [km/s]&  [km/s] & [km/s] &  [$^\circ$]& [km/s]& [km/s]&  & \\ 
        \hline 
                & 63 & $-0.40_{-1.34}^{+1.33}$ & $12.82_{-0.96}^{+1.07}$ & $4.14_{-0.44}^{+0.47}$ & $53.37_{-8.75}^{+9.81}$ & $-0.04_{-1.09}^{+1.09}$ & $9.91_{-0.79}^{+0.88}$ & $20.28_{-1.33}^{+1.35}$ & $62.21_{-5.42}^{+5.53}$ & $-1.14_{-1.87}^{+1.87}$ & $18.91_{-1.32}^{+1.46}$ & $-20.55_{-4.59}^{+4.85}$ & $59.09_{-7.20}^{+7.82}$ \\
                2    & 108 & $-1.58_{-1.34}^{+1.35}$ & $12.87_{-0.95}^{+1.07}$ & $3.84_{-0.46}^{+0.50}$ & $111.16_{-10.73}^{+9.27}$ & $-0.17_{-1.08}^{+1.08}$ & $9.76_{-0.79}^{+0.87}$ & $20.32_{-1.47}^{+1.49}$ & $108.19_{-5.27}^{+5.01}$ & $-2.33_{-1.79}^{+1.80}$ & $18.33_{-1.26}^{+1.42}$ & $-20.81_{-5.31}^{+5.85}$ & $54.33_{-7.62}^{+6.78}$ \\
                & 153 & $-1.69_{-1.24}^{+1.23}$ & $11.82_{-0.89}^{+0.99}$ & $4.61_{-0.68}^{+0.67}$ & $157.24_{-4.74}^{+4.20}$ & $-0.14_{-1.07}^{+1.07}$ & $9.74_{-0.78}^{+0.87}$ & $19.73_{-1.92}^{+1.94}$ & $153.19_{-3.94}^{+3.79}$ & $-2.09_{-1.48}^{+1.48}$ & $14.63_{-1.05}^{+1.17}$ & $-11.40_{-6.45}^{+4.26}$ & $32.41_{-5.91}^{+7.16}$ \\ \rowcolor{Gray} 
                & 63 & $0.09_{-1.13}^{+1.13}$ & $10.69_{-0.82}^{+0.92}$ & $2.08_{-0.37}^{+0.42}$ & $52.79_{-14.19}^{+17.43}$ & $0.29_{-1.09}^{+1.08}$ & $9.80_{-0.79}^{+0.87}$ & $10.14_{-1.34}^{+1.36}$ & $63.77_{-10.63}^{+11.38}$ & $-0.40_{-1.30}^{+1.31}$ & $12.65_{-0.94}^{+1.04}$ & $-4.27_{-3.58}^{+4.40}$ & $18.53_{-4.34}^{+6.32}$ \\ \rowcolor{Gray}
                1    & 108 & $-0.85_{-1.13}^{+1.14}$ & $10.62_{-0.83}^{+0.93}$ & $1.91_{-0.37}^{+0.42}$ & $111.09_{-18.59}^{+15.53}$ & $-0.23_{-1.08}^{+1.07}$ & $9.69_{-0.79}^{+0.86}$ & $9.81_{-1.44}^{+1.49}$ & $107.30_{-11.13}^{+10.05}$ & $-1.35_{-1.28}^{+1.29}$ & $12.39_{-0.92}^{+1.02}$ & $-4.43_{-3.34}^{+3.37}$ & $16.82_{-5.57}^{+5.90}$ \\ \rowcolor{Gray}
                & 153 & $-0.78_{-1.12}^{+1.12}$ & $10.58_{-0.82}^{+0.91}$ & $2.29_{-0.64}^{+0.63}$ & $158.97_{-9.24}^{+7.78}$ & $-0.03_{-1.11}^{+1.10}$ & $10.05_{-0.79}^{+0.88}$ & $9.71_{-1.97}^{+1.97}$ & $153.42_{-8.64}^{+7.79}$ & $-1.04_{-1.19}^{+1.21}$ & $11.41_{-0.87}^{+0.95}$ & $-2.13_{-2.64}^{+3.14}$ & $7.75_{-4.27}^{+4.60}$ \\
                & 63 & $-0.09_{-1.11}^{+1.11}$ & $10.42_{-0.82}^{+0.91}$ & $1.58_{-0.35}^{+0.38}$ & $59.14_{-18.94}^{+22.04}$ & $-0.02_{-1.09}^{+1.08}$ & $9.90_{-0.79}^{+0.88}$ & $7.85_{-1.35}^{+1.37}$ & $63.25_{-14.15}^{+14.64}$ & $-0.48_{-1.21}^{+1.21}$ & $11.55_{-0.87}^{+0.97}$ & $-1.23_{-3.21}^{+2.52}$ & $10.25_{-4.39}^{+4.81}$ \\
                0.75 & 108 & $-0.83_{-1.10}^{+1.09}$ & $10.16_{-0.80}^{+0.89}$ & $1.45_{-0.36}^{+0.39}$ & $108.35_{-24.02}^{+19.29}$ & $-0.30_{-1.08}^{+1.06}$ & $9.64_{-0.78}^{+0.87}$ & $7.32_{-1.42}^{+1.47}$ & $106.28_{-14.91}^{+13.01}$ & $-1.23_{-1.19}^{+1.17}$ & $11.18_{-0.86}^{+0.95}$ & $-1.84_{-2.82}^{+2.68}$ & $8.86_{-3.75}^{+5.61}$ \\
                & 153 & $-0.56_{-1.09}^{+1.08}$ & $10.00_{-0.80}^{+0.89}$ & $1.62_{-0.61}^{+0.62}$ & $157.24_{-14.17}^{+10.97}$ & $0.12_{-1.08}^{+1.08}$ & $9.83_{-0.78}^{+0.87}$ & $7.13_{-2.01}^{+1.98}$ & $151.62_{-11.38}^{+10.02}$ & $-0.75_{-1.13}^{+1.13}$ & $10.65_{-0.82}^{+0.90}$ & $-0.33_{-2.66}^{+2.14}$ & $3.38_{-2.83}^{+3.45}$ \\ \rowcolor{Gray}
                & 63 & $-0.30_{-1.06}^{+1.07}$ & $9.90_{-0.79}^{+0.86}$ & $1.03_{-0.35}^{+0.38}$ & $62.73_{-27.57}^{+31.56}$ & $-0.26_{-1.07}^{+1.06}$ & $9.62_{-0.78}^{+0.86}$ & $4.84_{-1.41}^{+1.41}$ & $65.80_{-20.73}^{+22.73}$ & $-0.54_{-1.10}^{+1.10}$ & $10.33_{-0.81}^{+0.90}$ & $-0.04_{-2.26}^{+1.59}$ & $3.19_{-2.32}^{+3.55}$ \\ \rowcolor{Gray}
                0.5  & 108 & $-0.39_{-1.07}^{+1.09}$ & $10.09_{-0.80}^{+0.89}$ & $1.06_{-0.38}^{+0.41}$ & $106.68_{-31.64}^{+26.60}$ & $-0.15_{-1.09}^{+1.09}$ & $9.93_{-0.79}^{+0.87}$ & $5.18_{-1.49}^{+1.50}$ & $104.52_{-22.07}^{+19.03}$ & $-0.64_{-1.12}^{+1.12}$ & $10.53_{-0.82}^{+0.91}$ & $-0.37_{-2.15}^{+2.29}$ & $3.02_{-3.01}^{+4.36}$ \\ \rowcolor{Gray}
                & 153 & $-0.54_{-1.07}^{+1.07}$ & $9.93_{-0.79}^{+0.88}$ & $0.95_{-0.58}^{+0.60}$ & $155.26_{-25.69}^{+18.46}$ & $-0.15_{-1.08}^{+1.06}$ & $9.79_{-0.78}^{+0.87}$ & $4.15_{-2.00}^{+2.00}$ & $154.80_{-20.86}^{+18.31}$ & $-0.70_{-1.09}^{+1.10}$ & $10.24_{-0.80}^{+0.89}$ & $0.28_{-1.65}^{+1.20}$ & $0.27_{-1.79}^{+2.76}$ \\
                & 63 & $0.18_{-1.06}^{+1.07}$ & $9.91_{-0.78}^{+0.88}$ & $0.59_{-0.43}^{+0.43}$ & $62.58_{-43.06}^{+42.54}$ & $0.22_{-1.07}^{+1.07}$ & $9.88_{-0.79}^{+0.87}$ & $2.60_{-1.60}^{+1.54}$ & $74.74_{-36.83}^{+38.19}$ & $0.01_{-1.09}^{+1.08}$ & $10.06_{-0.79}^{+0.88}$ & $-0.02_{-1.22}^{+0.75}$ & $-0.69_{-1.51}^{+2.73}$ \\
                0.25 & 108 & $-0.34_{-1.09}^{+1.07}$ & $9.99_{-0.79}^{+0.87}$ & $0.53_{-0.45}^{+0.44}$ & $118.05_{-45.41}^{+44.24}$ & $-0.09_{-1.09}^{+1.08}$ & $9.91_{-0.79}^{+0.87}$ & $2.61_{-1.68}^{+1.67}$ & $110.81_{-39.89}^{+34.31}$ & $-0.42_{-1.09}^{+1.09}$ & $10.08_{-0.80}^{+0.87}$ & $-0.25_{-1.37}^{+0.68}$ & $-1.05_{-1.13}^{+2.23}$ \\
                & 153 & $-0.35_{-1.06}^{+1.06}$ & $9.85_{-0.78}^{+0.87}$ & $0.43_{-0.48}^{+0.51}$ & $148.44_{-43.31}^{+39.82}$ & $-0.14_{-1.08}^{+1.08}$ & $9.81_{-0.79}^{+0.87}$ & $2.00_{-1.83}^{+1.88}$ & $143.13_{-38.86}^{+35.86}$ & $-0.40_{-1.06}^{+1.06}$ & $10.04_{-0.78}^{+0.87}$ & $-0.12_{-0.99}^{+0.65}$ & $-1.64_{-0.72}^{+2.28}$ \\\hline
        \end{tabular}
\end{sidewaystable}

\begin{sidewaystable}[] 
        \centering
        \caption{Output results for catalogs created according to the constant rotation law and following L07 spatial positions and error distribution. Columns as in Table \ref{targs1}.} 
        \label{lewis2}
        \begin{tabular}{cr|rrrr|rrrr|rr|rr}
        \hline
        n & $\theta$ & V$_{sys}$ & $\sigma_v$ & \textit{k} & $\theta_k$ & V$_{sys}$ & $\sigma_v$  & V$_c$ & $\theta_{V_c}$ & V$_{sys}$ & $\sigma_v$ &  $\textrm{ln}B_{lin,flat}$ & $\textrm{ln}B_{rot,disp}$ \\ 
        & [$^\circ$] & [km/s]& [km/s] & [km/s/$'$] & [$^\circ$]& [km/s]&  [km/s] & [km/s] &  [$^\circ$]& [km/s]& [km/s]&  & \\ 
        \hline 
                & 63 & $-3.44_{-2.31}^{+2.27}$ & $12.57_{-1.55}^{+1.78}$ & $4.54_{-1.04}^{+1.28}$ & $59.54_{-14.37}^{+20.86}$ & $0.47_{-2.09}^{+2.11}$ & $10.33_{-1.39}^{+1.59}$ & $20.45_{-2.79}^{+2.92}$ & $63.85_{-8.95}^{+10.41}$ & $-7.12_{-2.54}^{+2.56}$ & $17.54_{-1.85}^{+2.15}$ & $-3.81_{-3.25}^{+3.82}$ & $17.73_{-5.24}^{+4.53}$ \\
                2    & 108 & $-4.13_{-2.27}^{+2.29}$ & $12.93_{-1.56}^{+1.79}$ & $4.61_{-0.76}^{+0.89}$ & $117.36_{-21.94}^{+18.63}$ & $-0.28_{-2.06}^{+2.05}$ & $10.12_{-1.42}^{+1.60}$ & $20.21_{-2.22}^{+2.22}$ & $107.58_{-11.01}^{+10.86}$ & $-5.46_{-2.71}^{+2.73}$ & $18.55_{-1.97}^{+2.27}$ & $-4.89_{-4.77}^{+3.53}$ & $21.31_{-4.89}^{+5.69}$ \\
                & 153 & $-1.19_{-2.16}^{+2.15}$ & $11.79_{-1.50}^{+1.74}$ & $5.80_{-1.48}^{+1.52}$ & $164.20_{-10.67}^{+7.21}$ & $0.41_{-2.05}^{+2.02}$ & $9.94_{-1.40}^{+1.60}$ & $19.20_{-3.19}^{+3.36}$ & $153.51_{-9.44}^{+7.74}$ & $0.45_{-2.29}^{+2.28}$ & $15.09_{-1.73}^{+1.98}$ & $-2.01_{-3.26}^{+3.79}$ & $12.89_{-4.23}^{+4.61}$ \\ \rowcolor{Gray}
                & 63 & $-2.33_{-2.00}^{+2.03}$ & $10.75_{-1.44}^{+1.64}$ & $2.36_{-0.80}^{+1.00}$ & $61.83_{-26.15}^{+36.42}$ & $-0.47_{-2.10}^{+2.09}$ & $10.34_{-1.41}^{+1.62}$ & $9.81_{-2.74}^{+2.91}$ & $64.25_{-17.86}^{+22.93}$ & $-3.90_{-1.99}^{+1.98}$ & $12.56_{-1.51}^{+1.72}$ & $1.16_{-2.89}^{+1.68}$ & $4.85_{-2.86}^{+2.87}$ \\ \rowcolor{Gray}
                1    & 108 & $-1.96_{-2.05}^{+2.04}$ & $11.39_{-1.46}^{+1.68}$ & $2.38_{-0.72}^{+0.86}$ & $105.73_{-33.74}^{+32.36}$ & $-0.24_{-2.08}^{+2.09}$ & $10.44_{-1.43}^{+1.62}$ & $10.42_{-2.44}^{+2.53}$ & $103.67_{-20.94}^{+21.02}$ & $-2.71_{-2.02}^{+2.03}$ & $12.90_{-1.55}^{+1.78}$ & $0.10_{-2.44}^{+2.51}$ & $5.53_{-2.76}^{+3.61}$ \\ \rowcolor{Gray}
                & 153 & $-0.01_{-1.97}^{+1.98}$ & $10.62_{-1.44}^{+1.63}$ & $2.35_{-1.28}^{+1.45}$ & $162.19_{-28.59}^{+15.82}$ & $0.33_{-2.03}^{+2.03}$ & $10.05_{-1.40}^{+1.60}$ & $9.34_{-3.13}^{+3.34}$ & $149.88_{-22.72}^{+16.30}$ & $0.41_{-1.89}^{+1.87}$ & $11.57_{-1.46}^{+1.71}$ & $0.87_{-1.84}^{+1.36}$ & $2.69_{-1.87}^{+2.48}$ \\
                & 63 & $-1.66_{-1.98}^{+1.99}$ & $10.71_{-1.44}^{+1.64}$ & $1.85_{-1.02}^{+1.16}$ & $47.68_{-28.56}^{+39.59}$ & $-0.48_{-2.09}^{+2.10}$ & $10.49_{-1.42}^{+1.61}$ & $7.12_{-3.09}^{+3.20}$ & $59.19_{-24.25}^{+30.69}$ & $-2.90_{-1.87}^{+1.86}$ & $11.39_{-1.45}^{+1.65}$ & $1.42_{-1.85}^{+1.05}$ & $1.75_{-1.90}^{+2.48}$ \\
                0.75 & 108 & $-1.30_{-1.95}^{+1.97}$ & $10.42_{-1.43}^{+1.62}$ & $2.01_{-0.80}^{+0.93}$ & $104.04_{-38.23}^{+34.57}$ & $-0.02_{-2.06}^{+2.02}$ & $10.08_{-1.42}^{+1.60}$ & $8.22_{-2.51}^{+2.55}$ & $105.57_{-26.14}^{+25.26}$ & $-1.87_{-1.89}^{+1.86}$ & $11.77_{-1.47}^{+1.67}$ & $0.88_{-1.87}^{+1.63}$ & $3.27_{-2.50}^{+2.79}$ \\
                & 153 & $-0.76_{-1.95}^{+1.98}$ & $10.44_{-1.40}^{+1.63}$ & $1.77_{-1.22}^{+1.37}$ & $156.57_{-33.91}^{+21.93}$ & $-0.32_{-2.05}^{+2.01}$ & $10.29_{-1.41}^{+1.61}$ & $6.88_{-3.19}^{+3.34}$ & $149.72_{-31.14}^{+23.62}$ & $-0.31_{-1.84}^{+1.84}$ & $11.25_{-1.44}^{+1.64}$ & $1.63_{-1.78}^{+0.84}$ & $1.42_{-1.63}^{+2.42}$ \\ \rowcolor{Gray}
                & 63 & $-1.52_{-1.90}^{+1.92}$ & $10.59_{-1.41}^{+1.59}$ & $1.15_{-1.03}^{+1.00}$ & $54.31_{-41.70}^{+44.78}$ & $-0.80_{-2.03}^{+2.04}$ & $10.42_{-1.40}^{+1.59}$ & $5.01_{-3.04}^{+3.00}$ & $66.02_{-36.27}^{+40.91}$ & $-1.94_{-1.80}^{+1.78}$ & $10.83_{-1.41}^{+1.59}$ & $1.07_{-1.27}^{+0.60}$ & $-0.10_{-0.96}^{+1.86}$ \\ \rowcolor{Gray}
                0.5  & 108 & $-0.74_{-1.93}^{+1.94}$ & $10.57_{-1.42}^{+1.63}$ & $1.32_{-0.98}^{+0.96}$ & $122.71_{-43.13}^{+39.98}$ & $-0.10_{-2.02}^{+2.03}$ & $10.42_{-1.42}^{+1.62}$ & $5.97_{-2.78}^{+2.82}$ & $109.59_{-37.08}^{+34.12}$ & $-1.20_{-1.84}^{+1.82}$ & $11.02_{-1.42}^{+1.63}$ & $0.79_{-1.02}^{+0.80}$ & $0.42_{-1.32}^{+2.34}$ \\ \rowcolor{Gray}
                & 153 & $-0.37_{-1.88}^{+1.88}$ & $10.20_{-1.41}^{+1.58}$ & $0.64_{-1.13}^{+1.08}$ & $132.20_{-42.35}^{+38.03}$ & $-0.21_{-1.97}^{+1.97}$ & $10.22_{-1.41}^{+1.60}$ & $3.86_{-3.28}^{+3.18}$ & $136.71_{-42.10}^{+40.23}$ & $-0.24_{-1.74}^{+1.76}$ & $10.37_{-1.40}^{+1.58}$ & $1.21_{-0.76}^{+0.70}$ & $-0.31_{-0.86}^{+1.43}$ \\
                & 63 & $-0.80_{-1.89}^{+1.89}$ & $10.37_{-1.40}^{+1.59}$ & $0.49_{-1.06}^{+1.08}$ & $41.89_{-42.91}^{+42.86}$ & $-0.58_{-2.00}^{+1.99}$ & $10.28_{-1.40}^{+1.58}$ & $2.54_{-3.23}^{+3.27}$ & $64.70_{-46.50}^{+47.11}$ & $-1.07_{-1.75}^{+1.75}$ & $10.40_{-1.38}^{+1.56}$ & $0.19_{-0.50}^{+0.45}$ & $-1.27_{-0.47}^{+1.40}$ \\
                0.25 & 108 & $-0.18_{-1.89}^{+1.87}$ & $10.21_{-1.40}^{+1.59}$ & $0.58_{-0.99}^{+0.97}$ & $75.43_{-47.99}^{+46.13}$ & $0.24_{-2.00}^{+2.00}$ & $10.15_{-1.41}^{+1.60}$ & $2.82_{-3.04}^{+2.98}$ & $101.18_{-48.34}^{+47.40}$ & $-0.33_{-1.78}^{+1.73}$ & $10.49_{-1.40}^{+1.60}$ & $0.19_{-0.66}^{+0.39}$ & $-1.16_{-0.52}^{+1.04}$ \\
                & 153 & $0.23_{-1.92}^{+1.92}$ & $10.40_{-1.41}^{+1.61}$ & $0.56_{-1.06}^{+1.10}$ & $98.95_{-43.41}^{+39.65}$ & $0.14_{-2.00}^{+2.00}$ & $10.59_{-1.40}^{+1.61}$ & $2.25_{-3.36}^{+3.27}$ & $134.97_{-48.07}^{+44.17}$ & $0.08_{-1.79}^{+1.77}$ & $10.78_{-1.40}^{+1.59}$ & $0.26_{-0.48}^{+0.57}$ & $-1.24_{-0.40}^{+1.56}$ \\\hline
        \end{tabular}
\end{sidewaystable}

\end{appendix}

\end{document}